\documentclass[rmp,aps,twocolumn]{revtex4-1}

\usepackage{graphicx}
\usepackage{amsmath}
\usepackage{amssymb}
\usepackage{wasysym}
\usepackage{multirow}
\usepackage{meta-donnees}
\usepackage{upgreek}
\usepackage{hyperref}
\hypersetup{
	pdftex,
	pdfauthor={Guillaume Pignol},
	pdfstartview=FitH,
	pdfpagemode=UseNone,
	colorlinks,
	citecolor=[rgb]{0.2,0,0.6}, 
	linkcolor=blue
	}

\newcommand{\para}[1]{\vspace{0.2cm} \textbf{#1}}
\newcommand{\ket}[1]{\left| #1 \right>} 
\newcommand{\bra}[1]{\left< #1 \right|} 
\newcommand{\braket}[2]{\left< #1 \vphantom{#2} \right| \left. #2 \vphantom{#1} \right>} 
\newcommand{\micron}{\upmu {\rm m}}
\newcommand{\separe}{\bigskip \begin{center} \photon \end{center} \bigskip}

\begin{document}


\title{Probing Dark Energy models with neutrons}

\author{Guillaume Pignol}
\affiliation{LPSC, Universit\'e  Grenoble Alpes, CNRS/IN2P3, Grenoble, France}
\email{guillaume.pignol@lpsc.in2p3.fr}

\date{\today}

\begin{abstract}
There is a deep connection between cosmology -- the science of the infinitely large 
--and particle physics -- the science of the infinitely small. 
This connection is particularly manifest in neutron particle physics. 
Basic properties of the neutron -- its Electric Dipole Moment and its lifetime -- are intertwined with baryogenesis and nucleosynthesis in the early Universe. 
I will cover this topic in the first part, that will also serve as an introduction (or rather a quick recap) of neutron physics and Big Bang cosmology. 
Then, the rest of the manuscript will be devoted to a new idea: using neutrons to probe models of Dark Energy. 
In the second part, I will present the chameleon theory: a light scalar field accounting for the late accelerated expansion of the Universe, 
which interacts with matter in such a way that it does not mediate a fifth force between macroscopic bodies. 
However, neutrons can alleviate the chameleon mechanism and reveal the presence of the scalar field with properly designed experiments. 
In the third part, I will describe a recent experiment performed with a neutron interferometer at the Institut Laue Langevin that sets already interesting constraints on the chameleon theory. 
Last, the chameleon field can be probed by measuring the quantum states of neutrons bouncing over a mirror. 
In the fourth part I will present the status and prospects of the GRANIT experiment at the ILL. 
\end{abstract}

\maketitle
\setcounter{page}{1}
\tableofcontents


\section{The interplay between neutron particle physics and cosmology}

There is a profound connection between cosmology -- the science of the infinitely large --
and particle physics -- the science of the infinitely small. 
The primordial creation of the light nuclei is an example: 
a detailed knowledge of nuclear interactions is needed to predict the outcome of the Big Bang nucleosynthesis. 
Conversely, cosmological observations have a great impact on our understanding of the laws of Nature at the most fundamental level. 
The existence of Dark Matter, Dark Energy and the matter-antimatter asymmetry provide three glimpses of what is hiding beyond the Standard Model of particle physics. 
It is likely that future progress in particle physics will proceed from the solution of one of these cosmological puzzles, or vice versa. 

The deep connection between particle physics and cosmology is particularly manifest in neutron particle physics, 
when we use low energy neutrons to investigate the fundamental interactions and symmetries. 
The work presented in the next chapters builds on this tradition. 
We will contemplate the possibility to probe models of Dark Energy with neutron experiments. 
Before developing on this original idea, we review in the present chapter the traditional interplay between neutron particle physics and cosmology. 
A comprehensive treatment of this topic can be found in the excellent review of \textcite{Dubbers2011}. 
See also \textcite{Dubbers2014} for a shorter and more recent overview. 

There are two fundamental properties of the neutron connected to the physics of the early Universe. 
First, the value of the {\bf neutron lifetime} is a basic input for the theory of Big Bang nucleosynthesis. 
Second, the search for the {\bf neutron electric dipole moment} is key to elucidate the origin of the baryon asymmetry of the Universe. 
We will explain these two subjects, after a brief review of neutron particle physics and cosmology. 

\subsection{Free up neutrons}

Neutrons make up about half the mass of ordinary objects, like bicycles, cheese or elephants. 
Even an empty bottle, that is, full of air, contains more than $10^{20}$ neutrons per cm$^3$. 
These neutrons are bound into nuclei since billions of years. 
On the contrary free neutrons do not exist naturally because they decay with a radioactive period of ten minutes. 
It takes a significant effort to extract neutrons from nuclei; it is much more difficult than extracting electrons from atoms. 
Indeed the binding energy of neutrons in nuclei is of the order of 1~MeV whereas the chemical binding energy is only about a few electron-Volts. 
Thus, no chemical reaction would set neutrons free: a nuclear reaction is needed. 

\para{Neutron sources. } 
About twenty large nuclear installations dedicated to neutron production are presently being operated over the world. 
These installations fall into two categories: fission nuclear reactors and spallation sources. 

The reactor of the Institut Laue Langevin (ILL) in Grenoble belongs to the first category. 
Neutrons are produced by fission in a compact cylindrical core (containing 9 kg of weapons grade uranium, 93 \%  $^{235}$U) and moderated in heavy water. 
Each fission ejects on average 2.5 neutrons with an energy of 2~MeV. 
The reactor design optimizes the thermal neutron flux, reaching $10^{15}$~cm$^{-2}$~s$^{-1}$ with a reactor thermal power of 58~MW. 
In a typical nuclear reactor used to produce electricity with a thermal power of 3000~MW, the neutron flux is about $10^{14}$~cm$^{-2}$~s$^{-1}$. 
Among the 246 operating research reactors listed by the International Atomic Energy Agency, 
46 of them are high flux reactors (above $10^{14}$ neutrons cm$^{-2}$~s$^{-1}$), 
11 of them feature a cold neutron source and operate instruments available to external users (see Table \ref{listReactors}). 

\begin{table}
\center
\caption{
High flux reactors producing beams of cold neutrons. 
The number of instruments available for outside users is indicated. 
}
\scriptsize
\begin{tabular}{lllll}
\hline
Reactor & City 			& Started	& Th. Power & Instruments \\
\hline
ILL	& Grenoble (France)	& 1971		& 58 MW		& 45	\\
HFIR	& Oak Ridge (USA)	& 1965		& 85 MW		& 12 	\\
CARR	& Bejing (China)	& 2010		& 60 MW		& 12	\\
FRM II	& Munich (Germany)	& 2004		& 20 MW		& 19	\\
HANARO	& Daejon (Korea)	& 1995		& 30 MW		& 8	\\
WWR-M	& Gatchina (Russia)	& 1959		& 18 MW		& 19	\\
NIST	& Gaithersburg (USA)	& 1967		& 20 MW		& 28	\\
JRR-3M	& Tokai	(Japan)		& 1990		& 20 MW		& 26	\\
BRR	& Budapest (Hungary)	& 1959		& 10 MW		& 11	\\
OPAL	& Sydney (Australia)	& 2006		& 20 MW		& 7	\\
BER II	& Berlin (Germany)	& 1973		& 10 MW		& 22
\end{tabular}
\label{listReactors}
\end{table}

The ultracold neutron facility at the Paul Scherrer Institute (PSI) in Switzerland uses a neutron source based on the spallation process. 
A proton beam, with an energy of 600~MeV and an intensity of 2.5~mA is shot at a lead target. 
When a proton collides with a heavy nucleus (lead in this case), it ejects about 20 neutrons with an energy of about 20~MeV along with other nuclear fragments. 
The neutrons are then moderated in a heavy water tank surrounding the spallation target. 
Spallation sources are now competing with reactors as neutron factories, Table \ref{listSpallation} lists the spallation sources 
equipped with neutron instruments available to external users. 

\begin{table}
\center
\caption{
Neutron spallation sources serving instruments available to users. 
}
\scriptsize
\begin{tabular}{llll}
\hline
Spall. Source 	& City 		& Beam Power & Instruments \\
\hline
SINQ @PSI	& Villigen (Switzerland)	& 1.5 MW	& 20 \\
SNS @ORNL	& Oak Ridge (USA)		& 1.4 MW	& 25 \\
JSNS @KEK	& Tsukuba (Japan)		& 0.3 MW	& 17 \\
ISIS @RAL	& Oxford (UK)			& 0.2 MW	& 36 \\
Lujan @LANSCE	& Los Alamos (USA)		& 0.1 MW	& 13
\end{tabular}
\label{listSpallation}
\end{table}

What is the purpose of investing so much efforts and money for building sources of free neutrons? 
What is special about free neutrons that cannot be done with the neutrons inside nuclei? 
Beyond doubt the main desirable feature of the neutron is its electrical neutrality
\footnote{The particle data group \cite{PDG} evaluation of the neutron charge is $(-2 \pm 8) \times 10^{-22}$ electron charges, in agreement with electrical neutrality.}.
Even atoms, which have no net electric charge, interact mainly with electromagnetic interactions due to their relatively large polarisability. 
Atomic processes are essentially unaffected by nuclear or gravitational interactions and all phenomena we experience in everyday life are ultimately due to basic electromagnetic interactions (except for the rather trivial effect of weight). 
Using neutrons one can probe matter using non electromagnetic interactions. 
Studying the diffraction of neutrons at a sample material, 
solid-state physicists are able to obtain information about the structure of the sample 
which is complementary to the information they get using electromagnetic probes such as X-ray diffraction. 

Particle physicists also find neutron's sensitivity to non-electromagnetic forces quite useful to do fundamental physics. 
Studying free neutron beta decay we measure the fundamental parameters of the weak interaction. 
Measuring neutron bounces we explore gravity with quantum objects. 
The search for the neutron electric dipole moment constitutes a crucial test of the time reversal invariance in elementary processes. 
One can also use neutrons to search for new, yet undiscovered interactions. 
All these experiments would be impossible with charged particles for which the Coulomb force overwhelms any other force by many orders of magnitude. 
We shall come back to these subjects in the rest of this manuscript. 

At the ILL, 38 instruments are dedicated to the study of material structure, condensed matter and magnetism and 7 instruments are dedicated to quantum, nuclear and particle physics. 
For a broad exploration of the physics of slow neutrons, see \textcite{Byrne1994}. 

\para{Wave particle duality. }
Experiments very often benefit from using slow neutrons. 
Those neutrons result from the thermalisation of the primary fast neutrons in a thermal or cold moderator. 
In a moderator, a neutron will dissipate its energy by collision until it reaches the kinetic energy ($kT = 25$~meV at room temperature) of the molecules. 
The De-Broglie wavelength $\lambda$ of a neutron with kinetic energy $E$ is  
\begin{equation}
\lambda = \frac{2 \pi \hbar}{\sqrt{2 m E}} = 0.18 \ {\rm nm} \times \sqrt{\frac{25 \ {\rm meV}}{E}}, 
\end{equation}
where $m = 939.6$~MeV/c${^2}$ is the neutron mass. 
It is remarkable that the wavelength of thermal neutrons corresponds to the typical distance between atoms in solid matter. 
Thus, whereas fast neutrons behave essentially as particles, cold neutrons could behave like waves. 

E.~Fermi was the first to realize that slow neutrons could undergo optical phenomena such as reflection and refraction at surfaces. 
It can be shown (see for example \textcite{Golub1991}) that matter acts as a uniform medium with a potential energy for slow neutrons called the \emph{Fermi potential}. 
The Fermi potential is given by the following expression: 
\begin{equation}
V_F = 4 \pi \ \frac{\hbar^2}{2 m} \ b \, n, 
\label{FermiPot}
\end{equation}
where $b$ is the bound coherent neutron scattering length of the nuclei constituting the material and $n$ is its number density. 
If the material is heterogeneous one must sum the potentials of all nuclear species composing the material. 
For most materials the Fermi potential is positive, i.e. repulsive 
\footnote{
This is surprising because the strong nuclear interaction responsible for the Fermi potential is attractive: it holds neutrons inside nuclei. 
For an explanation of this apparent paradox, see for example \textcite{Pignol2009}.}; 
it is of the order of $10^{-7}$~eV, much smaller than the kinetic energy of thermal neutrons. 
For example, natural nickel is a material with a relatively high Fermi potential which amounts to $250$~neV. 

Neutrons approaching a surface at grazing incidence could be reflected by the Fermi potential of the material. 
Total reflection occurs when the kinetic energy associated with the velocity normal to the surface is smaller than the potential barrier \eqref{FermiPot}. 
In practice this phenomenon is at play in neutron guides. 
It is possible to transport neutrons from the core of the reactor where they are produced to an experimental hall situated at a distance of up to 100~m, 
using evacuated rectangular tubes, with a cross section of typically 100~cm$^2$, made up of plates coated with nickel or with a multilayer of nickel and titanium. 
Thermal neutrons with an energy of $E = 25$~meV are reflected off a nickel surface if the angle of incidence $\theta$ 
satisfies the condition $\sin^2 \theta < V_F / E$, that is, $\theta < 0.2 \deg$. 
For colder neutrons, the efficiency of the guiding is better and a beam of cold neutrons transported in a guide can have a larger angular divergency. 

\para{Ultracold neutrons. }
\textcite{Zeldovich1959} realized that neutrons with total kinetic energy lower than $250$~neV 
should be reflected at any angle of incidence and therefore could be stored in material bottles. 
These storable neutrons were called \emph{ultracold neutrons (UCNs)} by the pioneer workers in the field. 
It is not easy to get ultracold neutrons, because their proportion in the energy spectrum of thermal neutrons 
(when neutrons are thermalised in a moderator at room temperature) is only $10^{-11}$. 
The first storage of ultracold neutrons was reported by \textcite{Groshev1971}. 
The group of Russian physicists stored on average 2 neutrons at a time in a copper cylindrical bottle (diameter 14~cm and length 174~cm). 
The capacity of the bottle to store neutrons was quantified by a storage time of about 30~s (compare to the neutron beta decay lifetime of 880~s). 

The kinetic energy of ultracold neutrons corresponds to a temperature as low as a few mK. 
However, it is important to understand that ultracold neutrons can be stored in bottles at room temperature. 
The neutrons do not thermalise with the walls of the bottle, like photons of visible light do not thermalise when reflecting off a mirror. 
In fact the specular (i.e. mirror) reflection of ultracold neutrons and visible photons share many similarities, because the wavelength of UCNs is of the order of 100~nm, 
very close to the wavelength of visible light. 
In both cases the wavelength is much larger than the lattice spacing of atoms in the matter of the walls. 
It means that the particle interacts with a large number of atoms in the wall, it is almost blind to the thermal motion of individual atoms. 
Using appropriate wall materials, UCNs can undergo as many as $10^4$ specular reflections before being inelastically scattered or captured by a single nucleus. 

The ability to store neutrons for a long period was immediately recognized as a great opportunity for fundamental physics, for two main reasons. 
First, ultracold neutrons provide a direct method to measure the neutron lifetime by simply storing neutrons for a certain time and counting the survivors. 
Second, longer observation times enhance the sensitivity of detection schemes like the Rabi or Ramsey resonance techniques, following the uncertainty relation $\delta E \times T \geq \hbar$, 
where $\delta E$ is the precision of the measured energy and $T$ is the observation time. 
A 1~m long apparatus observes a neutron spin in a cold beam for typically 10~ms, whereas in a storage experiment observation times longer than 100~s could be achieved. 
The measurement of the neutron electric dipole moment (nEDM) could thus benefit from an improvement of the sensitivity by four orders of magnitude. 
Since 1970 the techniques to produce and handle ultracold neutrons were greatly improved to pursue these two goals: 
a precise measurement of the neutron lifetime and the search for the nEDM. 
We will come back to these two topics after taking a cosmological detour.

\subsection{The standard picture of the early Universe}

The modern picture of the Big Bang was initiated by G. Lema\^itre, supported by the measurement of the expansion of the Universe by \textcite{Hubble1929}. 
At that time, the modern theory of gravity, Einstein's general relativity, was just invented. 
Within this framework it was possible to consider the Universe as a physical system interacting with its content. 
Assuming the cosmological principle - the Universe is homogeneous and isotropic on large scales - 
the cosmic evolution of a spatially flat Universe is described by the Robertson-Walker metric
\begin{equation}
\label{RobertsonWalker}
ds^2 = dt^2 - a(t) \left( dx^2 + dy^2 + dz^2 \right)
\end{equation}
where $a(t)$ is the dimensionless scale factor defined in such a way that at the present time $t_0$ the scale factor is unity $a(t_0) = 1$. 
In our expanding Universe, the Hubble parameter $H(t) = \dot{a}(t)/a(t)$ is positive. 
In the cosmic past, all distances were smaller by the factor $a(t)$. 
For example, light emitted at time $t$ with a wavelength $\lambda$ is observed now with a larger wavelength $\lambda_0 = \lambda / a(t) = (z+1) \lambda$. 
The redshift $z = \frac{1}{a(t)} -1$, being actually observable, is often preferred to $t$ as a variable to indicate the ticking of the cosmic clock. 

Assuming that the Universe is filled with an homogeneous fluid with an energy density $\rho$ and a pressure $p$, 
Einstein's equations of general relativity (with a vanishing cosmological constant) reduce to the Friedmann equations 
\begin{eqnarray}
& & H^2 = \frac{8 \pi G}{3} \rho \label{F1} \\
& & \frac{d \rho}{dt} + 3 H (\rho + p) = 0. 
\label{F2}
\end{eqnarray}

In the early Universe, the expansion was dominated by radiation, i.e. photons and relativistic particles. 
In this case the pressure is related to the energy density as $p = \rho/3$. 
Then from the Friedmann equations we deduce  $\rho \propto a^{-4}(t)$, $a(t) \propto \sqrt{t}$ and $H(t) = \frac{1}{2 t}$. 
Going backwards in time (higher $z$) the energy density of the radiation $\rho \propto (z+1)^4$ was higher. 
Assuming that the radiation was in thermal equilibrium, one can associate a temperature $T$ to the radiation according to
\begin{equation}
\rho = \frac{\pi^2}{30} N(T) T^4, 
\end{equation}
where $N(T)$ counts the effective number of relativistic degrees of freedom ($+1$ per boson polarization state, $+7/8$ per fermion state). 
Combining with the Friedmann equations we can relate the age of the Universe $t$ to the temperature $T$ 
\begin{equation}
\label{Tvs_t}
\ t = \sqrt{\frac{45}{16 \pi^3 G N(T)}} \frac{1}{T^2} = \frac{2.4 \ {\rm s}}{\sqrt{N(T)}} \ \frac{1 \ {\rm MeV}^2}{T^2}. 
\end{equation}
The Universe was hotter in the past. 
For example, at the beginning of the Big Bang nucleosynthesis the temperature was $T \approx 1$~MeV, 
the effective numbers of degrees of freedom was $N(T) = 29/4$ (only photons and three families of neutrinos), 
thus the age of the Universe was $t \approx 0.5$~s. 
One can also relate the photon temperature to the redshift, 
\begin{equation}
\label{Tvs_z}
T = (z+1) T_0 \left( \frac{N(T_0)}{N(T)} \right)^{1/4}, 
\end{equation}
where $T_0 = 2.7255(6)$~K is the present temperature of the Cosmic Microwave Background (CMB). 
Note that Eq. \eqref{Tvs_t} is valid only in the radiation dominated epoch whereas Eq. \eqref{Tvs_z} holds for the whole thermal history of the Universe. 

\begin{table*}
\center
\caption{
The history of the Universe: standard scenario. 
Values in bold are taken from \textcite{PDG}. 
}
\begin{tabular}{lllll}
\hline 
 & Transition  			& Temperature 		  & Time 			& Redshift $z$ 		\\
\hline
\multirow{8}{*}{\includegraphics[width=0.07\linewidth]{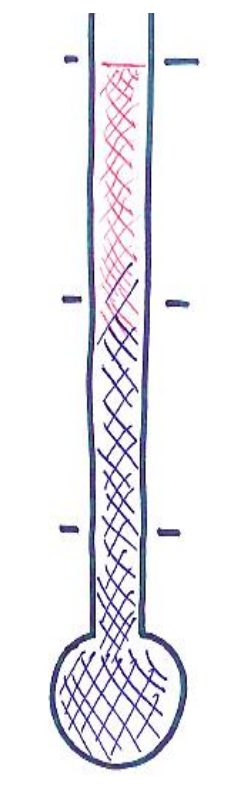}}
&  Inflation ends (reheating)	& $\sim 10^{15}$~GeV	  & $\sim 10^{-37}$~s		& $\sim 10^{28}$ 	\\
& Electroweak phase transition	& $170$~GeV		  & $10^{-11}$~s		& $10^{15}$		\\
& QCD phase transition		& $170$~MeV		  & $30 \, \mu$s		& $10^{12}$ 		\\
& Big Bang nucleosynthesis	& $80$~keV		  & $150$~s			& $3 \times 10^8$ 	\\
& Matter radiation equality	& $1$~eV		  & $30000$~Yr			& ${\bf 3360(70)}$ 	\\
& CMB decoupling (recombination)	& $0.3$~eV		  & ${\bf 372000}$~Yr	& ${\bf 1090.2(7)}$	\\
& First stars (reionization)	& $30$~K		  & ${\bf 460}$~MYr		& ${\bf 11}$	\\
& Today				& ${\bf 2.7255(6)}$~K & ${\bf 13.81(5)}$~GYr	& $0$
\end{tabular}
\label{historyUniverse}
\end{table*}

Very different energy scales of the primordial plasma were successively at play in the thermal history of the Universe. 
This leads to a fascinating and fruitful connection between cosmology and particle physics that relate phenomena at the highest energies with phenomena at the earliest epoch. 
Before exploring more specifically the relevance of low energy neutron physics, which is a subfield of particle physics, we shall shortly review the present standard scenario of the hot Big Bang (see Table \ref{historyUniverse}). 
For a more complete review on the status of the standard cosmological model, see \textcite{Bartelmann2010}. 

It is believed that the Universe started with a phase of accelerated expansion driven by a scalar field called the inflaton \cite{Guth1981}. 
What happened before is even more speculative because it is related with physics at the Planck scale and involves the great problem of unifying gravity with quantum mechanics. 
The inflaton field is supposed to be coupled to Standard Model particles and the energy density of the scalar field would be converted into particles by the end of the inflation. 
This \emph{reheating} process would start the thermal history of the Universe with an initial temperature of presumably about $10^{15}$~GeV. 

Then the Universe cooled down while expanding and a succession of phase transitions occurred. 
At temperatures corresponding to the electroweak scale ($T \approx 200$~GeV), 
the Higgs boson field expectation value condensed from the symmetric phase $\langle \phi \rangle = 0$ down to the non-symmetric phase $\langle \phi \rangle = 246$~GeV. 
We will come back to the electroweak phase transition in section \ref{EDM}. 
Next, at a temperature below $T = 100$~MeV, quarks and gluons condensed into nucleons during the QCD phase transition. 
The Universe was then populated by photons, neutrinos, relativistic electrons and positrons, and traces of protons and neutrons. 
The nucleons are leftovers from a tiny asymmetry between quarks and antiquarks, the so-called baryon asymmetry of the Universe (BAU) which was generated earlier. 
Further condensation of nucleons into nuclei became possible at the temperature of $T \approx 80$~keV during the Big Bang nucleosynthesis (BBN). 
We will explain the connection between the BBN and the measurement of the neutron lifetime in the next section. 

The radiation dominated period of the Universe ended at a temperature of $T = 1$~eV, 
when the energy density of dark matter became larger than that of radiation. 
Shortly after, at $T = 0.3$~eV, electrons and protons combined into hydrogen atoms and the Universe became transparent to photons. 
When we observe the CMB today we see in fact photons emitted during the recombination at the redshift of $z\approx 1000$. 
In the matter dominated Universe, the primordial density fluctuations grew until star formation began at the very recent redshift of $z=11$. 
In the late luminous Universe the expansion is again accelerating, perhaps driven by a new substance called Dark Energy. 
We will develop in the next chapters some possible ways to address the Dark Energy problem with neutron experiments. 

\subsection{Big Bang nucleosynthesis and the neutron lifetime}

\para{The Big Bang nucleosynthesis. }
\textcite{Gamow1948} formulated the hypothesis that nuclear fusion in the early Universe is responsible for the formation of nuclei. 
He argued that the primordial nucleosynthesis started at a temperature of $T_{\rm BBN} = 100$~keV, when the radiative dissociation of deuterium stopped. 
The age of the Universe was then a few minutes, which he considered to be the timescale associated with nucleosynthesis. 
Conveniently, this is also the order of magnitude of the neutron lifetime, which was poorly known in 1948. 
Then, he required the mean time for the reaction $p + n \rightarrow d + \gamma$ to be about 2 minutes. 
Indeed, if the mean time was much longer then no complex nuclei would have been formed during the BBN. 
On the other hand if the mean time was much shorter then there would be no deuterium left (subsequent fusion reactions happen to be faster than deuterium formation). 
Quantitatively this educated guess, sometimes referred to as the Gamow criterion, can be expressed as
\begin{equation}
n_b \ \sigma v \times 2 \ {\rm min } \approx 1, 
\end{equation}
where $\sigma \approx 10^{-29}$~cm$^2$ is the fusion cross section, $v \approx 5 \times 10^8$~cm/s is the thermal velocity of the protons and neutrons at the temperature $T_{\rm BBN}$ and $n_b$ is the number density of protons and neutrons. 
With the above criterion Gamow got $n_b \approx 10^{18}$~cm$^{-3}$ corresponding to a mass density of $\rho_b(t_{\rm BBN}) \approx 10^{-6}$~g/cm$^3$. 

Following this line of argument one can predict the present temperature of the CMB, $T_0 = T_{\rm BBN} \sqrt[3]{\rho_b(t_0)/\rho_b(t_{\rm BBN})}$, 
by using the scaling laws $T(t) \propto a(t)^{-1}$ and $\rho_b(t) \propto a(t)^{-3}$. 
Using the estimate $\rho_b(t_0) \approx 10^{-30}$~g/cm$^3$ obtained from the Hubble rate, 
\textcite{Alpher1949} predicted $T_0 \approx 5$~K. 
The Big Bang model was unambiguously confirmed in 1965 when Penzias and Wilson discovered the CMB. 

Building on the seminal work of Gamow, a more sophisticated theory of the BBN was worked out. 
A key quantity is the ratio of baryon to photon number densities $\eta = n_b/n_\gamma$. 
For a given value of $\eta$ the theory can predict the relative abundance of the four light elements produced during the BBN, namely D, $^3$He, $^4$He and $^7$Li. 
From the observed relative abundances of each of these elements one can extract $\eta$. 
The most sensitive probe of $\eta$ comes from the measurement of the deuterium to proton ratio in extragalactic clouds, D/H$ = (2.53 \pm 0.04) \times 10^{-5}$, 
from which the following result is obtained \cite{Cooke2014}
\begin{equation}
\label{etaBBN}
\eta_{\rm BBN} = (6.0 \pm 0.1) \times 10^{-10}. 
\end{equation}

The study of the temperature anisotropies in the CMB provides 
an independent determination of the baryon density of the Universe at the recombination. 
The Planck satellite \cite{Planck2014} has produced the result 
\begin{equation}
\label{etaCMB}
\eta_{\rm CMB} = (6.05 \pm 0.07) \times 10^{-10}, 
\end{equation}
in excellent agreement with the BBN value. 

In order to compute the abundances from the BBN theory, the neutron lifetime $\tau_n$ needs to be known with sufficient accuracy, 
because the number of neutrons available for fusion at $T_{\rm BBN}$ depends on $\tau_{\rm n}$. 
According to state-of-the-art calculations by \textcite{Coc2014}, the sensitivity of the deuterium fraction to the values of the input parameters is 
\begin{equation}
\frac{\Delta ({\rm D}/{\rm H})}{{\rm D}/{\rm H}} = -1.6 \, \frac{\Delta \eta}{\eta} + 0.4 \, \frac{\Delta \tau_{\rm n}}{\tau_{\rm n}}. 
\end{equation}
The parametric uncertainty due to the error on the neutron lifetime becomes negligible compared to the observational error if $\Delta \tau_{\rm n} \ll 30$~s. 
We will explain in the next paragraph how the needed sub-percent accuracy has been achieved in the 1990's.

\para{The measurement of the neutron lifetime. }
The neutron decays due to the weak interaction into a proton, an electron and an antineutrino with a period of about ten minutes. 
It is the simplest case of nuclear beta decay. 
Given its importance in cosmology and particle physics, the neutron lifetime has been measured by more than 20 experiments. 
There are two distinct experimental approaches to measure the neutron lifetime $\tau_{\rm n}$. 
\begin{enumerate}
\item {\bf The beam method}. 
A detector records the decay products in a well-defined part of a neutron beam. 
A neutron beam is indeed radioactive due to beta decay and the rate of appearance in the beam of the decay products is 
\begin{equation}
- \frac{d N}{dt} = \frac{N}{\tau_{\rm n}}.
\end{equation}
This method requires (i) a determination of the number of neutrons $N$ in the beam, 
that is, an absolute determination of the neutron flux and (ii) a detector for decay products with a well calibrated registration efficiency to measure $dN/dt$. 
\item {\bf The bottle method}. 
A bottle is filled with UCNs. After a certain waiting time $t$, the trapped neutrons are emptied into a detector to count the number of remaining neutrons $N(t)$. 
One repeats this operation for various times $t$ and the storage time $\tau_{\rm stor}$ is extracted from a fit of the storage curve
\begin{equation}
N(t) = N(0) \, e^{-t/\tau_{\rm stor}}. 
\end{equation}
The efficiency of the neutron counter does not need to be known accurately, but the losses due to absorption or inelastic scattering during collisions with the walls must be carefully controlled in order to determine $\tau_{\rm n}$ from the storage time $\tau_{\rm stor}$. 
\end{enumerate}

\begin{figure}
\centering
\includegraphics[width=0.97\linewidth]{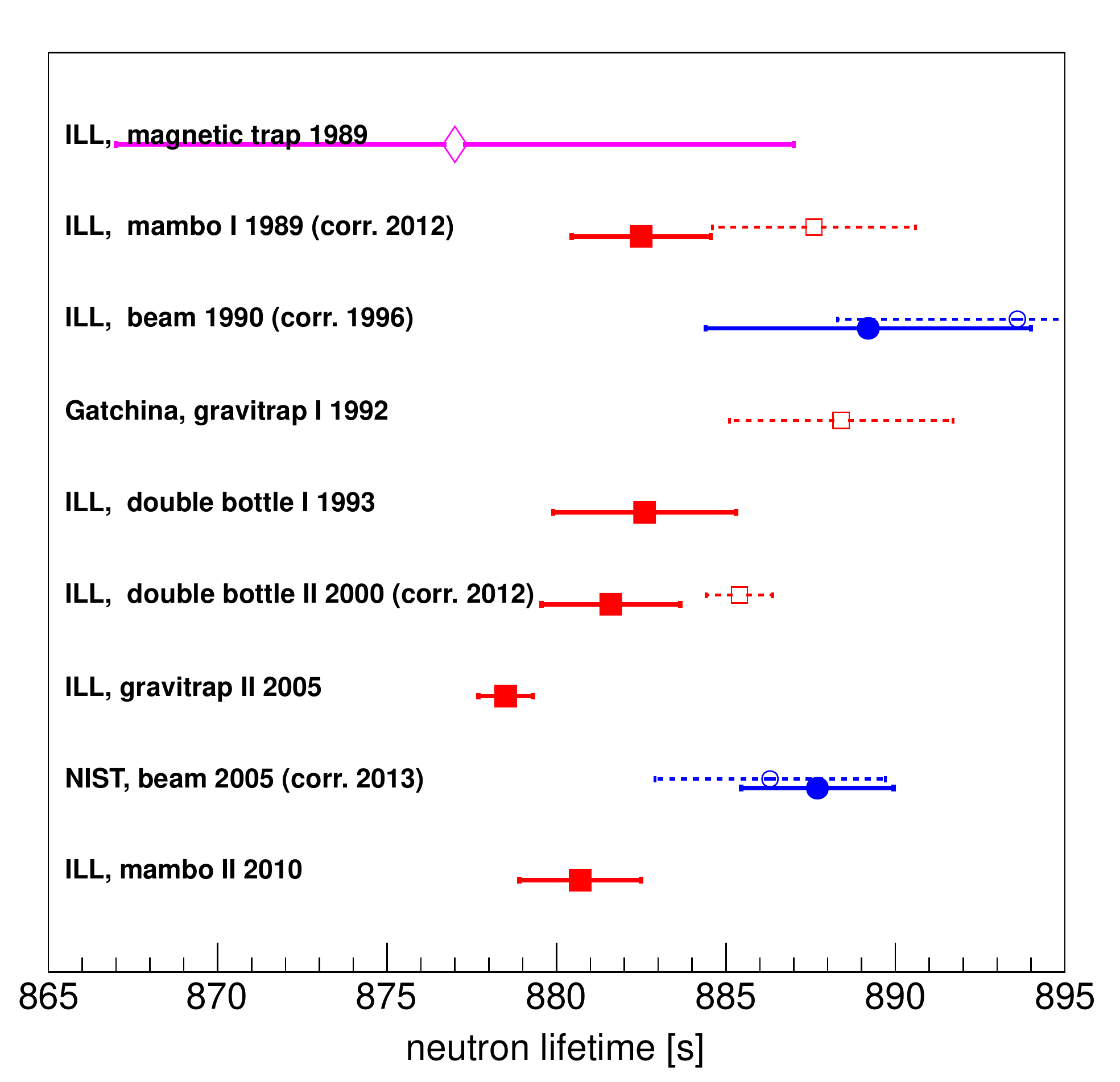}
\caption{
Results of the latest neutron lifetime measurements with the beam method (\textcolor{blue}{$\CIRCLE$}), material bottle method (\textcolor{red}{$\blacksquare$}) and magnetic bottle method (\textcolor{magenta}{$\diamondsuit$}). 
Dashed lines correspond to measurements which have been withdrawn or corrected later on. 
}
\label{neutronLifetime}
\end{figure}

The current situation of the measurements is summarized in Fig. \ref{neutronLifetime}. 
References to the publications of each experiment can be found in the recent review by \textcite{Wietfeldt2011}, 
except for the recent reevaluation of the latest beam result by \textcite{Yue2013}. 

Concerning the beam method, one can choose to measure either the electron or the proton activity of the beam (it would be silly to try to measure the neutrino activity). 
Although early experiments (not represented in Fig. \ref{neutronLifetime}) detected the electrons coming out of the beam, 
modern measurements count the protons. 
Combining the two most recent measurements we obtain
\begin{equation}
\tau_{\rm n}^{\rm beam} = ( 888.0 \pm 2.1 ) \, {\rm s}.
\label{lifetimeBeam}
\end{equation}

Let us now come to the storage method where one measures the storage time $\tau_{\rm stor}$ of UCNs in a material bottle, 
which is a combination of the beta decay lifetime and the lifetime due to losses at wall collisions : $1/\tau_{\rm stor} = 1/\tau_{\rm n} + 1/\tau_{\rm wall}$. 
Most of the time neutrons are specularly reflected at wall collisions, but there is a small probability of capture or up-scattering. 
The loss probability can be of the order of $\mu \approx 10^{-5}$, using walls coated with hydrogen-free oil such as fomblin for example. 
In a typical material bottle with mean distance $\lambda \approx 30$~cm, UCNs with velocity $v \approx 3$~m/s will collide with the walls at a frequency $f = v/\lambda \approx 10$~Hz. 
Then we expect $\tau_{\rm wall} = 1/f \mu \approx 10^4$~s, which corresponds to a 10~\% correction of the beta decay lifetime. 
The usual strategy to control this effect is to measure the storage time for various geometries of the bottle, with different volume to surface ratio, 
and extrapolate to the ideal case of vanishing wall collision frequency. 
In some experiments the bottle is cooled down to reduce the neutron losses due to up-scattering. 
The five UCN measurements performed at the ILL produced the combined value 
\begin{equation}
\tau_{\rm n}^{\rm bottle} = ( 879.6 \pm 0.8 ) \, {\rm s}. 
\label{lifetimeBottle}
\end{equation}

There is a $3.8 \sigma$ discrepancy between the beam result \eqref{lifetimeBeam} and the bottle result \eqref{lifetimeBottle} that needs to be resolved. 
To alleviate the issue of the losses at wall collisions which are not fully understood yet, new projects aim at confining UCNs in a magnetic bottle. 
The neutron is a spin 1/2 particle with a magnetic moment $\mu_n = - 60$~neV/T. 
Thus a magnetic ``wall'' of the order of 1~T acts as a repulsive potential of height 60~neV for low field seekers neutrons 
(i.e. neutrons with spin parallel to the magnetic field). 
The measurement of the neutron lifetime is still an active field of research. 
Several teams in Europe and in the US are currently attempting to obtain a reliable measurement with an accuracy of 0.1~s. 

\subsection{The matter antimatter asymmetry and the neutron electric dipole moment}
\label{EDM}

Apparently our Universe is made up of matter, not antimatter. 
Not a single complex antinucleus like antihelium has been detected in cosmic rays. 
No excess of gamma radiation resulting from the annihilation of antimatter with matter is observed. 
Most cosmologists believe that the matter dominance over antimatter extends to at least the whole visible Universe. 
In fact, the imbalance is tiny. 
When the baryons were in equilibrium with the rest of the plasma in the early Universe just before the QCD phase transition, 
for every billion baryon-antibaryon pair there was one spare baryon. 
It should be noted that we quantify the matter-antimatter asymmetry by the baryon asymmetry. 
For sure there are also more electrons than antielectrons in the Universe today. 
However, an excess of antineutrinos over neutrinos in the cosmic neutrino background could perhaps compensate for the electron-positron asymmetry. 
Therefore we do not know for certain that an excess of matter over antimatter exists for leptons. 
As we have seen, the asymmetry parameter $\eta = n_b/n_\gamma$ deduced from considerations about the primordial nucleosynthesis \eqref{etaBBN} 
agrees with the one deduced from the microwave background \eqref{etaCMB}. 
Given that the two methods rely on two completely separated epochs in the Universe, the agreement is remarkable. 

\para{The Sakharov conditions to generate the baryon asymmetry. } 
We have solid evidence that the baryon asymmetry exists since before the Big Bang nucleosynthesis. 
Is it merely an initial condition of the Big Bang tuned to allow intelligent life to emerge? 
In the context of inflation this idea is no longer plausible, because inflation would have tremendously diluted the initial baryon density. 
Well before inflation was invented, \textcite{Sakharov1967} proposed that the asymmetry could have been generated dynamically from an initially symmetric state. 
He outlined three conditions that are necessary for this to be possible: 
\begin{enumerate}
\item The baryon number should not be conserved. 
\item The Universe should at some time depart from thermal equilibrium. 
\item The discrete symmetries C (charge conjugation) and CP (charge conjugation combined with parity transformation) should be violated. 
\end{enumerate}
See \textcite{Bernreuther2002} for a pedagogical introduction to baryogenesis, the hypothetical process in the early Universe that meets all these conditions to generate the baryon asymmetry. 

\para{Electroweak baryogenesis. }
The second Sakharov condition suggests that the baryon asymmetry was generated during a phase transition in the early Universe before the primordial nucleosynthesis. 
Electroweak baryogenesis, a mechanism that biases the baryon number during the electroweak phase transition, 
is one of the most attractive among the many proposed realizations of baryogenesis. 
In fact, all three Sakharov conditions are qualitatively fulfilled by the standard electroweak theory although it fails quantitatively to predict the correct baryon asymmetry. 
Nevertheless, electroweak baryogenesis is still a viable scenario in extensions of the Standard Model (SM) of particle physics. 
New physics is required at or close to the electroweak scale, this makes the scenario testable - and falsifiable - by current or planned experiments. 
Let us take a closer look at the Sakharov conditions in the context of the electroweak phase transition. 

Surprisingly, the SM does accommodate baryon number violation, 
induced by non-perturbative effects associated with the nontrivial structure of the SU(2) gauge fields vacuum. 
We do not observe baryon number violation in the laboratory because the process requires a quantum tunneling through a large energy gap with an extremely small probability. 
However, \textcite{Kuzmin1985} discovered that baryon number violation processes called \emph{sphalerons} were frequent in the early Universe when the temperature was high enough to overcome the energy barrier. 
In fact, the first Sakharov condition does not demand physics beyond the Standard Model. 

\begin{figure}
\centering
\includegraphics[width=0.97\linewidth]{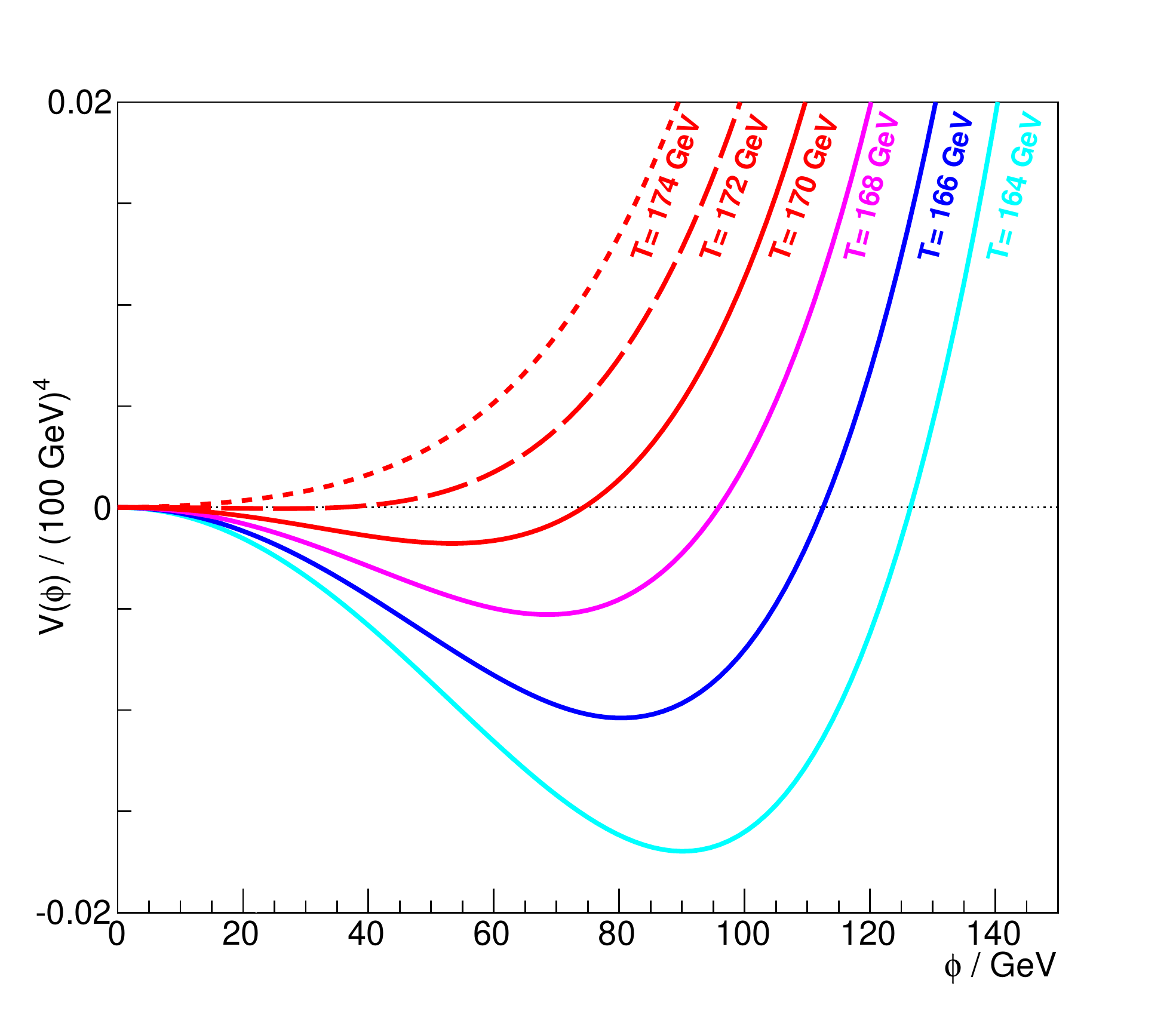}
\includegraphics[width=0.97\linewidth]{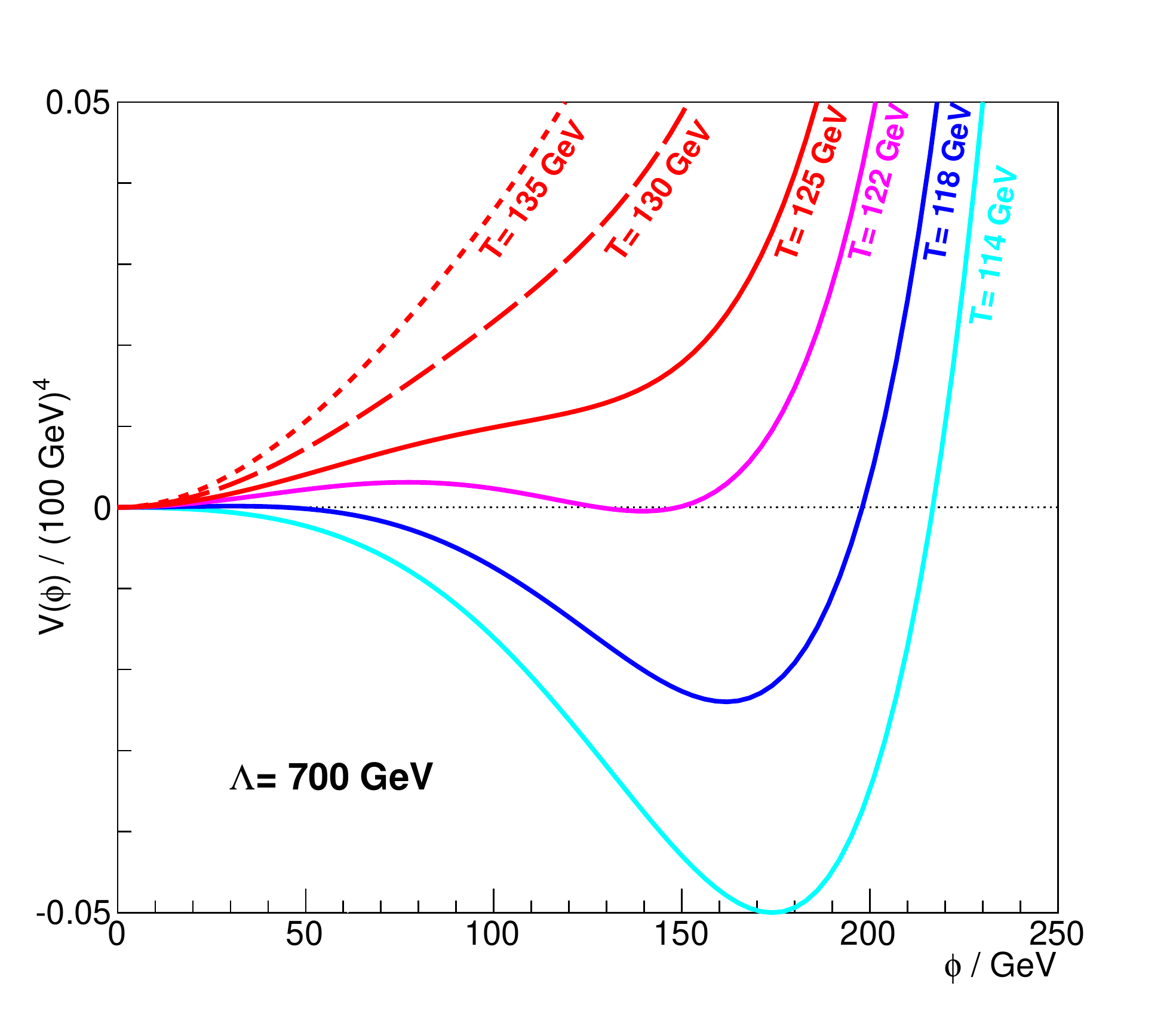}
\caption{
Effective Higgs potential for different temperatures at the electroweak phase transition. 
Top: calculation within the SM, a smooth cross-over occurs at $T = 170$~GeV. 
Bottom: calculation within the $\phi^6$ extension, a first order phase transition occurs at $T = 120$~GeV. 
}
\label{EWPT}
\end{figure}

Next, the second Sakharov condition deserves a discussion about the electroweak phase transition, when the Higgs field acquired a non-zero expectation value. 
When coupled to the thermal bath at temperature $T$, the effective potential of the SM Higgs field $\phi$ can be written 
as a sum of the bare potential and a thermal correction: 
\begin{equation}
\label{HiggsPotential}
\mathcal{V}_{\rm eff}(\phi, T) = - \frac{\mu^2}{2} \phi^2 + \frac{\lambda}{4} \phi^4 + \mathcal{V}_{\rm th}(\phi, T). 
\end{equation}
At zero temperature the thermal correction to the potential is absent ($\mathcal{V}_{\rm th} = 0$) and the potential \eqref{HiggsPotential} has the famous Mexican hat shape 
with a minimum at $\phi = \sqrt{\mu^2/\lambda} = 246$~GeV. 
At high temperature the thermal correction $\mathcal{V}_{\rm th} \neq 0$ modifies the potential in such a way that the minimum of the potential is $\phi = 0$. 
When the temperature decreased, the field $\phi$ condensed from the symmetric state $\phi=0$ to the state $\phi = 246$~GeV, breaking spontaneously the electroweak symmetry. 
Successful baryogenesis requires that the phase transition be first order: the field must change discontinuously from $\phi = 0$ to $\phi >0$. 
Figure \ref{EWPT} (Top) shows the modification of the potential close to the phase transition at $T = 170$~GeV. 
The formulas for the effective potential are taken from the thesis of \textcite{Fromme2006}, 
where I specified the value of the Higgs mass to $m_h = 126$~GeV conforming with the recent discovery at the Large Hadron Collider. 
We see that the minimum of the potential smoothly moves away from zero, there is no sufficient departure from thermal equilibrium and no baryogenesis in the Standard Model. 

Now, a non-standard Higgs potential could lead to a first order electroweak phase transition. 
As an example, we could add to the potential \eqref{HiggsPotential} a non-renormalisable operator of the form $\frac{1}{8 \Lambda} \phi^6$. 
We plot in Fig. \ref{EWPT} the modification of the effective potential in the case $\Lambda = 700$~GeV, using again the formulas in \textcite{Fromme2006}. 
In this case the minimum of the potential suddenly changes from $\phi = 0$ to $\phi = 140$~GeV at the critical temperature of $T_{\rm cr} = 120$~GeV, this is a first order phase transition. 
If such a transition did take place in the early Universe, bubbles of true vacuum started to nucleate and expand in a sea of false vacuum. 
In such a boiling environment, the Universe could have been driven sufficiently out of equilibrium for baryogenesis to be possible. 

A first order electroweak phase transition can be realized in more sophisticated extensions of the scalar sector of the SM, 
such as the two-Higgs-doublet model or supersymmetric extensions. 
Since these models modify the physics at the electroweak scale, they are testable at particle colliders. . 

Last, electroweak baryogenesis requires CP violation at the electroweak scale. 
It turns out that the CP violation contained in the SM, from phase $\delta$ of the quark mixing matrix, is not strong enough to generate the baryon asymmetry. 
Thus CP-violating new physics is required to satisfy the third Sakharov condition as well. 
Such new physics is best probed by low energy precision experiments searching for electric dipole moments (EDM) of particles. 

In summary, the failure of the SM to allow for the electroweak baryogenesis is a hint toward the presence of new physics lying just above the electroweak scale that may be discovered both by collider experiments and EDM searches. 

\begin{figure}
\centering
\includegraphics[width=0.77\linewidth]{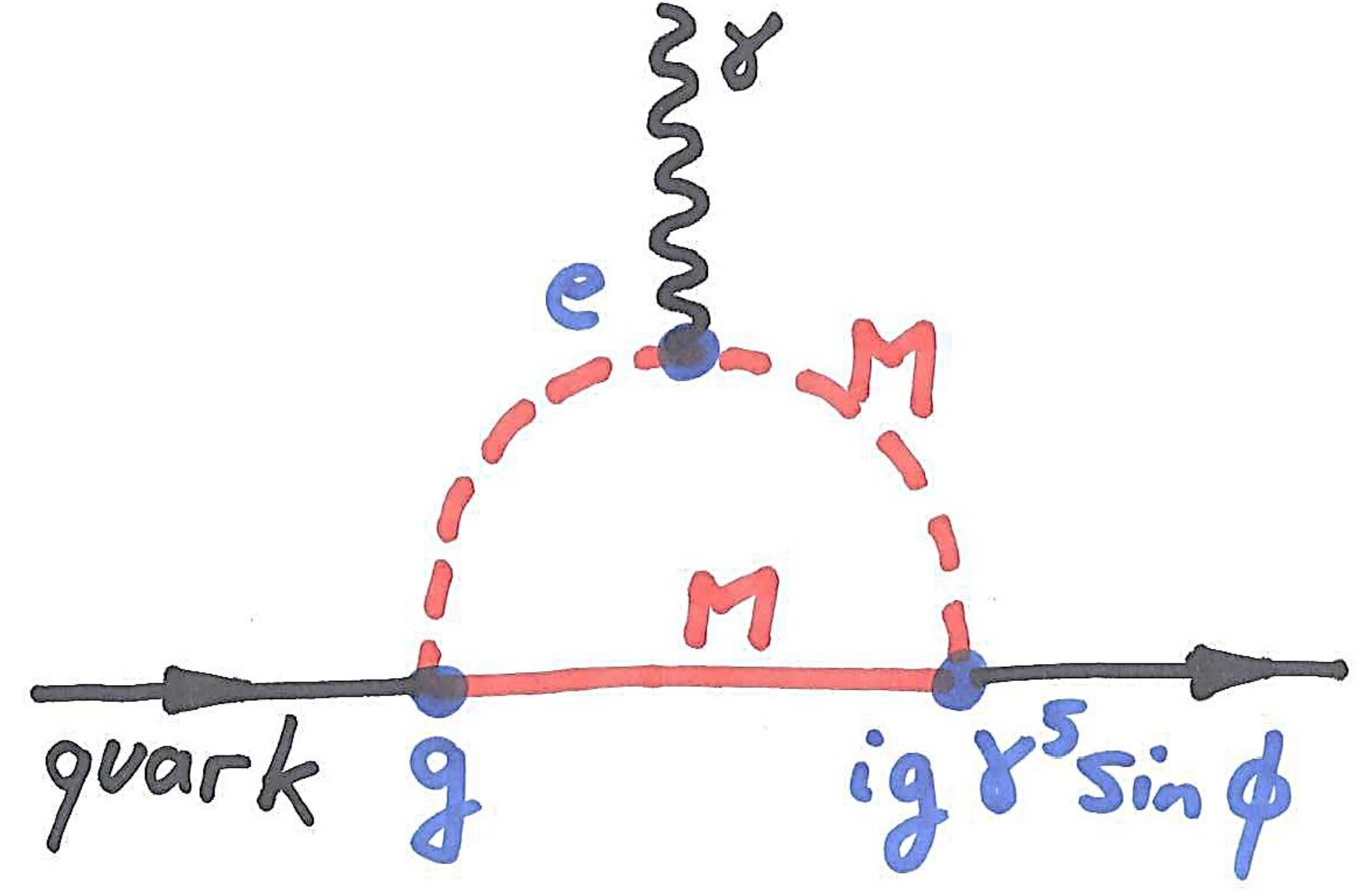}
\caption{
One loop diagram contributing to the quark EDM. 
The quark EDMs then transfer almost directly to the neutron EDM.
}
\label{EDM_loop}
\end{figure}

\para{CP violation and neutron EDM. }
The existence of a nonzero EDM for a spin 1/2 particle such as the neutron would imply the violation of the CP symmetry. 
In the Standard Model, the induced neutron EDM expected from the $\delta$ phase is tiny, $d_n \approx 10^{-32} \ e$~cm. 
This value is to be compared to the current best limit obtained at the ILL by \textcite{Baker2006}: 
\begin{equation}
|d_n| < 3 \times 10^{-26} \ e \, {\rm cm \ (90 \, \% \ C.L.)}. 
\end{equation}
Therefore, improvements of the neutron EDM measurement are motivated by the potential discovery of a new source of CP violation beyond the SM. 
As for any low energy observable, the new physics at a high energy scale manifests itself via quantum loops involving virtual particles. 
Figure \ref{EDM_loop} shows a Feynman diagram contributing to a quark EDM via the CP-violating vertices of heavy scalars and fermions with masses $M$.

Generically the neutron EDM induced by such a loop amounts to \cite{Pospelov2005} 
\begin{equation}
d_n \approx \left( \frac{1 \, {\rm TeV}}{M} \right)^2 \times \sin \phi \times 10^{-25} \ e \, {\rm cm}
\end{equation}
where $M$ is the mass of the particles running in the loop. 
In this case the heavy particles couple strongly with the quark 
(such as the SUSY coupling between quark, squark and gluinos) with a CP-odd vertex multiplied by $\sin \phi$. 
The CP-odd part usually originates from the imaginary part of some parameter in the Lagrangian and $\phi$ would then correspond to the CP-violating phase of that specific parameter. 
Thus, considering natural CP violation ($\sin \phi \approx 1$) in the new heavy sector, the neutron EDM is sensitive to new physics at the multi-TeV scale. 

Searches for permanent electric dipole moments of other particles (protons, deuterons, muons, atoms, molecules) are complementary probes of CP violation above the electroweak scale. 
Many experimental efforts are under way, they have been compiled recently by \textcite{Kirch2013}. 
Below we concentrate on the search for the neutron EDM. 

\para{Measuring the neutron EDM. }
The neutron EDM measurement is based on the analysis of the Larmor precession frequency of neutrons, stored in a volume permeated with electric and magnetic static fields either parallel or antiparallel. 
For such configurations, the precession frequency $f_n$ reads
\begin{equation}
h f_n = \mu_n B \pm d_n E
\end{equation}
where $\mu_n$ and $d_n$ are the magnetic and electric dipole moments of the neutron and $h$ is Planck's constant. 
The frequency difference of these two configurations gives directly access to the neutron EDM: $d_n = h \Delta f_n / 4E$. 
To measure the precession frequency, we use Ramsey's method of separated oscillatory fields which provides a statistical precision per cycle of 
\begin{equation}
\sigma(d_n) = \frac{\hbar}{2 \alpha E T \sqrt{N}}
\end{equation}
where $T$ is the precession time, $\alpha$ is the visibility (related to the polarization of the neutrons) and $N$ the total number of detected neutrons. 
Using stored ultracold neutrons, the precession time can be set to $T = 200$~s, five orders of magnitude 
longer as compared to the first experiment performed by \textcite{Smith1957} using a neutron beam! 

The main experimental challenge in current experiments consists in achieving a magnetic field homogeneity at the
level of $10^{-4}$ over a volume of typically 20 $\ell$ 
while maintaining a temporal stability of about $10^{-7}$ over 100 s. 
These requirements are necessary to control the subtle systematic effects (see for example \textcite{Pignol2012}). 
Atomic magnetometry and magnetic shielding techniques are therefore at the core of such a measurement. 

\begin{figure}
\centering
\includegraphics[width=0.97\linewidth]{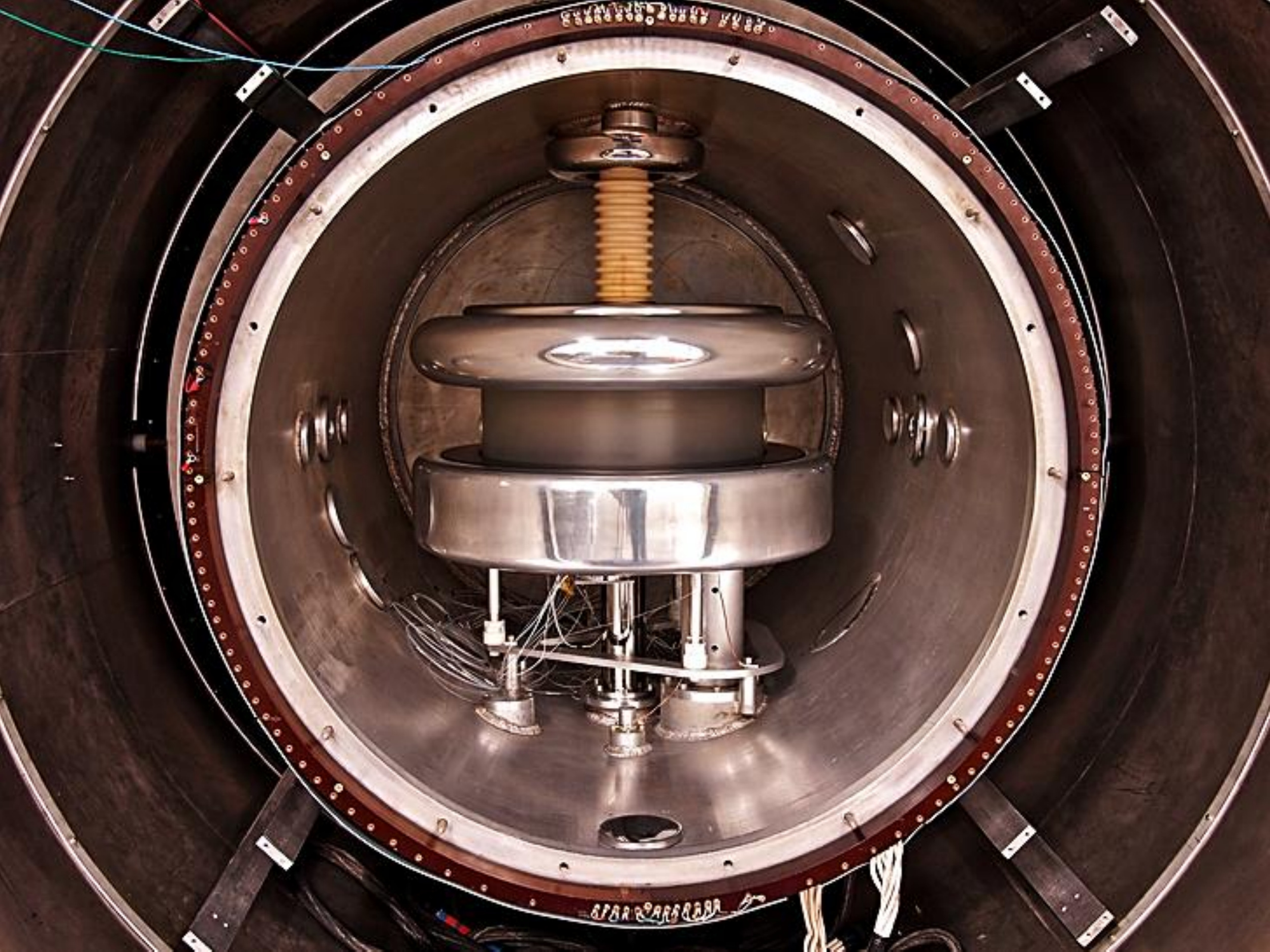}
\caption{
The nEDM apparatus installed at the PSI. 
The precession chamber sits in a large cylindrical vacuum chamber (with front end open on the picture), surrounded by a four-layer mumetal magnetic shield. 
The UCN guide injects the neutrons from the bottom. 
High voltage is applied using the HV feed-trough connected to the top electrode. 
}
\label{OILL}
\end{figure}

As of 2015 there are six projects worldwide competing for improving the measurement of the neutron EDM, all at different stages. 
\begin{enumerate}
\item A PNPI group is currently operating a room-temperature double-chamber spectrometer at the ILL \cite{Serebrov2014}. 
Later the apparatus will be moved back to PNPI where a new UCN source will be built. 

\item Another PNPI group pursues a completely different experimental method. 
It exploits inter-atomic electric fields in crystals that are $10^4$ times stronger than electric fields produced by charging electrodes as in all other experiments. 
However the interaction time is much shorter because it is a neutron scattering technique. 
The result obtained recently at the ILL \cite{Fedorov2010} is two orders of magnitude less precise than the conventional UCN method. 

\item A German project at Munich \cite{Altarev2012} is waiting for the construction of a new UCN source at the reactor FRM-2. 
The newly built magnetically shielded room performs very well \cite{Altarev2014}. 
This large room is designed to host a double-chamber room-temperature experiment. 

\item A Canada-Japan collaboration plans to build a superthermal UCN source and a room-temperature EDM apparatus at TRIUMF Vancouver, 
based on the prototype operated at RNCP Osaka \cite{Masuda2012,Matsuta2013,Masuda2014}. 

\item An ambitious project is pursued by the US community based on the original proposal of \textcite{Golub1994}. 
The idea is to run the experiment in superfluid $^4$He, that serves both as a superthermal UCN source and a precession chamber. 
Traces of spin-polarized $^3$He will be injected in the volume to both detect the neutrons and act as a comagnetometer \cite{Ye2009}. 
The project is still in the R\&D phase and will start real data taking at SNS Oak Ridge in 2020 at best. 

\item A European collaboration (Belgium, France, Germany, Poland, Switzerland and UK) is running an experiment at the PSI UCN source \cite{Baker2011}. 
We are currently operating an upgraded version of the RAL-Sussex spectrometer (see picture Fig. \ref{OILL}) which has produced the lowest experimental limit for the neutron EDM 
and that we moved from the ILL to the PSI in 2009. 
In the same time we have started the design of n2EDM, a next generation room-temperature double-chamber apparatus in view of its delivery around 2018. 
With the current apparatus we aim at slightly improving the present limit on the neutron EDM whereas with n2EDM we aim at a precision of better than $10^{-27} \, e$~cm. 
In addition, we have produced several spin-off scientific results with the apparatus 
(i) a search for neutron to mirror-neutron oscillations \cite{Ban2007,AltarevPRD2009} 
(ii) a sensitive test of Lorentz invariance \cite{Altarev2009,Altarev2010,Altarev2011} 
(iii) a measurement of the neutron magnetic moment with an uncertainty of 0.8 ppm \cite{Afach2014} 
(iv) a search for Axionlike particles \cite{Afach2015}.
\end{enumerate}

Although the experiments are getting more and more difficult 
(in the 1950s and 1960s two or three persons could run an experiment, nowadays author lists rarely count less than 20 people) 
prospects are good to improve the accuracy on the neutron EDM by more than an order of magnitude in the next decade. 

\separe

In conclusion, experiments with neutrons address the cosmological question of the origin of the baryonic matter. 
Today we understand how protons and neutrons combined in the early Universe to form nuclei. 
Experiments measuring the neutron lifetime contributed a fare share of the successful prediction of the Big Bang nucleosynthesis. 
However we still do not know how the neutrons and protons were created in the first place, because we do not understand the baryon asymmetry of the Universe. 
Planned measurements of the neutron EDM will contribute to solve this question. 

\clearpage
\section{Chameleon Dark Energy}
\label{Chameleon}

The 2011 Nobel Prize in physics was awarded to S. Pelmutter, B. Schmidt and A. Riess 
for the discovery of the late acceleration of the expansion of the Universe. 
Although the acceleration is now established as a fact, it remains a great puzzle. 
We do not know the nature of the Dark Energy responsible for the acceleration. 

The two other cosmological puzzles, the origin of the baryon asymmetry and the nature of the Dark Matter, are thought to be related to physics at high energy, beyond the electroweak scale. 
On the contrary, the energy scale associated to the Dark Energy has the peculiar value of 2~meV. 
It suggests that there is still new physics in the infrared to be discovered, perhaps with low energy precision experiments using neutrons? 

\subsection{The accelerated expansion of the Universe}

\begin{figure}
\centering
\includegraphics[width=0.88\linewidth]{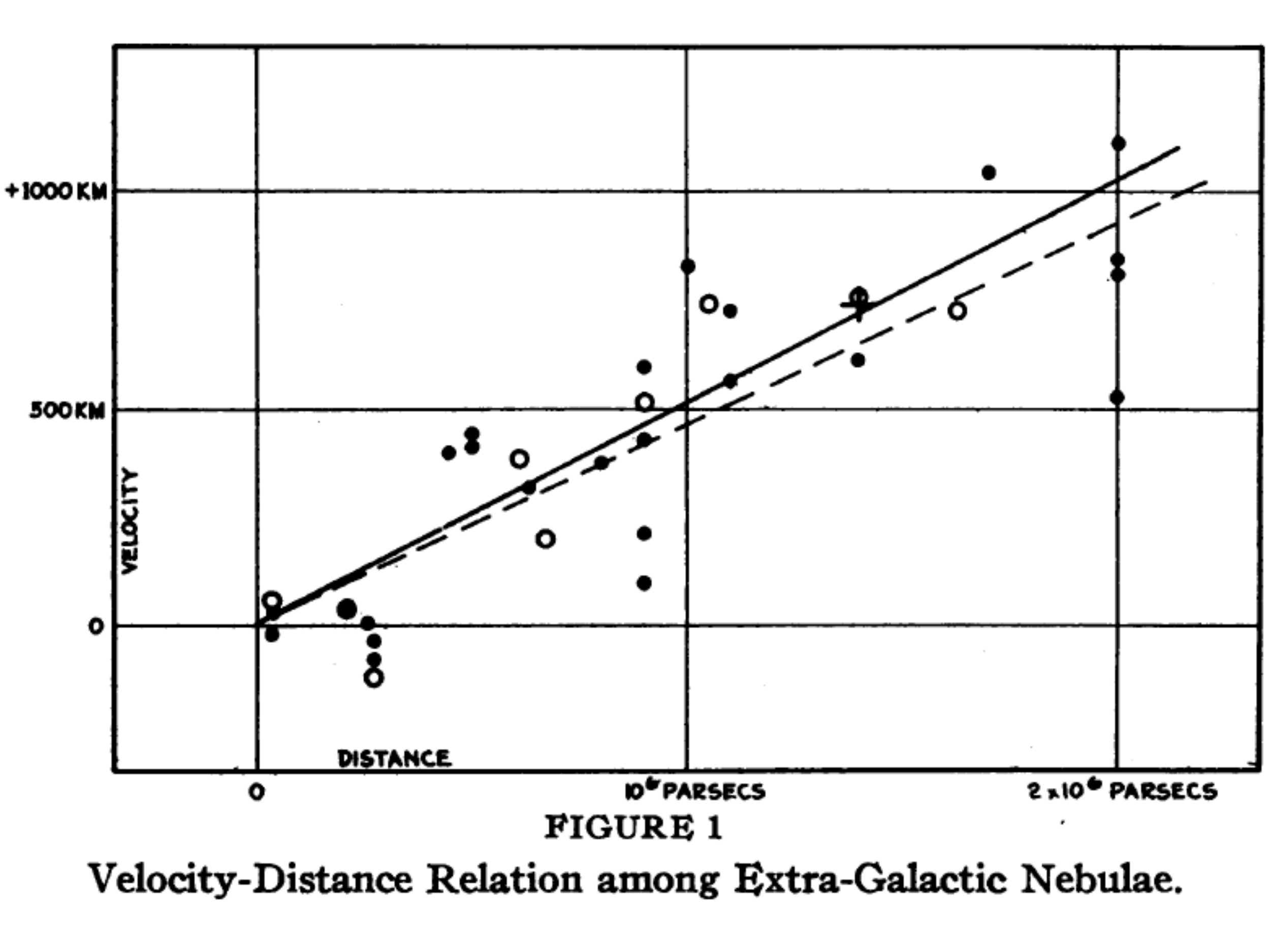}
\includegraphics[width=0.99\linewidth]{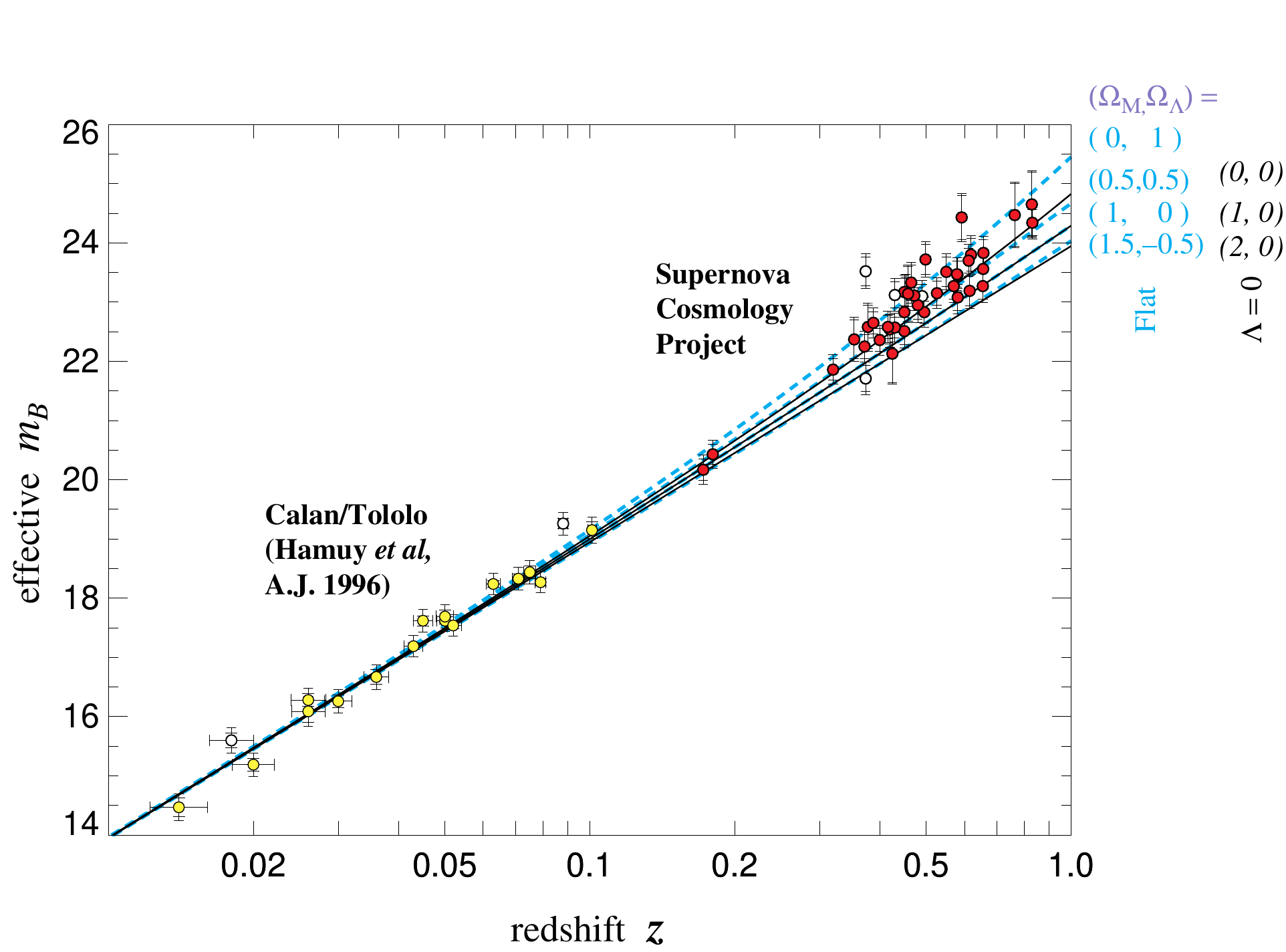}
\caption{
Top: diagram taken from \textcite{Hubble1929} representing the velocity (deduced from the redshift) of nearby galaxies as a function of distance. 
Bottom: supernovae Hubble diagram taken from \textcite{Perlmutter1999} representing the magnitude (indicating the distance) as a function of the redshift. 
Notice that the highest velocity of $1000$~km/s in the 1929 diagram corresponds to the redshift $z = 0.003$ which is smaller than the smallest redshift in the supernovae diagram.}
\label{Hubble}
\end{figure}

A linear relation between the distance and radial velocity among galaxies, $v = H_0 d$, was first established by \textcite{Hubble1929}. 
The first Hubble diagram is shown in Fig. \ref{Hubble}. 
Incidentally, due to an error in the distance evaluations, Hubble derived an expansion rate of $500$~km/s/Mpc which is much larger than the most recent determination by \textcite{Planck2014}: 
\begin{equation}
H_0 = 67.3 \pm 1.2 \quad {\rm km/s/Mpc}. 
\end{equation}

Hubble's linear law was discovered using observations of nearby galaxies. 
At much larger distances, the relation is not expected to be linear anymore. 
The expansion of the Universe was thought to slow down because of the attractive power of gravity. 
The deceleration of the Universe was then searched for by drawing a Hubble diagram with more distant objects. 
Explosions of type IA supernovae are 
(i) very bright, they can be seen at cosmological distances 
(ii) very reproducible, they all have almost the same intrinsic brightness. 
A sufficient number of supernovae observations using the Hubble space telescope and ground based telescopes were reported by two teams \cite{Riess1998,Perlmutter1999}. 
The supernovae Hubble diagram of one team is shown in Fig. \ref{Hubble}. 

Surprisingly, it was found that the expansion of the Universe accelerates. 
Let us analyse this incredible finding in the framework of a flat Universe 
whose expansion is driven by an homogeneous substance with an energy density $\rho$ and a pressure $p$ (i.e. a perfect fluid). 
We assume an \emph{equation of state} of the substance of the form $p = w \rho$. 
From the Friedmann equations \eqref{F1},\eqref{F2} one derives the acceleration: 
\begin{equation}
\frac{\ddot{a}}{a} = - \frac{4 \pi G}{3} \rho \, (1+3 w). 
\end{equation}
A substance made up of non-relativistic matter ($w=0$) or radiation ($w=1/3$) would decelerate the expansion. 
Instead, a positive acceleration requires $w < -1/3$. 
The mysterious substance driving the expansion must have a negative pressure. 
This substance is generically called Dark Energy. 

\subsection{The $\Lambda$CDM concordance model}

In the standard model of cosmology, the acceleration of the expansion is attributed to a perfect fluid with $w = -1$. 
According to the second Friedmann equation \eqref{F2}, 
the energy density of this fluid is constant. 
It does not dilute with the expansion, it is a cosmological constant. 

\begin{figure}
\centering
\includegraphics[width=0.93\linewidth]{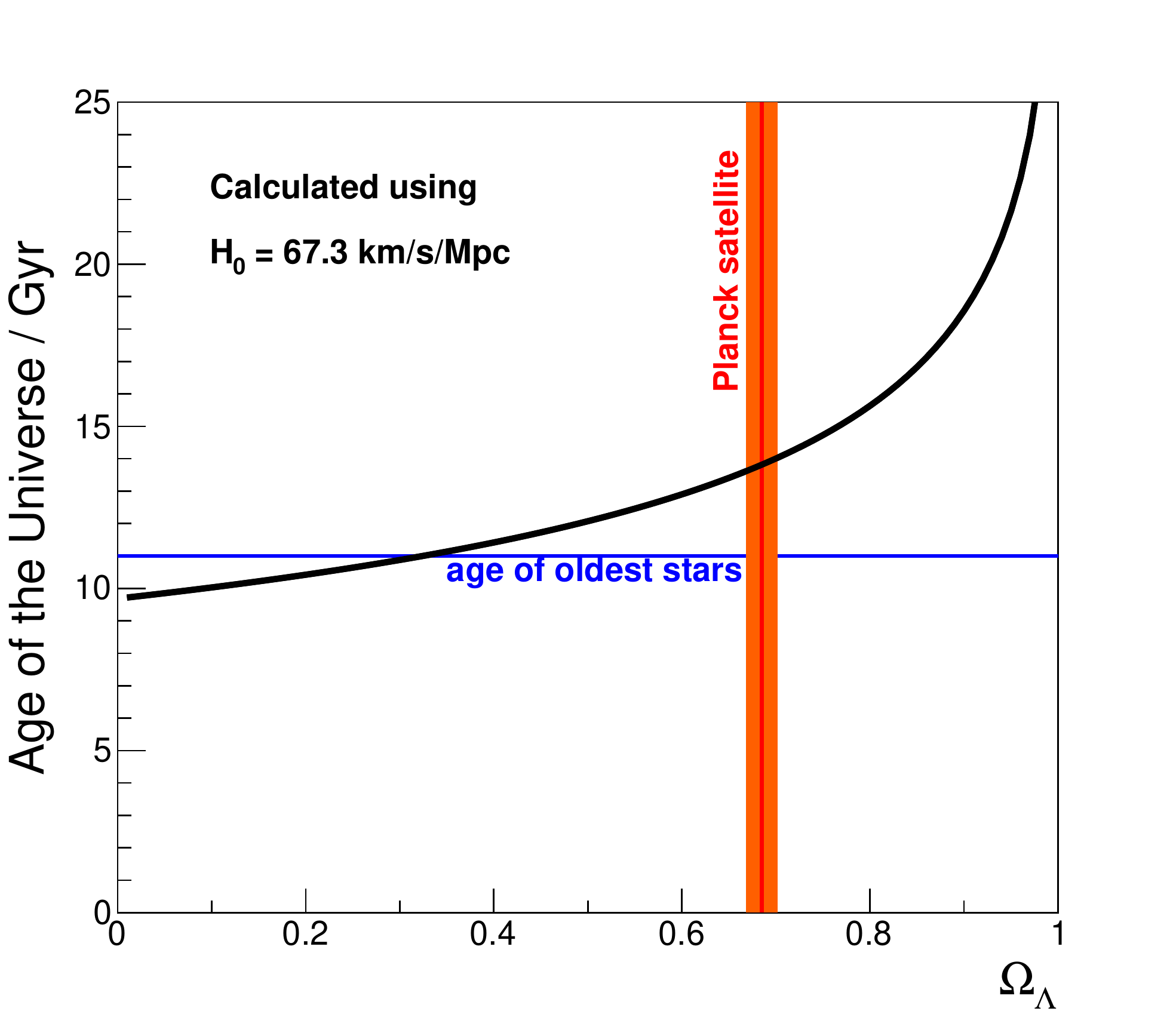}
\caption{
The age of the Universe calculated in the flat $\Lambda$CDM model as a function of $\Omega_\Lambda$. 
The vertical orange band represents the Dark Energy fraction reported by \textcite{Planck2014}. 
\label{ageUniverse}
}
\end{figure}

In the late Universe, the expansion is driven by two distinct perfect fluids: the cosmological constant and some non-relativistic matter (radiation became unimportant a few million years after the Big Bang). 
Hence the name ``$\Lambda$CDM'' for the standard model, which stands for a cosmology with a cosmological constant and cold dark matter (baryonic matter accounts for only a fraction of the non-relativistic matter). 

We note $\rho_{M,0}$ and $\rho_{\Lambda,0}$ the present value of the energy density associated with the matter and the cosmological constant, 
and $\rho_{c,0} = \frac{3 H_0^2}{8 \pi G}$ the critical density. 
We use the standard notation $\Omega_M = \rho_{M,0} / \rho_{c,0}$ and $\Omega_\Lambda = \rho_{\Lambda,0}/\rho_{c,0}$, 
which represent the fraction of the total energy density in the form of matter and cosmological constant. 
For a flat Universe, Eq. \eqref{F1} considered at the present epoch implies $\Omega_M + \Omega_\Lambda = 1$. 
From the second Friedmann equation \eqref{F2} the past values of the densities are 
\begin{equation}
\rho_M(t) = \frac{\rho_{M,0}}{a(t)^3} \quad , \quad \rho_\Lambda(t) = \rho_{\Lambda,0}.
\end{equation}
Then, Eq. \eqref{F1} can be brought to the form
\begin{equation}
\left( \frac{\dot{a}}{a} \right)^2 = H_0^2 \left( \frac{\Omega_M}{a^3} +\Omega_\Lambda \right). 
\end{equation}
This is the equation describing the expansion rate in the flat $\Lambda$CDM model. 
Let us calculate the age of the Universe, by transforming the previous equation into: 
\begin{equation}
1 = \frac{\dot{a} \, \sqrt{a}}{H_0 \sqrt{\Omega_M + a^3 \Omega_\Lambda}}. 
\end{equation}
Next, we integrate the equation from $t=0$ to $t=t_0$ ($t_0$ is the age of the Universe). 
Recall that in our conventions $a(t_0) = 1$. 
We get
\begin{equation}
t_0 = \frac{1}{H_0} \int_0^1 \frac{\sqrt{a} \, da}{\sqrt{\Omega_M + a^3 \Omega_\Lambda}} = \frac{2}{3 H_0} \frac{ {\rm asinh} \sqrt{\Omega_\Lambda / \Omega_M }}{\sqrt{\Omega_\Lambda}}.
\end{equation}
The age of the Universe is plotted in Fig. \ref{ageUniverse} as a function of $\Omega_\Lambda$. 
We can compare it to the age of the oldest stars, estimated to at least 11~Gyr \cite{Bartelmann2010}. 
The Einstein-de Sitter model, i.e. the flat matter dominated Universe with $\Omega_M = 1$ and $\Omega_\Lambda = 0$, 
is excluded because the Universe cannot be younger than the stars. 
The long-standing disagreement between the Hubble expansion rate and the age of the oldest stars has finally been resolved in the $\Lambda$CDM model with the introduction of the cosmological constant. 

In addition to the supernovae and the age of oldest stars, the acceleration of the Universe is now supported by several other independent cosmological observations including baryon acoustic oscillations, weak gravitational lensing and counts of galaxy clusters. 
These observational probes of cosmic acceleration were recently reviewed in \textcite{Weinberg2013} and future prospects are described. 

Since the discovery of the accelerated expansion, the precision of the cosmological data increased spectacularly, in particular the measurements of the CMB. 
The $\Lambda$CDM model, which emerged at the end of the last century as a concordance cosmology, is still able to describe all observations. 
$\Lambda$CDM fits indicate that the energy budget of the Universe today is: 
\begin{equation}
\Omega_b = 5 \, \% \quad \Omega_{\rm DM} = 27 \, \% \quad \Omega_\Lambda = 68 \, \%, 
\end{equation}
where the non-relativistic matter has a baryonic and non-baryonic component ($\Omega_M = \Omega_{\rm DM} + \Omega_b$). 
These three components driving the expansion of the Universe today are associated with the three major puzzles in cosmology. 
It is often said that the baryonic content of the Universe is the only part which is well understood. 
In fact, the very presence of baryons results from the initial matter-antimatter asymmetry of the Universe, it cannot be explained by the standard theory. 
Next, the microscopic description of the non-baryonic Dark Matter is lacking. 
It could be made up of massive weakly interacting particles a.k.a. WIMPS. 
It could also well be something completely different like a Bose-Einstein condensate of very light scalar particles like Axions. 
Last, the microscopic nature of the Dark Energy is completely unknown. 
Contrary to Dark Matter, no well motivated extensions of the Standard Model of particle physics provide a natural candidate for the Dark Energy. 

\subsection{The cosmological constant problem}

In natural units ($\hbar = c = 1$),  the energy density associated with the Dark Energy is
\begin{equation}
\label{LAMBDA}
\rho_\Lambda = \Omega_\Lambda \, \rho_{c,0} = (2.2 \, {\rm meV})^4. 
\end{equation}
The fact that the energy scale associated with the Dark Energy is so curious constitutes the cosmological constant problem. 
In the $\Lambda$CDM model, the Dark Energy is simply the energy density of the vacuum, or equivalently a cosmological constant. 
In principle the vacuum energy is expected to receive contributions from several sources. 

At the classical level, any scalar field $\phi$ contributes to the vacuum energy by the amount $\mathcal{V}(\phi_0)$ where $\mathcal{V}(\phi)$ is the energy density function of the field and $\phi_0$ is the vacuum value of the field. 
One could argue that the zero energy is arbitrary, we can choose to define it as the state of lowest energy for all fields in the Universe. 
All local phenomena do not depend on this choice of the zero energy, there is no reason that it should have such a dramatic gravitational effect on cosmological scales. 
However, in the case of phase transitions, the effect of zero energy cannot be ignored by simply declaring that the zero energy cannot gravitate. 
During the electroweak symmetry breaking mentioned in section \ref{EDM}, the energy density associated with the Higgs field changes, by an amount as large as $(100 \, {\rm GeV})^4$. 
The vacuum energy was much larger before the electroweak phase transition. 
This difference in vacuum energy is definitely a physical meaningful quantity, according to straightforward application of the general relativity theory in a purely classical context. 
If differences in vacuum energy certainly gravitate, what determines the actual value of the vacuum energy? 
It could perfectly correspond to an intrinsic property of the Universe that should be treated as another free parameter.  
Why then is it adjusted so that the vacuum energy after the electroweak phase transition is almost zero, but not quite? 
It seems that Nature decided to ``lift'' the Mexican hat potential of the Higgs field in a very precise, fine-tuned amount. 

In addition, the vacuum energy receives contributions from quantum fluctuations induced by virtual particles. 
Quantum field theories predict that every bosonic degree of freedom of frequency $\omega$ has a zero-point energy of $\hbar \omega /2$. 
Since there are in principle an infinite number of degrees of freedom in the quantum fields, 
the total vacuum energy is technically infinite, 
unless the contribution from the fermionic and bosonic degrees of freedom compensate each other. 
If physics can be described by an effective local quantum field theory up to the Planck scale, 
\begin{equation}
m_{\rm Pl} = \sqrt{\frac{\hbar c}{8 \pi G}}= 2.4 \times 10^{18} \, {\rm GeV},
\end{equation}
dimensional analysis indicate that the vacuum energy induced by quantum fluctuations should be of the order of $m_{\rm Pl}^4 \approx 10^{120} \rho_\Lambda$. 
This is definitely the worse theoretical prediction ever. 
Supersymmetric compensation between the fermionic and bosonic quantum fluctuations above the SUSY breaking scale 
could reduce the disagreement between the expected and observed vacuum energy from $10^{120}$ down to $10^{58}$, still enormous. 

In summary, the vacuum energy can be separated as a sum of two terms: $\rho_\Lambda = \rho_{\Lambda, \rm class} + \rho_{\Lambda, \rm quant}$. 
The term induced by quantum fluctuations $\rho_{\Lambda, \rm quant}$ cannot be calculated in a self consistent way in quantum field theory but it is expected to be huge from dimensional analysis. 
The classical term $\rho_{\Lambda, \rm class}$ is the energy density of the scalar fields in the ground state, it is a free parameter of the theory. 
At the end, the vacuum energy can perfectly be considered as a free parameter in the present incomplete theories of particle physics and gravity. 
However, there is a naturalness problem very similar to the hierarchy problem of particle physics 
in the sense that ``bare'' value has to be extraordinarily fine-tuned to almost compensate for the huge quantum contribution. 

Well before the discovery of the accelerating Universe, \textcite{Weinberg1987} discussed an anthropic explanation to the cosmological constant problem. 
He gave an upper bound on the vacuum energy density: 
\begin{equation}
\label{Weinberg}
\rho_\Lambda \lesssim (10 \, {\rm meV})^4. 
\end{equation}
The argument goes as follows. 
If the vacuum energy is too large, the accelerated phase of the expansion starts too early. 
Then the inhomogeneities of the mass density do not have time to grow sufficiently before getting stretched away by the accelerated expansion. 
Only if the Weinberg bound \eqref{Weinberg} is satisfied can the perturbations grow to form stars and galaxies. 
Otherwise the Universe would consist in an homogeneous fluid in accelerated expansion, quite unfriendly for life. 
Imagining that many universes exist, with a different value for the cosmological constant in each, 
it is not surprising that we happen to live in a seemingly very special universe where life is possible at all. 
Indeed the observed vacuum energy density \eqref{LAMBDA} is not particularly fine-tuned with respect 
to the range of possible values in the anthropic sense given by \eqref{Weinberg}. 

This anthropic explanation is further supported by the landscape of string theory: 
there is a huge number of possible vacua of the string theory associated with the many many different ways to compactify the extra dimensions. 
One might believe that all these versions of string theory ``exist'' somehow, 
and intelligent life can emerge only in a tiny fraction of a vast number of possible versions. 
The fine-tuning of the parameters could then be just an illusion. 

In addition, the eternal inflation scenario supports the plausibility that many worlds can ``exist'' in a multiverse. 
In these theories, the Universe is inflating due to the energy density of a scalar field -- the inflaton -- which is in a false vacuum. 
The inflaton decays in the true vacuum, stopping the inflationary phase. 
But this process happens only locally, forming a bubble of true vacuum surrounded by the rest of the Universe which is still inflating. 
In some versions of the theory, the part of the Universe which is still inflating grows fast enough to accommodate for the continuous formation of bubbles 
(the speed of the bubble walls could be slower than the exponential expansion of the false vacuum space). 
In this case the inflationary phase of the Universe lasts forever in some regions of the Universe and causally disconnected bubbles are produced on and on. 
Different bubbles having different cosmological constants is not strictly speaking a prediction of eternal inflation, 
but if inflation is due to Planck-scale physics or stringy effects, who knows? 
We might be living in one out of a vast number of bubbles, where the vacuum energy density is such that it does not forbid the appearance of intelligent life. 

\subsection{The quintessence as Dark Energy}

The anthropic explanation of the cosmological constant problem is both elegant and fascinating. 
However it rests on speculative ideas, to say the least. 
This explanation is a source of great inspiration for imagination and philosophy, 
unfortunately it gives poor insight for further observations and experiments. 
It is thus necessary to explore other possibilities. 
We can take the view that Dark Energy is a dynamical process revealing new physics at the energy scale $\rho_\Lambda^{1/4} \approx 1$~meV. 
A review of the possible dynamics of Dark Energy and its observational consequences can be found in \textcite{Copeland2006}. 
Although none of these models really solve the cosmological constant problem in a top-down way, 
they are useful bottom-up approaches to guide future observations and experiments to refine our knowledge of Dark Energy. 

The most popular routes to a dynamical extension of the simple $\Lambda$CDM parametrization are quintessence models. 
Akin to inflationary models, the accelerated expansion is due to the energy density of a scalar field. 
The scalar field responsible for the acceleration of the Universe during the inflation era is called the inflaton, 
whereas the hypothetical scalar field making up the Dark Energy driving the late acceleration is called the quintessence. 
There are even models explaining both the early inflation and the late acceleration of the Universe using a single scalar field \cite{Peebles1999}. 

Let us review the theory of the gravitational dynamics of a scalar field. 
The action for gravity and the quintessence field $\varphi$ with the so-called \emph{minimal coupling} is
\begin{equation}
S = \int d^4 x \sqrt{-g} \left( - \frac{m_{\rm Pl}^2}{2} \, \mathcal{R} + \mathcal{L}_\varphi \right), 
\end{equation}
where $g = \det(g_{\mu \nu})$ is the determinant of the metric tensor, $\mathcal{R} = g^{\mu \nu} R_{\mu \nu}$ is the trace of the Ricci curvature tensor $R_{\mu \nu}$ and
\begin{equation}
\mathcal{L}_\varphi = \frac{1}{2} g^{\mu \nu} \partial_\mu \varphi \partial_\nu \varphi - \mathcal{V}(\varphi)
\end{equation}
is the Lagrangian of the scalar field with a generic potential $\mathcal{V}(\varphi)$. 

Einstein's equations governing the response of the metric to the scalar field 
are obtained from the variational principle $\delta S / \delta g^{\mu \nu} = 0$: 
\begin{equation}
G_{\mu \nu} = 8 \pi G \, T_{\mu \nu}^\varphi, 
\end{equation}
where $G_{\mu \nu} = R_{\mu \nu} -1/2 g_{\mu \nu} \mathcal{R}$ is Einstein's tensor and 
\begin{eqnarray}
T_{\mu \nu}^\varphi & = & 2 \frac{\delta \mathcal{L}_\varphi}{\delta g^{\mu \nu}} - g_{\mu \nu} \mathcal{L}_\varphi \\
& = & \partial_\mu \varphi \partial_\nu \varphi - \frac{1}{2} g_{\mu \nu} g^{\alpha \beta} \partial_\alpha \varphi \partial_\beta \varphi + g_{\mu \nu} \mathcal{V}(\varphi) 
\end{eqnarray}
is the stress-energy tensor of the scalar field. 

In the homogeneous Universe, the scalar field is uniform in a Robertson-Walker metric \eqref{RobertsonWalker}. 
The problem reduces to only two degrees of freedom $a(t)$ and $\varphi(t)$ and 
tensors take a diagonal form: 
\begin{eqnarray}
g_{\mu \nu}         & = & {\rm diag} \left( 1, - a^2 \right) \\
g^{\mu \nu}         & = & {\rm diag} \left( 1, - 1/a^2 \right) \\
G_{\mu \nu}         & = & {\rm diag} \left( 3H^2, a^2 [-3H^2 - 2\dot{H} ] \right) \\ 
T_{\mu \nu}^\varphi & = & {\rm diag} \left( \frac{1}{2} \dot{\varphi}^2 + \mathcal{V}(\varphi) , a^2 [ \frac{1}{2} \dot{\varphi}^2 - \mathcal{V}(\varphi) ] \right)
\end{eqnarray}

The $00$ component of Einstein's equations reduces then to the Friedmann equation
\begin{equation}
\label{FriedmannPhi}
H^2 = \frac{8 \pi G}{3} \left( \frac{1}{2} \dot{\varphi}^2 + \mathcal{V}(\varphi) \right). 
\end{equation}
In addition, the evolution of the scalar field is given by the variational principle $\delta S / \delta \varphi = 0$, which in the homogeneous Universe reduces to 
\begin{equation}
\label{KleinGordonPhi}
\ddot{\varphi} + 3 H \dot{\varphi} + \mathcal{V}'(\varphi) = 0.
\end{equation}

The coupled dynamics of the expansion (i.e. the time evolution of $H = \dot{a}/a$ and $\varphi$)
 is completely specified by the two equations \eqref{FriedmannPhi} and \eqref{KleinGordonPhi}. 
It is useful to notice that these two equations are equivalent to the two Friedman equations \eqref{F1},\eqref{F2} with 
$\rho = \dot{\varphi}^2/2 + \mathcal{V}(\varphi)$ the energy density of the quintessence and $p = \dot{\varphi}^2/2 - \mathcal{V}(\varphi)$ its pressure. 
The equation of state parameter 
\begin{equation}
w = \frac{p}{\rho} = \frac{\dot{\varphi}^2/2 - \mathcal{V}(\varphi)}{\dot{\varphi}^2/2 + \mathcal{V}(\varphi)}
\end{equation}
is a dynamical quantity that can evolve with time. 

Different cosmologies can be built depending on the potential $\mathcal{V}(\varphi)$. 
If $\varphi$ corresponds to a standard massive scalar field, such as the Axion, the potential $\mathcal{V}(\varphi) = \frac{1}{2} \mu^2 \varphi^2$ is parabolic. 
For the harmonic potential there is equipartition between the kinetic energy $\dot{\varphi}^2/2$ and the potential energy $\mathcal{V}(\varphi)$ and therefore $w=0$. 
In this case $\varphi$ is a candidate for pressurless cold Dark Matter. 
It is also possible to build models which mimic a cosmological constant with $w \approx -1$, where the field does not develop significant kinetic energy. 
Such a model has been first proposed by \textcite{Ratra1988}, with the potential given by
\begin{equation}
\label{RatraPeebles}
\mathcal{V}(\varphi) = M_\Lambda^4 \left( \frac{M_\Lambda}{\varphi} \right)^n
\end{equation}
where $M_\Lambda$ is a new mass scale and $n$ is a positive exponent called the Ratra-Peebles index. 

As a generic feature of quintessence models, a departure from $w=-1$ is expected. 
Therefore there is a chance that future supernovae surveys could detect a deviation from the $\Lambda$CDM model where $w=-1$. 
Now, is there another way to reveal the presence of the quintessence field $\varphi$? 
If the field were coupled to ordinary matter, then it would mediate a fifth force. 
It is then very appealing to design laboratory experiments searching for this new force. 

\subsection{The chameleon}

A quintessence field coupled with matter mediates a new force with a range extending up to cosmological scales. 
At first it was generally assumed that the coupling must be very small, much smaller than the gravitational strength, 
in order to satisfy the stringent tests of the equivalence principle, as well as the tests of general relativity in the Solar system. 
However, \textcite{KhouryPRL2004,KhouryPRD2004} discovered that the coupled quintessence scalar theory 
automatically features a very efficient mechanism to suppress the force called the chameleon mechanism. 
Then, \textcite{Brax2004} explored the cosmological consequences of the chameleon theory to find that it is a nice viable Dark Energy candidate. 

Let us explain the chameleon screening mechanism. 
We start by considering the Lagrange density of the scalar field $\varphi$ coupled with a fermion $\psi$ (with a mass $m$) in a Minkowski spacetime:
\footnote{The term $\beta m/m_{\rm Pl} \varphi \ \bar{\psi} \psi$ 
originates from a \emph{conformal coupling} of the scalar field to the fermion. 
The conformal coupling consists in replacing the metric $g_{\mu \nu}$ in the fermion Lagrange density by a function of the scalar field $A^2(\varphi) g_{\mu \nu}$ : $\mathcal{L}_m (A^2(\varphi)g_{\mu \nu}, \psi)$. 
A popular choice of the conformal coupling function is $A(\varphi) = e^{\beta \varphi / m_{\rm Pl}}$. 
Since the excursions of the field always satisfies $\beta \varphi / m_{\rm Pl} \ll 1$ in concrete cosmological or laboratory situations, we can approximate $A(\varphi) = 1 + \beta \varphi / m_{\rm Pl}$. 
It finally produces the Yukawa coupling term $\beta m/m_{\rm Pl} \varphi \ \bar{\psi} \psi$. 
} 
\begin{equation}
\label{lagrangien}
\mathcal{L} = \frac{1}{2} \partial_\mu \varphi \ \partial^\mu \varphi - \mathcal{V}(\varphi) - \beta \frac{m}{m_{\rm Pl}} \ \varphi \ \bar{\psi} \psi. 
\end{equation}
The dimensionless constant $\beta$ corresponds to the strength of the coupling relative to gravity. 
From the Lagrange density we can deduce in principle how $\varphi$ affects a particle, and conversely, how the presence of particles generates a field. 
\begin{enumerate}
\item A fermion evolving in a field $\varphi({\bf r})$ is affected by the potential 
\begin{equation}
V({\bf r}) = \beta \frac{m}{m_{\rm Pl}} \ \varphi({\bf r}), 
\end{equation}
corresponding to a force:
\begin{equation}
{\bf F} = - \beta \frac{m}{m_{\rm Pl}} \ \boldsymbol{\nabla} \varphi({\bf r}). 
\end{equation}
\item The field $\varphi({\bf r})$ generated by a distribution of fermions is given by the Euler-Lagrange equation
\begin{equation}
\partial_\mu \left( \frac{\partial \mathcal{L}}{\partial(\partial_\mu \varphi)} \right) - \frac{\partial \mathcal{L}}{\partial \varphi} = 0.
\end{equation}
\end{enumerate}
More specifically, the Euler-Lagrange equation deduced from the Lagrange density \eqref{lagrangien} reads
\begin{equation}
\label{KG1}
\partial_\mu \partial^\mu \varphi + \mathcal{V}'(\varphi) + \beta \frac{ \rho }{m_{\rm Pl}} = 0, 
\end{equation}
where $\rho$ is the mass density of the fermions (we have assumed the non-relativistic limit $m \bar{\psi} \psi = \rho$). 

\begin{figure}
\centering
\includegraphics[width=0.93\linewidth]{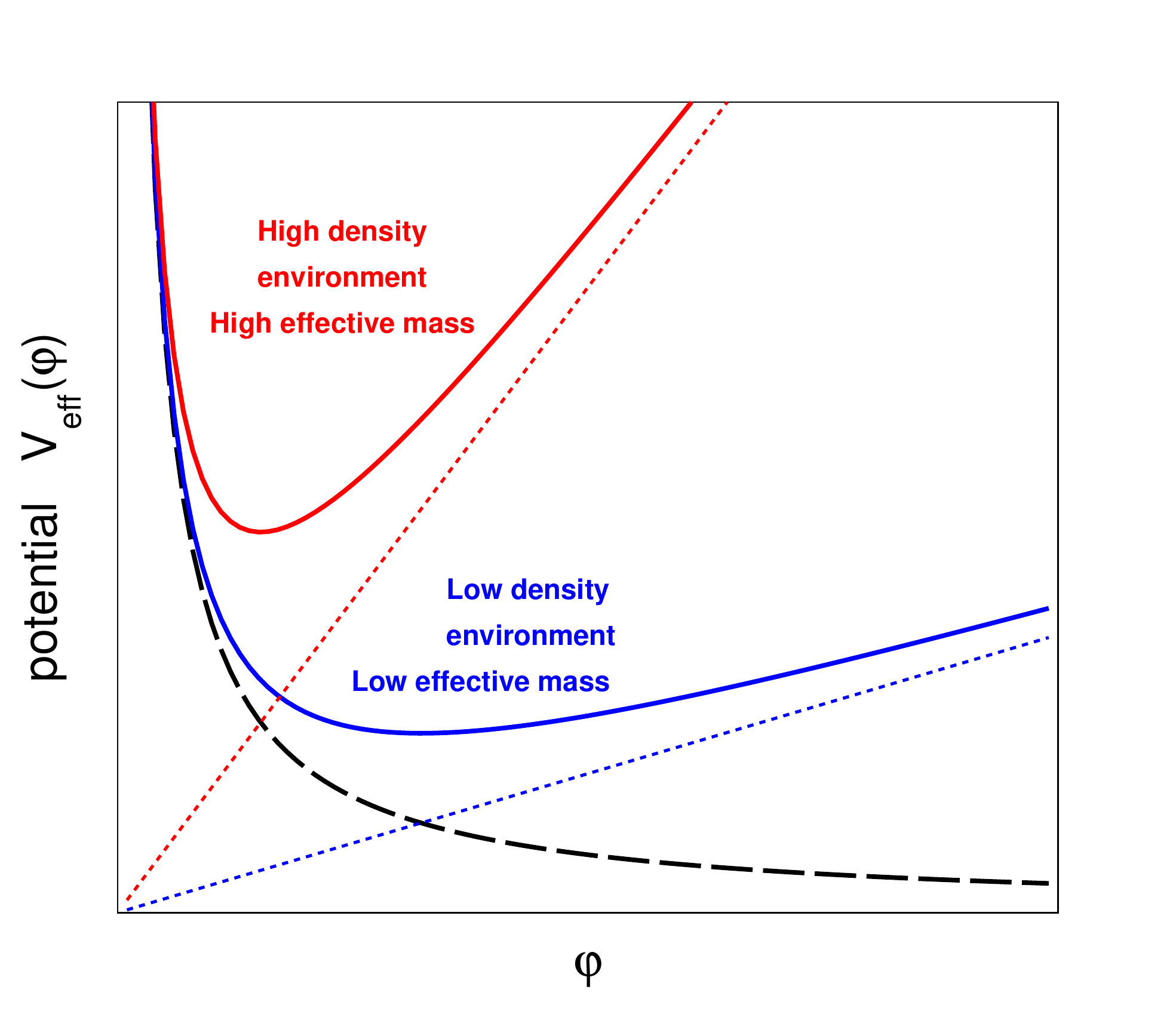}
\caption{
The black dashed line represents the runaway potential $\mathcal{V}(\varphi)$. 
The solid lines represent the effective potential $\mathcal{V}_{\rm eff}(\varphi)$ 
where the environment density $\rho$ is low (blue) and where the density is high (red). 
 \label{chameleonPotential}
}
\end{figure}

It is then apparent that, within an environment of density $\rho$, 
the dynamics of the chameleon is not governed by $\mathcal{V}(\varphi)$, 
but rather by the effective potential
\begin{equation}
\mathcal{V}_{\rm eff} = \mathcal{V}(\varphi) + \beta \frac{\rho}{m_{\rm pl}} \ \varphi, 
\end{equation}
in the sense that Eq. \eqref{KG1} can be written in the form $\partial_\mu \partial^\mu \varphi + \mathcal{V}_{\rm eff}'(\varphi) = 0$. 

Let us recall that the equation governing a basic massive scalar field (with a potential $\frac{1}{2} \mu^2 \varphi^2$) is the Klein-Gordon equation 
$\partial_\mu \partial^\mu \varphi + \mu^2 \varphi = 0$, where $\mu$ is the mass of the elementary excitations of the field. 
In this well-known case, the field generated by a static point source (if the field is coupled to fermions, this corresponds to a source term in the right hand side of the Klein-Gordon equation) 
takes the form of a Yukawa potential $\varphi(r) \simeq e^{-\mu r}/r$. 
It means physically that the field mediates an interaction between fermions with a range $1/\mu$ (in S.I. units,  $\lambda = \frac{\hbar c}{\mu c^2}$). 
A very massive field will mediate a short-range force, whereas a light field can travel long distances. 

We now come back to the more complicated situation of the coupled quintessence field. 
We will discuss the consequences of the specific features of the effective potential depicted in Fig. \ref{chameleonPotential}. 
Assuming that the density $\rho$ is uniform everywhere, then there is a static and uniform solution to Eq. \eqref{KG1}, 
given by the minimum of the effective potential $\mathcal{V}_{\rm eff}'(\varphi_{\rm min}) = 0$. 
In the case of the Ratra-Peebles potential \eqref{RatraPeebles} the minimum of the effective potential reads
\begin{equation}
\label{phiMin}
\varphi_{\rm min} = M_\Lambda \left( \frac{n M_\Lambda^3 m_{\rm pl}}{\beta \rho} \right)^{\frac{1}{n+1}}. 
\end{equation}
For small perturbations of the field around this uniform value 
(induced by an extra mass on top of the environment density $\rho$ for example), 
we can approximate the effective potential around the minimum by
\begin{equation}
\mathcal{V}_{\rm eff}(\varphi) \approx \mathcal{V}_{\rm eff}(\varphi_{\rm min}) + \frac{1}{2} (\varphi - \varphi_{\rm min})^2 \ \mathcal{V}_{\rm eff}''(\varphi_{\rm min}).
\end{equation}
We can attribute an effective mass $\mu$ for the perturbations of the field, with 
\begin{equation}
\label{effectiveMass}
\mu^2 = \mathcal{V}_{\rm eff}''(\varphi_{\rm min}) = \beta \frac{\rho}{m_{\rm pl}} \frac{n+1}{\varphi_{\rm min}}. 
\end{equation}
The effective mass is associated with the curvature of the effective potential at the minimum. 
Clearly, what is happening here is that the mass of the field is density-dependent, 
in such a way that the range of the force mediated by the field shrinks in a high density environment. 
For this reason, the scalar field is called ``chameleon'': 
it adapts its properties to the environment to evade observation. 

Consider for example the Sun as a source of the chameleon field. 
Since the density in the interior of the Sun is quite high, the chameleon is effectively massive in the Sun. 
Therefore, a test particle situated outside the Sun is not affected by the inner part of the Sun because the field 
created by the mass inside the Sun cannot propagate to the outside. 
In fact, it is shown in \textcite{KhouryPRL2004,KhouryPRD2004} that when the coupling $\beta$ is large enough, 
then only a thin shell at the surface of a large body such as the Sun or the Earth contributes to the force on a test particle outside the body. 
Then, in a more elaborate analysis, \textcite{Mota2007} concluded that strongly coupled chameleons, 
i.e. $\beta \gg 1$, are not ruled out by terrestrial and solar system tests of gravity. 

\subsection{1D solutions of the chameleon equation}

\begin{figure}
\centering
\includegraphics[width=0.93\linewidth]{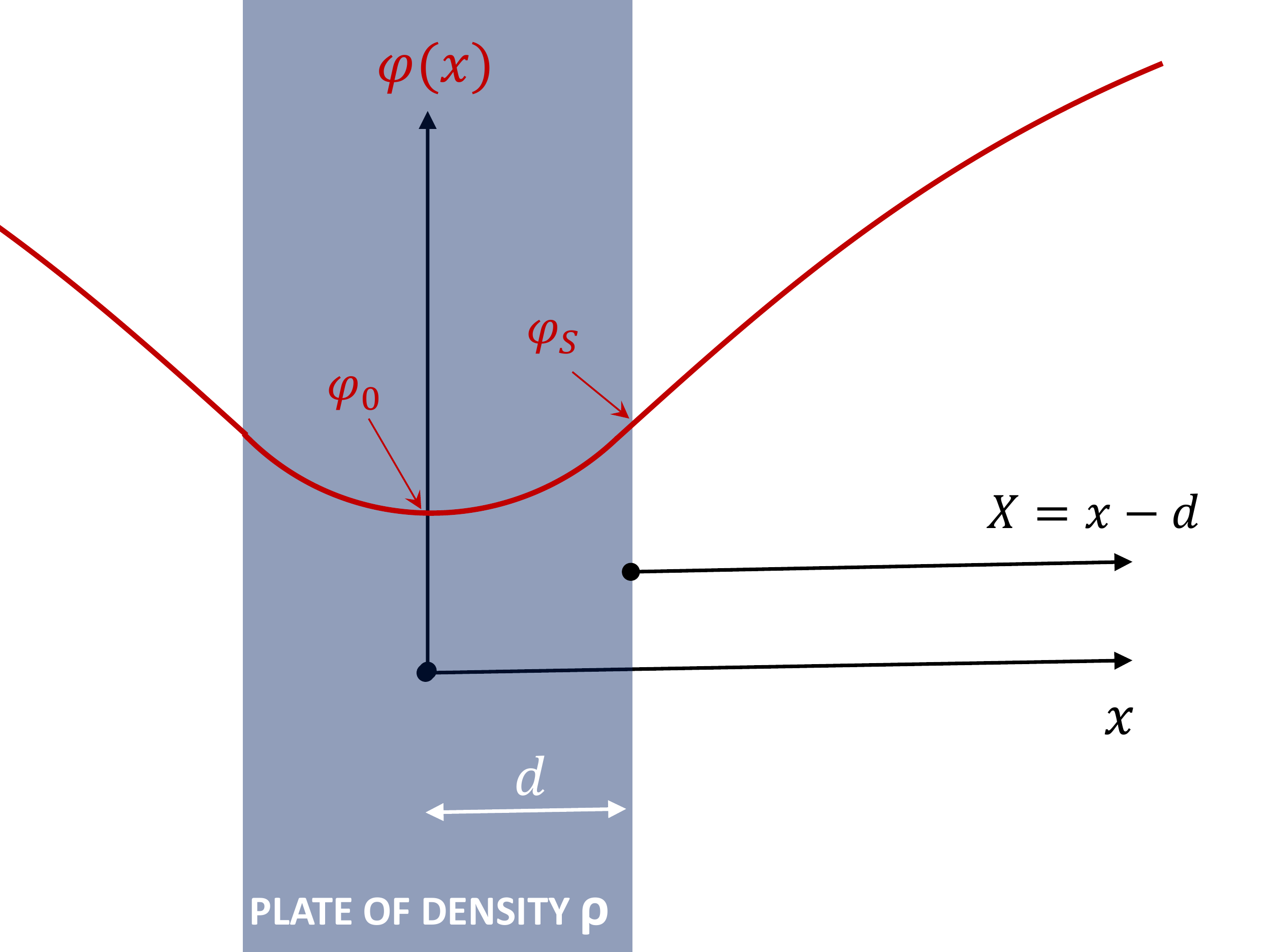}
\caption{
Sketch of the one-dimensional problem. 
 \label{chameleon1D}
}
\end{figure}

In order to gain more insight about the properties of the chameleon field, 
we shall now solve explicitly the chameleon equation in the case of the simple one-dimensional problem 
illustrated on Fig. \ref{chameleon1D}. 
We consider a plate of material (to be specific, a plate of aluminium with thickness $2 d = 1$~mm) surrounded by perfect vacuum. 

We consider a static solution in one dimension, therefore 
$\partial_\mu \partial^\mu \varphi = \frac{\partial^2 \varphi}{\partial t^2} - \Delta \varphi = - \frac{d^2 \varphi(x)}{dx^2}$. 
In this case the chameleon equation \eqref{KG1} becomes
\begin{equation}
\frac{d^2 \varphi}{dx^2} = \mathcal{V}_{\rm eff}'(\varphi)
\end{equation}
In addition, the symmetry of the problem implies $\varphi(-x) = \varphi(x)$. 
We will therefore solve for $x>0$ with the boundary condition $\varphi'_0 = 0$. 
For simplicity we treat the case with $n=2$ but other Ratra-Peebles indices can be treated similarly. 

\para{Solution in vacuum. } 
In the vacuum ($X>0$), the chameleon equation reads
\begin{equation}
\frac{d^2 \varphi}{dX^2} = - 2 \ \frac{M_\Lambda^6}{\varphi^3}. 
\end{equation}
we can find a family of solutions with different values $\varphi_S$ at the boundary: 
\begin{equation}
\label{vacSol1}
\varphi(X) = \varphi_S \sqrt{ 1 + X/X_S }
\end{equation}
with $X_S$ satisfying
\begin{equation}
\label{vacSol2}
X_S = \frac{\sqrt{2}}{4} \frac{\varphi_S^2}{M_\Lambda^3}. 
\end{equation}
Notice the remarkable expression of the derivative at the boundary
\begin{equation}
\varphi_S' = \sqrt{2} \ \frac{M_\Lambda^3}{\varphi_S}.
\end{equation}

Next, we deal with the inside of the plate ($0<x<d$). 
No exact solution exists for the problem, 
we need to consider two asymptotic regimes. 

\para{Solution inside the plate: linear regime. } 
We start by considering what we call the linear regime, 
where $\varphi_0 \gg \varphi_{\rm min}$. 
We will derive {\rm a posteriori} the corresponding condition for the parameter $\beta$. 
We recall that $\varphi_{\rm min}$ is the value of the field 
minimizing the effective potential inside the plate, it is given by Eq. \eqref{phiMin}. 
In this regime, the effective potential is dominated by the coupling term and we can neglect the self-interaction term $\mathcal{V}(\varphi)$. 
Therefore, the chameleon equation reduces to a Poisson equation :
\begin{equation}
\frac{d^2 \varphi}{dx^2} \approx \beta \frac{\rho}{m_{\rm pl}} = \frac{2 M_\Lambda^6}{\varphi_{\rm min}^3}.
\end{equation}
with the solution
\begin{equation}
\varphi(x) = \varphi_0 + \frac{M_\Lambda^6}{\varphi_{\rm min}^3} x^2
\end{equation}
We now need to connect this solution to the vacuum solution at the boundary by equalizing the value of the field and the derivative. 
We get the solution
\begin{equation}
\frac{\varphi_0}{M_\Lambda} =  \frac{1}{\sqrt{2} (M_\Lambda d)}\left( \frac{\varphi_{\rm min}}{M_\Lambda} \right)^3 - (M_\Lambda d)^2 \left( \frac{M_\Lambda}{\varphi_{\rm min}} \right)^3. 
\end{equation}
Now we can work out {\it a posteriori} the domain of validity of the linear approximation. 
We see from the solution that the condition $\varphi_0 \gg \varphi_{\rm min}$ is equivalent to the condition
$M_\Lambda d \ll (\varphi_{\rm min}/M_\Lambda)^2$, or, expressed in terms of the basic parameters: 
\begin{equation}
\label{linearRegime}
{\rm linear \ regime: } \quad \quad  \beta \ll \frac{m_{\rm pl} M_\Lambda^3}{\rho} \ \frac{1}{(M_\Lambda d)^{3/2}}. 
\end{equation}

Let us evaluate numerically this condition in the case of our aluminium plate. 
Equations are expressed in natural units, so we will express all masses in eV and all distances in eV$^{-1}$. 
We will consider the energy scale of the chameleon potential to correspond to the Dark Energy scale: 
\begin{equation}
M_\Lambda = \rho_\Lambda^{1/4} = 2.4 \times 10^{-3} \ {\rm eV}. 
\end{equation}
The Planck mass amounts to $m_{\rm pl} = 2.4 \times 10^{27}$~eV. 
The half-thickness of the plate is $d = 0.5 \ {\rm mm} = 2.5 \times 10^{3} \ {\rm eV}^{-1}$ 
and the density of aluminium is $\rho = 2.7 \ {\rm g/cm}^3 = 1.2 \times 10^{19} \ {\rm eV}^4$. 
We find that the solution is described by the linear regime when 
\begin{equation}
\beta \ll 0.2.
\end{equation}
The linear regime does not describe strongly coupled chameleons (i.e. $\beta \gg 1 $) in the presence of ordinary objects of normal size! 

\para{Solution inside the plate: saturated regime. } 
In the opposite regime, which we call the saturated regime, 
the field inside the plate is very close to $\varphi_{\rm min}$. 
We recall that $\varphi_{\rm min}$ would be the uniform value of the field 
in the situation where the plate with density $\rho$ fills the entire space uniformly. 
In the saturated regime we can write $\varphi(x) = \varphi_{\rm min} + \phi$, 
with $\phi \ll \varphi_{\rm min}$. 
In this case, the chameleon equation reduces to the Klein-Gordon equation
\begin{equation}
\frac{d^2 \phi}{dx^2} = \mu^2 \phi
\end{equation}
where the effective mass $\mu$ satisfies Eq. \eqref{effectiveMass}, that is: 
\begin{equation}
\mu^2 = \beta \frac{\rho}{m_{\rm pl}} \frac{3}{\varphi_{\rm min}} = \frac{6 M_\Lambda^6}{\varphi_{\rm min}^4}.
\end{equation}
At this point it is useful to notice that the opposite of 
the linear regime condition Eq. \eqref{linearRegime} can be written in the form: 
\begin{equation}
\label{saturatedRegime}
{\rm saturated \ regime: } \quad \quad  \mu d \gg 1. 
\end{equation}
Physically, it means that in the saturated regime the range of the force mediated by the chameleon is much shorter than the thickness of the plate. 

The solution of the Klein-Gordon equation is:
\begin{equation}
\phi(x) = \phi_0 \cosh (\mu x).
\end{equation}
The continuity of $\varphi$ and $\varphi'$ at the boundary provides a quadratic equation for $\phi_0$:
\begin{equation}
\phi_0 \mu \sinh(\mu d) \left( \varphi_{\rm min} + \phi_0 \cosh(\mu d) \right) = \sqrt{2} M_\Lambda^3, 
\end{equation}
or equivalently, we get a quadratic equation for $\varphi_S = \varphi_{\rm min} + \phi_0 \cosh \mu d$: 
\begin{equation}
(\varphi_S - \varphi_{\rm min} ) \varphi_S = \sqrt{2} \frac{M_\Lambda^3}{\mu} \  \frac{\cosh \mu d}{\sinh \mu d}. 
\end{equation}

\begin{figure}
\centering
\includegraphics[width=0.93\linewidth]{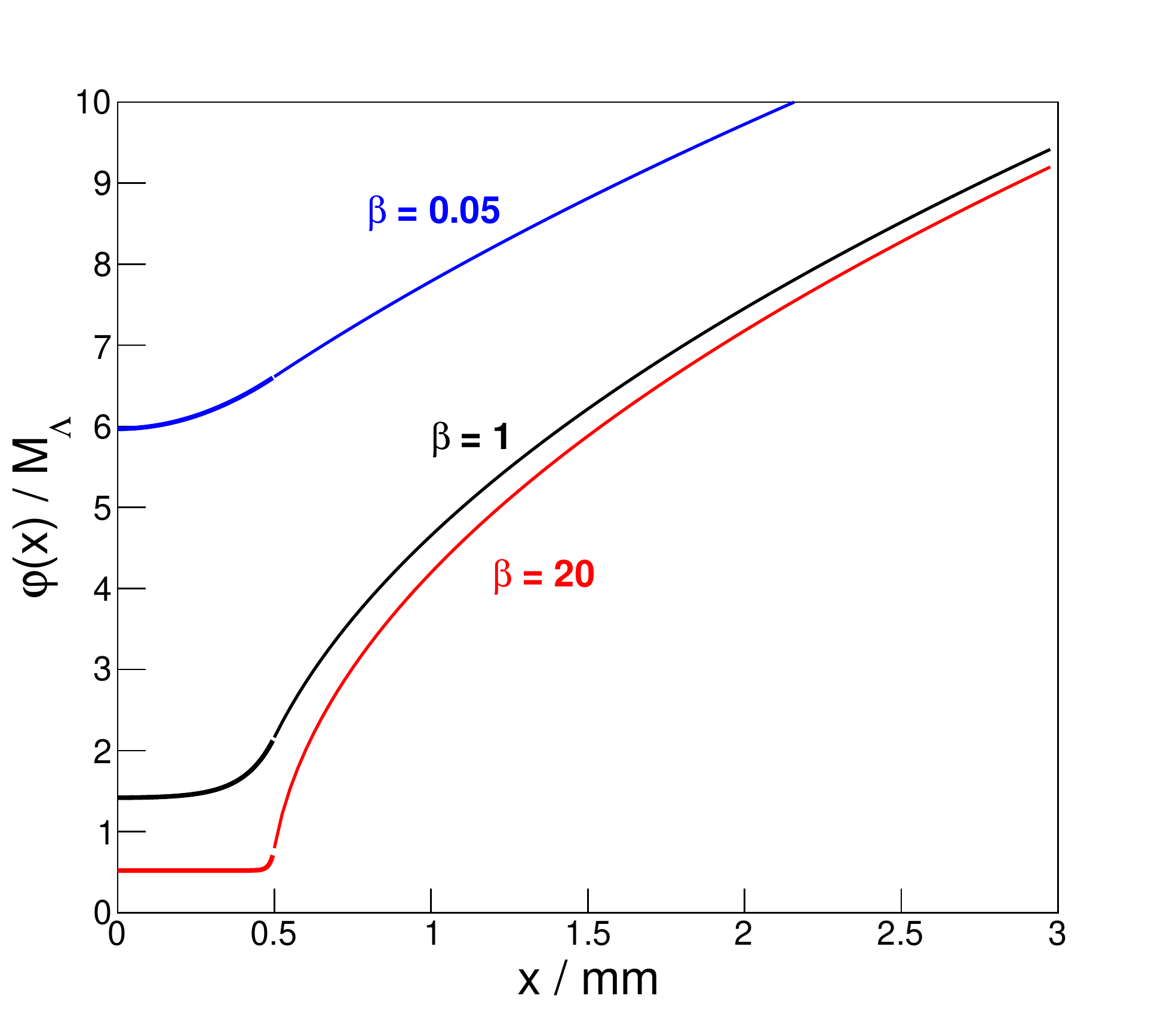}
\caption{
Solution of the 1D chameleon equation with $n=2$. 
The half-thickness of the aluminium plate is $d = 0.5$~mm. 
The blue curve ($\beta = 0.05$) is calculated using the linear approximation. 
The black ($\beta=1$) and red ($\beta=20$) curves are calculated using the saturated approximation. 
 \label{solution1D}
}
\end{figure}

\para{Interpretation of the solution. }
The solution of the chameleon equation on the right side of the plate 
is plotted on Fig. \ref{solution1D}, for three different values of the coupling constant $\beta$. 
The field inside the plate is ``attracted'' to small values. 
Outside the plate, the field wants to grow to minimize the potential $\mathcal{V}(\varphi)$. 

For strongly coupled chameleons, the field inside the plate takes a nearly uniform value $\varphi = \varphi_{\rm min}$. 
When the field is saturated inside the bulk, the plate acts as a screen: 
if a source mass moves on the left side of the plate, the field will be modified on the left side but it will remain saturated to $\varphi_{\rm min}$ inside the plate. 
Therefore the solution on the right side of the plate will remain unaffected. 
The plate shields the right side from what happens on the left side. 
This screening mechanism makes the chameleon field difficult to detect by laboratory experiments. 

\begin{figure}
\centering
\includegraphics[width=0.93\linewidth]{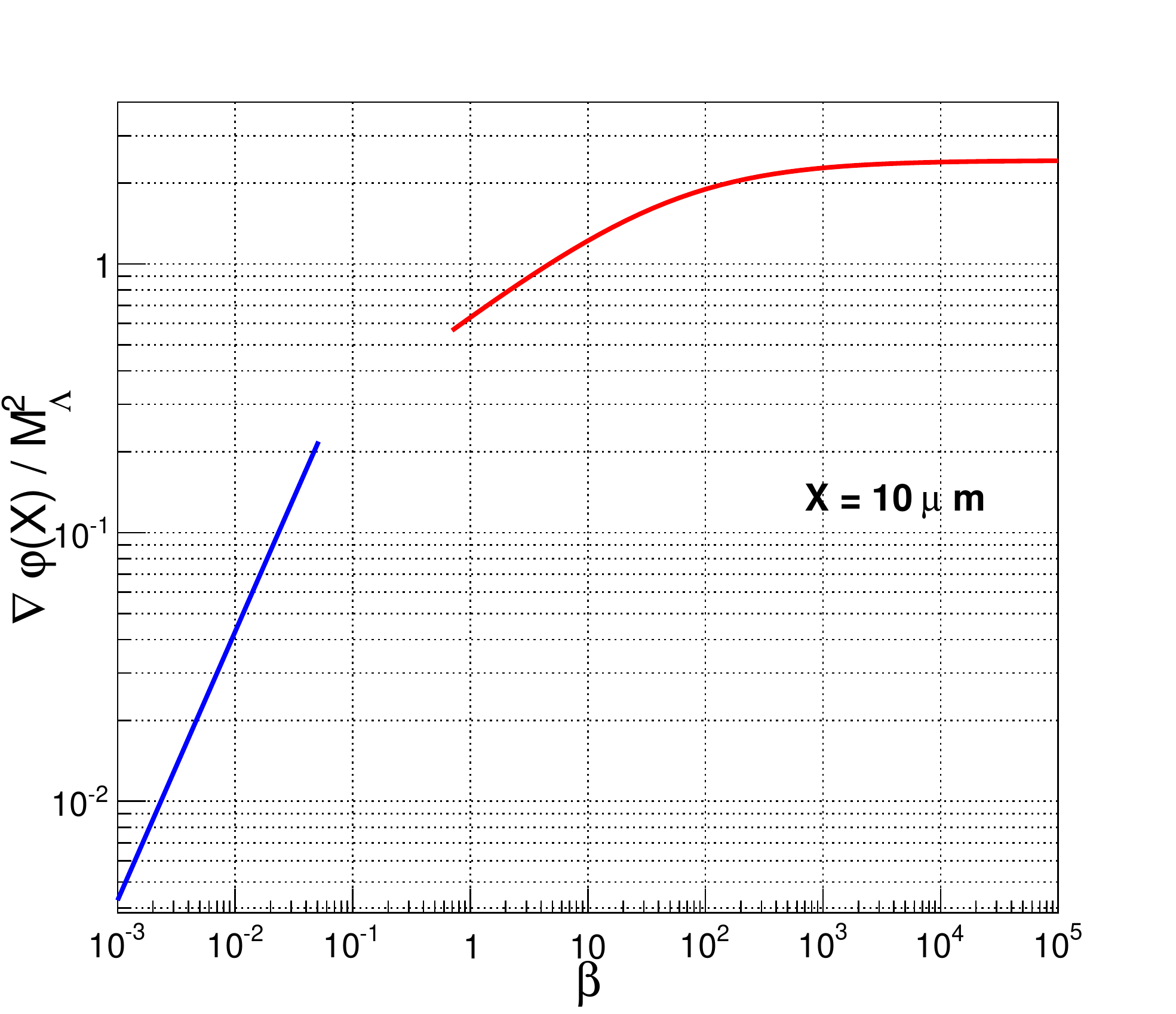}
\caption{
Gradient of the chameleon field at $X=10 \ \micron$ from the surface of the aluminium plate, as a function of $\beta$. 
The blue curve is calculated using the linear approximation, the red curve is calculated using the saturated approximation. 
 \label{solution1Dgrad}
}
\end{figure}

Another important feature of the chameleon is the saturation of the field. 
We see in Fig. \ref{solution1D} that the solution outside the plate becomes independent of $\beta$ for large values of $\beta$. 
To better understand this, let us discuss the gradient of the field $\frac{d \varphi}{dx}$ at the immediate vicinity of the plate, say $X = 10 \ \mu$m. 
The gradient of the field is analogous to an electric field (which is the gradient of the electric potential). 
The force acting on a test mass is proportional to $\frac{d \varphi}{dx}$ in the same way as the force acting on a charged particle is proportional to the electric field. 
There is no saturation effect for the electric field, in the sense that an electrically charged plate produces at the vicinity of the surface an electric field proportional to the charge density of the plate. 
Fig. \ref{solution1Dgrad} shows the gradient of the chameleon field as a function of $\beta$. 
In the linear regime, the gradient grows linearly with $\beta$ (or $\rho$) as for the electric field (hence the name of the linear regime). 
On the contrary, in the saturated regime, the gradient saturates to a finite asymptotic value. 

As a matter of fact, when searching for strongly coupled chameleons, 
one should think of solid matter as a zero boundary condition for the field $\varphi$. 
In this case the field in the vicinity of the surface of any plate has been derived by \textcite{Brax2011} for any Ratra-Peebles index $n$. 
In S.I. units it reads: 
\begin{equation}
\label{OnePlate}
\varphi(X) = M_\Lambda \left( \frac{2+n}{\sqrt{2}} \ \frac{M_\Lambda}{\hbar c} \ X \right)^{2/(2+n)}. 
\end{equation}
In the particular case $n=2$ we recover Eq. \eqref{vacSol1} in the limit $\varphi_S = 0$. 
This solution is independent of the coupling strength $\beta$, because of the saturation property we just discussed. 

The problem of the ``capacitor'', i.e. two parallel plates separated by a thickness $2 R$ of vacuum, 
has been considered by \textcite{Ivanov2013}. 
They proposed the approximate analytical solution 
\begin{equation}
\label{TwoPlates}
\varphi(x) = M_\Lambda \left( \frac{2+n}{2 \sqrt{2}} \ \frac{R M_\Lambda}{\hbar c} \ \left[ 1 - \frac{x^2}{R^2} \right] \right)^{2/(2+n)}
\end{equation}
which is exact for $n=2$. 
In this equation $x=0$ refers to the middle position of the capacitor. 

\begin{figure}
\centering
\includegraphics[width=0.93\linewidth]{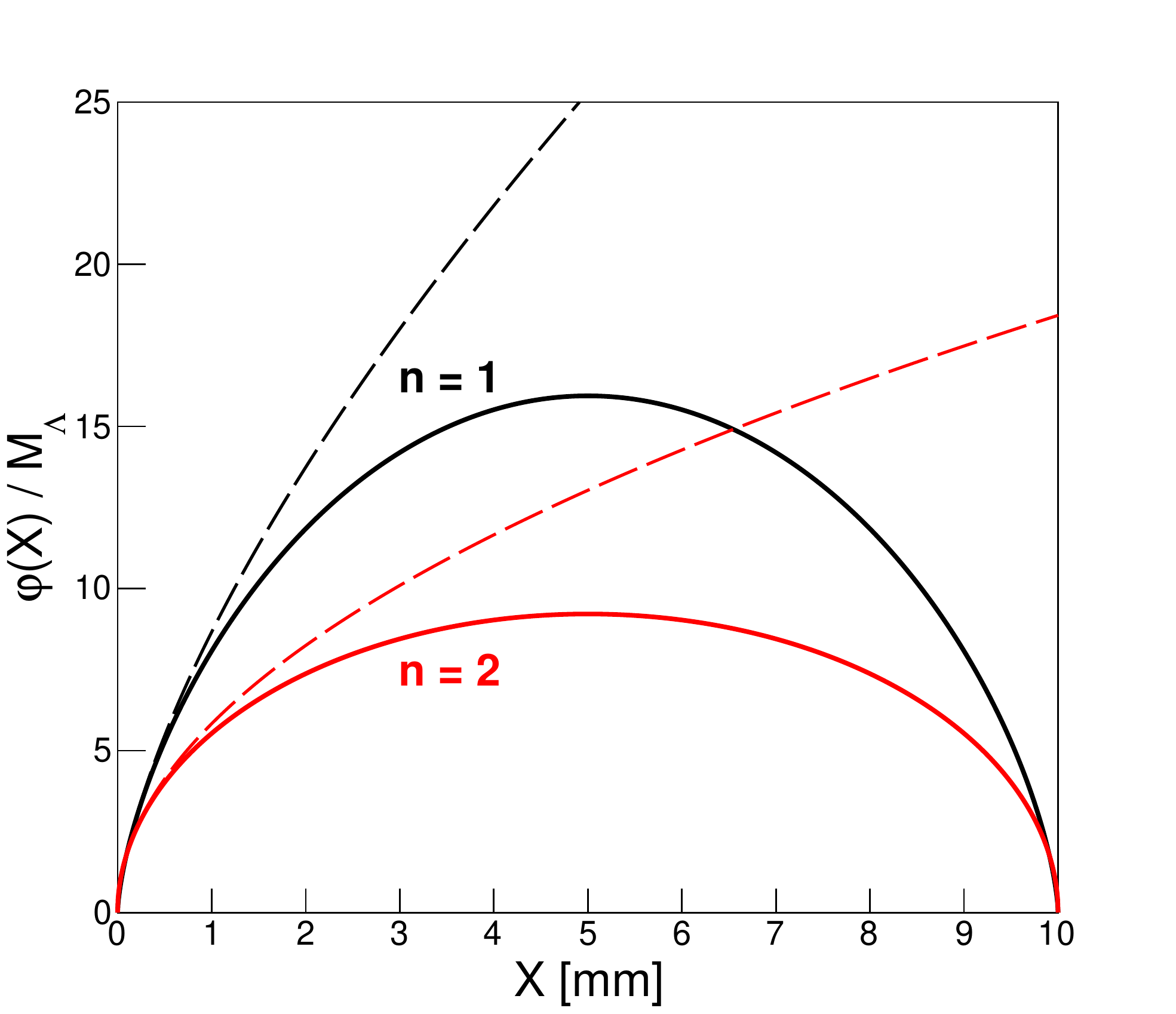}
\caption{
Solution of the chameleon equation for $n=1$ (black curves) and $n=2$ (red curves). 
Solid lines correspond to the two-plates solution Eq. \eqref{TwoPlates} with $2 R = 1$~cm. 
Dashed lines correspond to the one-plate solution Eq. \eqref{OnePlate}. 
 \label{solution1Dbubble}
}
\end{figure}

The one-plate and two-plate chameleon profiles are displayed in Fig. \ref{solution1Dbubble}. 
What is the physical reality of those chameleon profiles? 
Consider a test particle of mass $m$ in the vicinity of a plate, or inside a two-plate capacitor, 
the chameleon profile generates a potential energy $\beta m / m_{\rm pl} \varphi(x)$. 
We will discuss later suitable neutron experiments to probe these chameleon fields. 

It is important to note that a macroscopic test mass will not be subjected to the same potential energy. 
For strong couplings $\beta$, the core of the test mass will be shielded from the external chameleon field. 
It is sometimes useful to think of experiments searching for a new force as made of two components: a source (sometimes called an attractor) and a probe (sometimes called a detector). 
In the case of strongly coupled chameleons, a macroscopic source such as a plate produces a saturated chameleon field, independent of $\beta$. 
Also, the response of a macroscopic probe will show some saturation effect, 
whereas the response of a particle such as a neutron will be proportional to the coupling strength $\beta$.

\subsection{Searching for the chameleon in the lab}

\para{Probing new forces with neutrons. } 
Since the existence of new interactions are generic predictions of theories beyond the standard model, 
they are actively searched for in a great variety of experiments probing scales from subatomic to astronomical distances (see \textcite{Antoniadis2011} for a recent review). 

It is however quite recently that slow neutrons were recognized as interesting probes of new forces. 
In particular, neutron scattering experiments were found to be sensitive to new Yukawa forces of nanometric range \cite{Nesvizhevsky2008,Pokotilovski2006}. 
At the micrometer scale, the measurement of quantum states of neutrons bouncing over a mirror provides some constraints, 
however they are not as stringent as those derived from fifth-force searches using macroscopic bodies. 

In addition, neutrons can be sensitive to spin-dependent forces in the sub-millimeter range (induced by new bosons of mass $\mu > 10^{-3} \ {\rm eV}$). 
Limits on CP violating spin-dependent forces induced by the exchange of light Axionlike particles were first set by bouncing neutrons 
\cite{Baessler2007,Baessler2009,Jenke2014}. 
These limits were quickly superseded by experiments measuring the spin-precession of ultracold neutrons 
\cite{Serebrov2010,Afach2015}. 
Now the neutron limits compete with experiments using hyper-polarized $^3$He \cite{Petukhov2010}. 
At shorter range, a neutron diffraction experiment provides an interesting constraint on Axionlike particles \cite{Voronin2009}. 
There are also neutron beam experiments probing new spin-dependent interactions mediated by spin 1 bosons 
\cite{Piegsa2012,Yan2013}. 

\para{Neutrons probing strongly coupled chameleons. } 
Due to the peculiarity of the force induced by the chameleon field, 
the constraints on generic short range Yukawa force do not generally apply. 
In this case, it is quite attractive to consider the neutron as a probe of strongly coupled chameleons 
because of the absence of the saturation effect. 

The case of the neutron bouncer is particularly interesting. 
Ultracold neutrons can bounce above an horizontal plate because 
they are specularly reflected by the Fermi potential when they fall on the plate (which we call a mirror). 
It is possible to prepare neutrons almost at rest (concerning the vertical motion), so that the bouncing height is only a fraction of a millimeter. 
In this case the energy of the vertical motion is quantized, 
as demonstrated experimentally for the first time at the Institut Laue Langevin by \textcite{Nesvizhevsky2002}. 
We \cite{Brax2011} argued that the chameleon field Eq. \eqref{OnePlate} would be detectable in this system and deduced the upper limit $\beta < 10^{11}$. 
Next, \textcite{Jenke2011} realized the first spectroscopy of the quantum states by inducing resonant transitions. 
The chameleon field would shift the frequency of the resonances and the upper limit $\beta < 6 \times 10^8$ was deduced \cite{Jenke2014}. 
In part \ref{sectionGRANIT} we will dwell on this topic and give the prospects of the GRANIT experiment. 

Besides quantum states of bouncing neutrons, there is a second route: neutron interferometry. 
\textcite{Pokotilovski2013} proposed to build a Llyod's mirror interferometer for very cold neutrons to probe the chameleon. 
Such a method could in principle be sensitive to chameleon couplings down to $\beta \sim 10^{7}$ but the cold neutron interferometer
technique has yet to be developed. 
As an alternative, we \cite{BraxPignol2013} proposed to use triple Laue-case interferometers with slow
neutrons that have been operated routinely for decades in several neutron facilities. 
A first experiment has been performed in summer 2013 at the Institut Laue Langevin \cite{Lemmel2015}. 
We will describe it in part \ref{sectionInterferometry}. 

The constraints in the chameleon parameter space (Ratra-Peebles index $n$ and matter coupling $\beta$) from these neutron experiments 
are shown in Fig. \ref{exclusion} together with constraints from non-neutron experiments. 

\para{Other laboratory searches for chameleons. } 
Aside from neutron experiments, there are of course a few other means to probe the chameleon. 
\begin{itemize}
\item Using torsion pendulum with exquisite sensitivity 
(such as \textcite{Kapner2007}) one searches for an anomalous torque between a rotating attractor disk and a torsion pendulum separated by a fraction of a millimeter. 
The detector and the attractor are both disks with a diameter of a few centimeters, they are then subjected to the chameleon screening mechanisms in the case $\beta \gg 1$. 
\textcite{Upadhye2012} undertook the very complicated task of deriving the limits on the chameleon couplings from this experiment. 
For $n=1$ the excluded range is $10^{-2} < \beta < 10$. 
Couplings larger than 10 are allowed!
\item Measurements of the Casimir force could also be sensitive to chameleons \cite{Brax2007}. 
Existing experiments are sensitive only in the region of the parameter space where $M_\Lambda > 0.1~$eV. 
Future experiments could be sensitive to the interesting case $M_\Lambda = 2.4 \times 10^{-3}$~eV for a wide range of coupling $\beta$. 
\item Atomic precision tests in the hydrogen atom can reveal the presence of a scalar field coupled to protons and electrons. 
\textcite{BraxBurrage2011} derived the limit $\beta < 10^{14}$ from existing data. 
\item An atom-interferometry experiment using ultracold cesium atoms \cite{Hamilton2015} has been reported (while this manuscript was nearly finished). 
They obtained the limit $\beta < 2 \times 10^4$ for $n=1$. 

\item In addition to the coupling to matter (the term $\beta m / m_{\rm Pl} \varphi \bar{\psi} \psi$), the chameleon field can be coupled to photons with the Lagrange density term 
\begin{equation}
\frac{1}{4} \frac{\beta_\gamma}{m_{\rm Pl}} \ \varphi \ F_{\mu \nu} F^{\mu \nu}. 
\end{equation}
This case leads to a unique experimental signature. 
Due to the $\gamma \gamma \varphi$ vertex, chameleons can be produced by shining an intense laser in a region permeated by an intense magnetic field (take a spare dipole magnet from a giant particle accelerator). 
In this context, a chameleon is an elementary quantum excitation of the field $\varphi$. 
The production region is a vacuum chamber, the produced chameleon has a very low mass. 
The walls of the vacuum chamber act as a repulsive potential for the chameleon, because the mass of the chameleon is large inside the walls. 
Therefore the chameleons can be stored in the chamber (provided that the coupling to matter is strong enough, in practice it requires $\beta > 10^4$). 
After switching off the laser, the trapped chameleons can be converted back to photons  through the reverse of the process that formed them, one would then measure an \emph{afterglow}. 
A dedicated experiment called CHASE (Chameleon Afterglow Search) was performed at Fermilab \textcite{Steffen2010}. 
No afterglow was observed, the experiment sets the limit $\beta_\gamma < 10^{11}$. 
\end{itemize}

\para{Modified gravity. } 
We conclude this part by mentioning that the chameleon theory is one 
amongst several known (theoretically) screening mechanisms (see \textcite{Khoury2013} for a pedagogical overview). 
In the general framework of modified gravity, the Einstein-Hilbert action of general relativity is modified by 
adding a scalar field $\varphi$. 
There are many possible choices for the potential $\mathcal{V}(\varphi)$, for the coupling to the standard model $A(\varphi)$, for non-standard kinetic terms, etc. 
To understand the different possibilities we follow \textcite{Brax2013} and write the Lagrange density linearised around a background configuration $\varphi_0$: 
\begin{equation}
\mathcal{L} = \frac{Z(\varphi_0)}{2} \partial_\mu \varphi \partial^\mu \varphi - \frac{1}{2} \mu^2(\varphi_0) \varphi^2 + \frac{\beta(\varphi_0)}{m_{\rm Pl}} \varphi \rho. 
\end{equation}
In general the background configuration $\varphi_0$ depends on the environment density $\rho$, 
a property which could bring about screening. 
We can see that there are three general classes of screening mechanisms. 
\begin{itemize}
\item The mass $\mu(\varphi_0)$ becomes large in a dense environment. This is the chameleon mechanism. 
\item $Z(\varphi_0)$ becomes large in a dense environment. 
A special case of this is the \emph{Galileon} field. 
\item The coupling $\beta(\varphi_0)$ becomes small in a dense environment. 
The Damour-Polyakov mechanism belongs to this class. 
The \emph{symmetron} is another example. 
\end{itemize}

\separe

\begin{figure}
\centering
\includegraphics[width=0.93\linewidth]{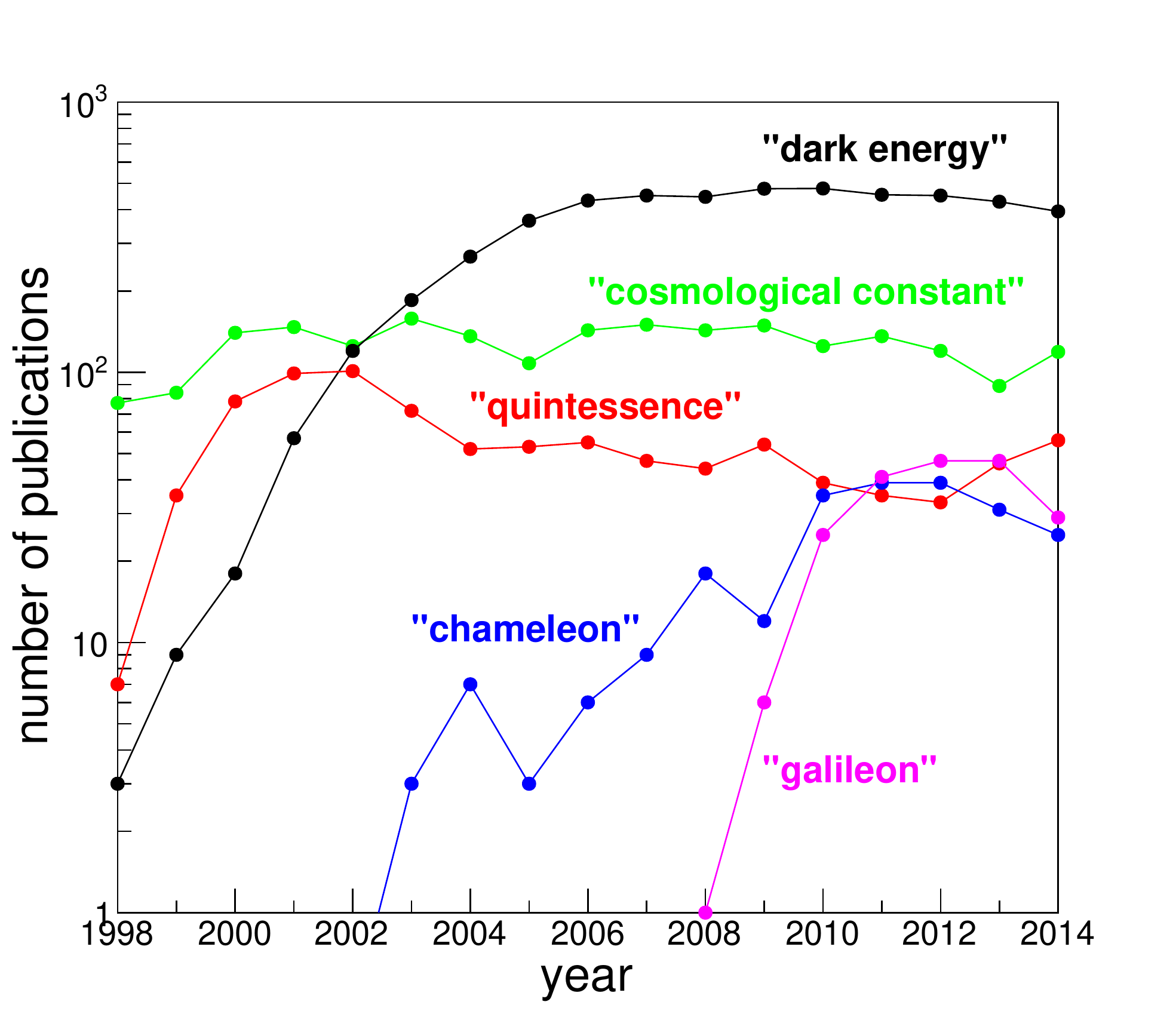}
\caption{
Number of publications per year containing in the title the keywords 
``dark energy'' (black), ``cosmological constant'' (green), ``quintessence'' (red), ``chameleon'' (blue), ``galileon'' (pink). 
The counting was done using inspirehep.net.
 \label{spiresCount}
}
\end{figure}

The comprehensive study of these theories, their cosmological implications, and the design of laboratory experiments 
is a vivid ongoing field of research (see Fig \ref{spiresCount}). 
The hunt is on for the non-gravitational interaction of dark energy. 

\clearpage
\section{Neutron interferometry constrains the chameleon}
\label{sectionInterferometry}

This part describes a search for the chameleon using neutron interferometry. 
The experiment has been reported also in \textcite{Lemmel2015}. 

\subsection{Neutron interferometry}

\begin{figure}
\centering
\includegraphics[width=0.93\linewidth]{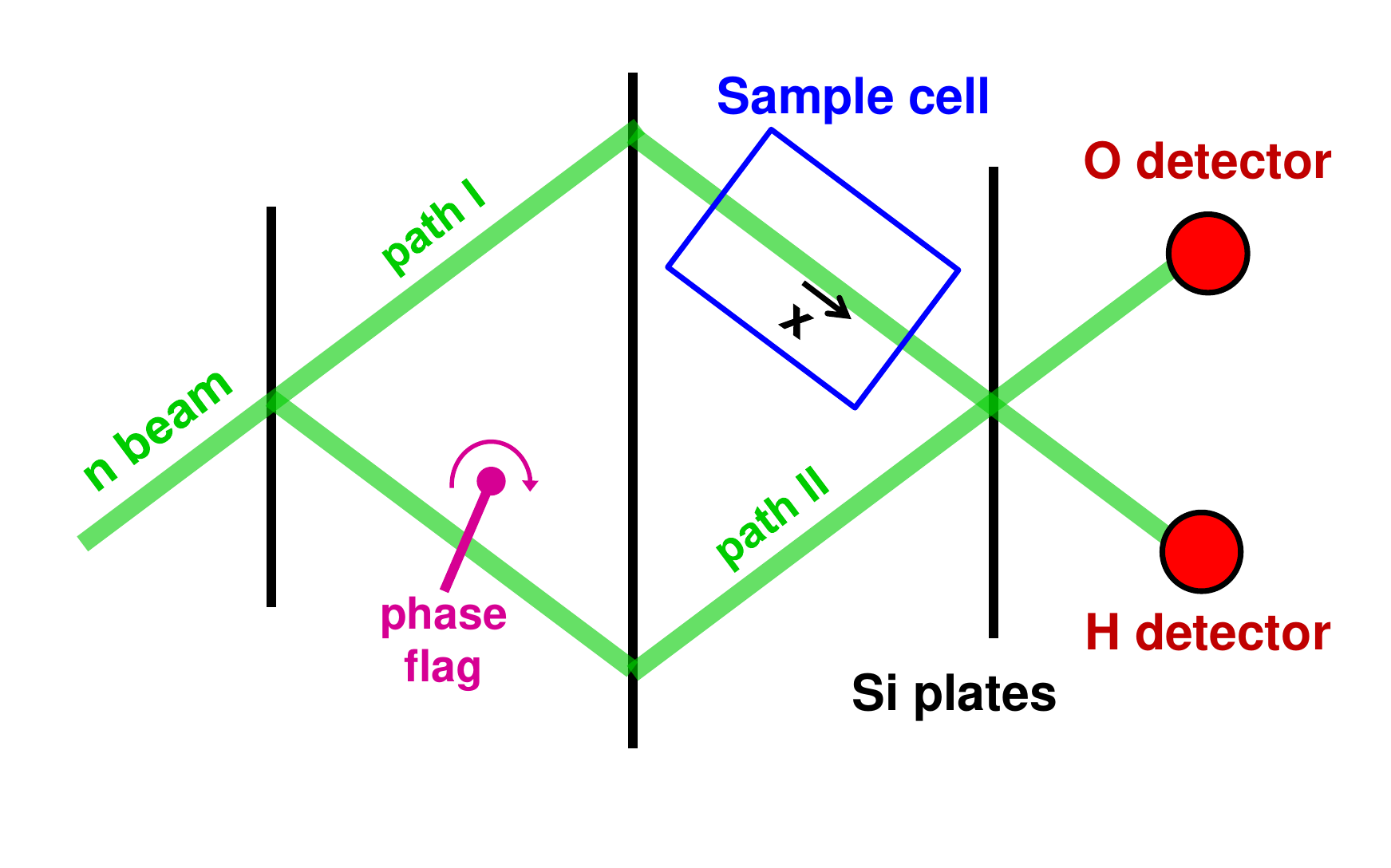}
\caption{
General principle of neutron interferometry. 
 \label{interferometerSketch}
}
\end{figure}

The first neutron interferometer, analogous to a Mach-Zehnder interferometer for light, has been realized by \textcite{Rauch1974}. 
The principle of neutron interferometry is depicted in Fig. \ref{interferometerSketch}. 
A monochromatic neutron beam with a wavelength of $\lambda = 0.27$~nm (or wavenumber of $k = 2 \pi / \lambda = 23$~nm$^{-1}$) 
is split by Bragg diffraction into two coherent beams using a single-crystal silicon plate. 
Then part of these two beams are recombined using two additional similar parallel plates. 
The neutron detectors measure the flux resulting from the interference of neutrons going through path I and path II. 
A phase flag (usually an aluminium plate with variable angle) 
is introduced in path II to record the interferogram, 
i.e. the neutron flux in detectors H and O as a function of the phase flag position:  
\begin{equation}
F_O = A_O + B_O \cos (\xi_{\rm flag}) ; \  F_H = A_H - B_H \cos (\xi_{\rm flag}), 
\end{equation}
where $\xi_{\rm flag}$ is a function of the phase flag position. 
Then, a sample is introduced in path I and a new interferogram is recorded: 
\begin{equation}
F_O = A_O + B_O \cos (\xi_{\rm flag} + \xi) ; \  F_H = A_H - B_H \cos (\xi_{\rm flag} + \xi).
\end{equation}
The phase $\xi$ due to the sample is expressed as an integral of the potential $V(x)$ along the neutron beam in the sample \cite{Rauch2000}: 
\begin{equation}
\xi = - \frac{m}{k \hbar^2} \int V(x) dx. 
\label{phase}
\end{equation}
Recording interferograms with and without sample one can extract the phase shift due to the sample. 

In order to probe the chameleon, 
we \cite{BraxPignol2013} proposed to use a rectangular vacuum cell as a sample. 
When evacuated, a chameleon field $\varphi$ with a bubble-like profile will appear in the cell. 
Since the neutron has an energy $V = \beta m/m_{\rm pl} \ \varphi$ in the chameleon field, 
the phase associated with the chameleon bubble reads: 
\begin{equation}
\xi = - \frac{m}{k \hbar^2}  \beta \frac{m}{m_{\rm pl}} \int \varphi(x) dx . 
\end{equation}

The cell needs to be evacuated in order to build up a chameleon field ``bubble'' in the cell, 
because even air at atmospheric pressure is too dense. 
Indeed, from the calculations presented in the previous part of this document, 
we can estimate that the chameleon field is in the saturated regime in a few cm of air for strong coupling ($\beta \gg 1$). 
The bubble solution \eqref{TwoPlates} cannot build up in air, instead, the solution would be nearly uniform in the cell. 
Note that the neutron interferometers are generally operated in air. 
The search for the chameleon field calls for a dedicated experiment with a vacuum cell. 

Interestingly, it is possible in principle to switch off and on the chameleon bubble by adding or removing a small amount of gas in the cell. 
We can profit from this effect in an experiment, because this is a rather unique signature which is easy to exploit in a practical setup. 
If $\beta = 10^8$, a pressure of $10^{-2}$~mbar of helium (about a million times less dense than air!) is enough to suppress the bubble, as we will calculate later.

\subsection{The experiment}

\begin{figure}
\centering
\includegraphics[width=0.93\linewidth]{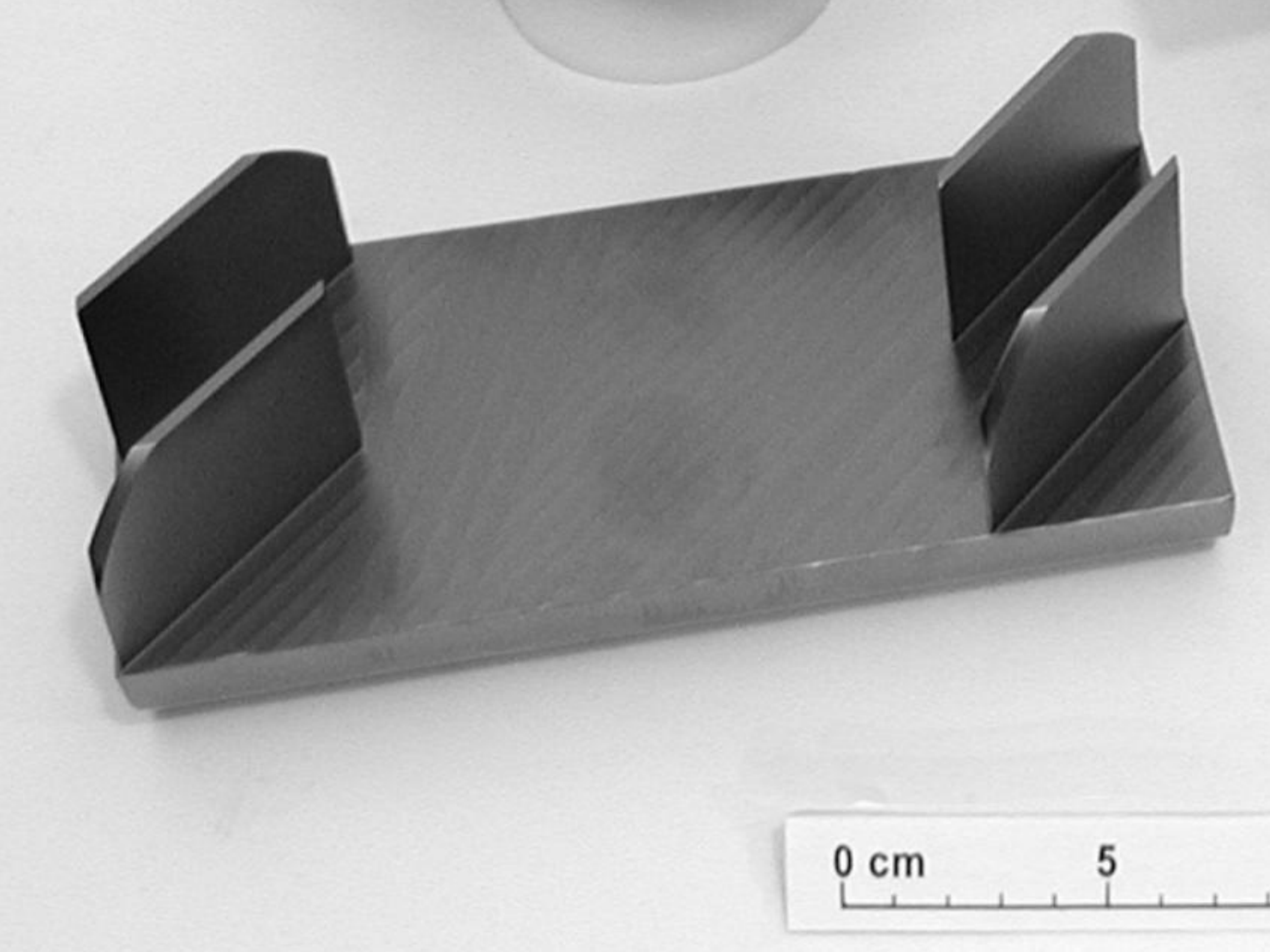}
\caption{
Picture of the large interferometer made out of a single-crystal silicon used in the S18 experiment. 
 \label{interferometer}
}
\end{figure}

A dedicated experiment to search for the chameleon field was performed at the S18 instrument (see \textcite{Kroupa2000} for a description) at ILL in summer 2013. 
We used the largest single-crystal interferometer available (see picture Fig. \ref{interferometer}) 
with a beam separation of $50$~mm. 

\begin{figure}
\centering
\includegraphics[trim=5cm 8cm 5cm 8cm,clip=true,width=0.93\linewidth]{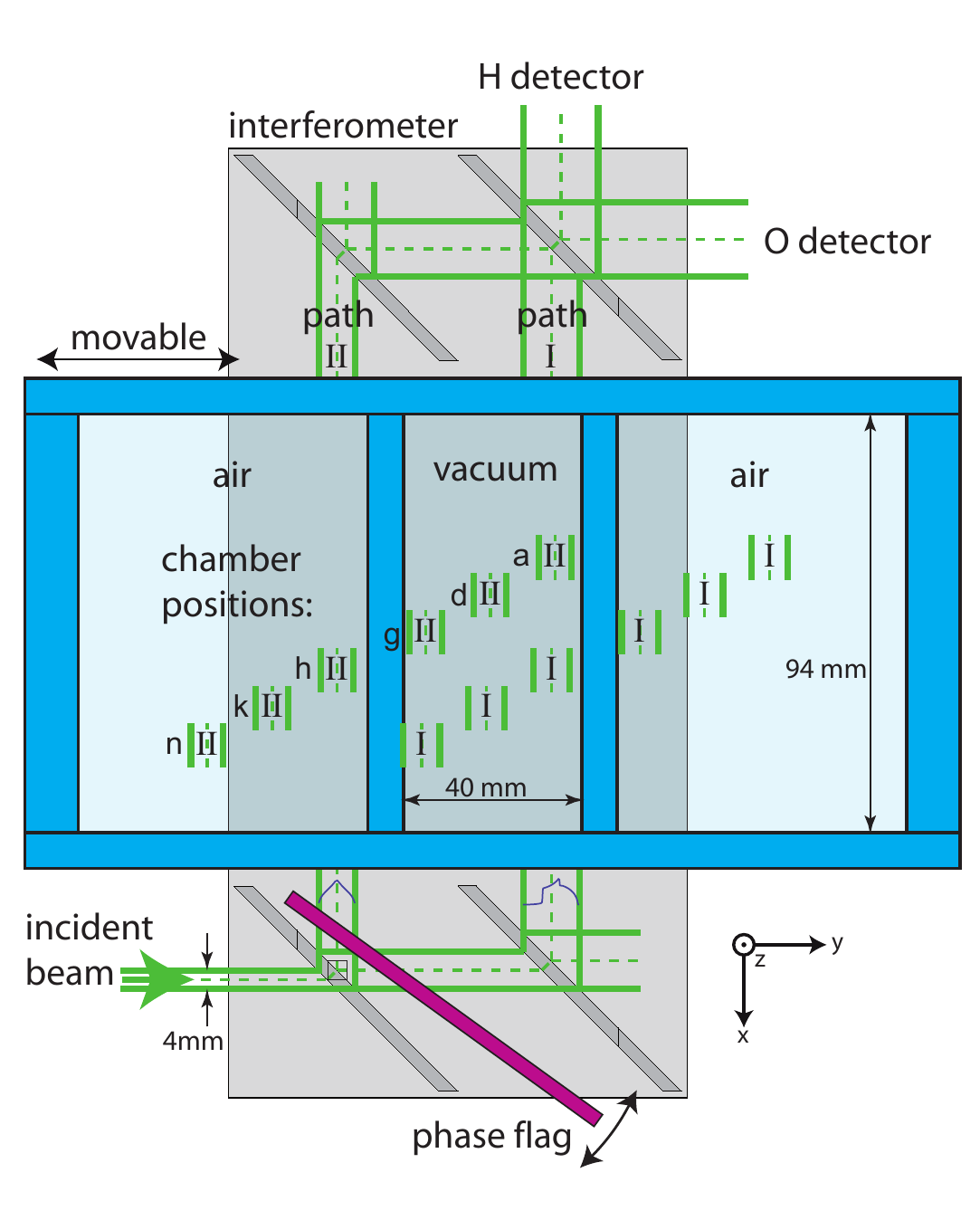}
\caption{
Top view of the setup. 
The chamber box (blue) can be moved by a remote controlled robot, 
allowing the beams to pass at different positions labeled by 'a' to 'n'. 
 \label{sketchS18}
}
\end{figure}

A vacuum chamber specially designed for the chameleon search was built by the Atominstitut group in Vienna. 
The device, depicted in Fig. \ref{sketchS18} consists of an aluminium vacuum chamber with inner dimensions $40 \times 40 \times 94 \ {\rm mm}^3$ and two air 
chambers sitting alongside the vacuum chamber. 
The whole box can be moved sidewards for swapping the vacuum cell between beam path I and beam path II and to probe different beam trajectories within the vacuum cell. 
The vacuum chamber is connected to a gas system through a hole of diameter $5$~mm on the top of the cell. 
The gas system comprises a pressure gauge, a motorized leak valve connected to a helium bottle and pumps. 
A turbomolecular pump is running continuously while a controlled amount of helium is let in through the leak valve. 
This way, the pressure of helium in the cell can be controlled in the range $10^{-4} \ {\rm mbar} < P < 10^{-2} \ {\rm mbar}$. 
To get higher pressure the turbomolecular pump needs to be disconnected. 

During data-taking, we recorded many interferograms while varying the parameters of the cell (helium pressure and beam position). 
It takes about half an hour to record one interferogram. 
A major difficulty concerns the phase drifts due to environmental factors such as temperature gradients, air flow and vibrations. 
The use of such a big single-crystal interferometer is very delicate since it is more sensitive to these nuisances compared to smaller interferometers. 
To compensate for environmental phase drifts, interferograms were recorded in parallel. 
It means that the phase flag was rotated to the first angular position
and neutrons were counted for a certain amount of time for each parameter setting. 
Then the phase flag is rotated to the next position and neutrons are counted again for all parameter settings etc.
We recorded data with two different modes: the ``profile mode'' and the ``pressure mode''. 

\begin{figure}
\centering
\includegraphics[width=0.93\linewidth]{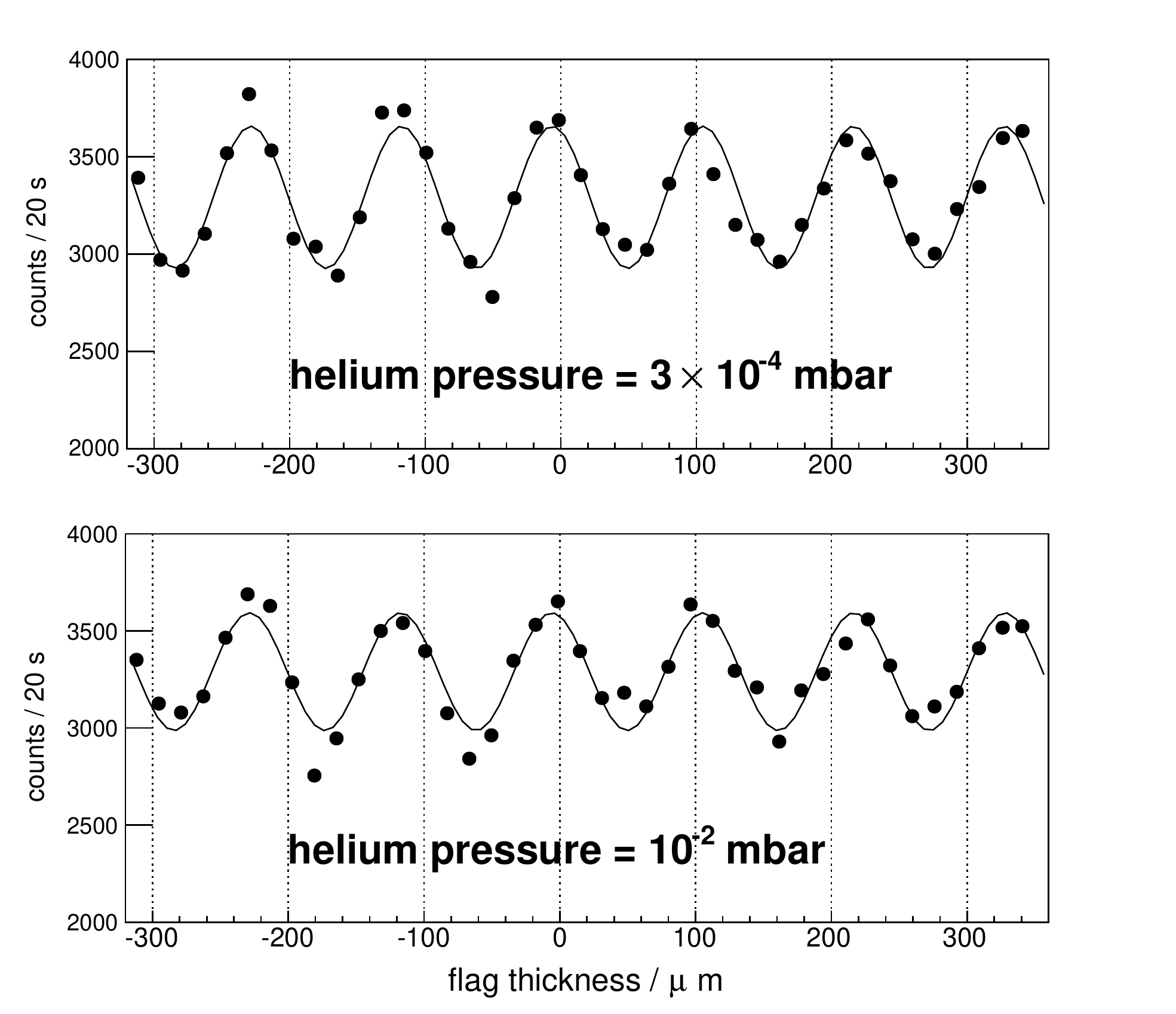}
\caption{
Interferograms recorded in position 'd'. 
The plots show the intensity in the detector O as a function of the path length difference created by the rotating phase flag. 
 \label{interferogram}
}
\end{figure}

\para{Pressure mode. }
In the pressure mode, the beam II passes in the middle of the cell (position 'd'). 
We recorded four interferograms in parallel with four different helium pressures ($2.4 \times 10^{-4}, \ 7.1 \times 10^{-4}, \ 2.7 \times 10^{-3}, \ 1.1 \times 10^{-2}$~mbar). 
A subset of these interferograms is shown in Fig. \ref{interferogram}. 

\para{Profile mode. } 
In the profile mode, a fixed value of the helium pressure is established in the cell. 
Interferogram with different positions of the cell are recorded in parallel, to look for the bubble profile. 
More specifically, we recorded interferograms at positions 'a', 'd' and 'g', that we will refer later as
 \emph{center position} $(y=0,z = -4 \ {\rm mm})$ and \emph{side position} $(y = \pm 15 \ {\rm mm}, z = -4 \ {\rm mm})$. 
This mode allows to probe a wider domain of pressure because the pressure needs not be changed rapidly. 
However, there is a possible systematic effect in this mode associated with the variation of the wall thickness of the cell. 
The data in the profile mode require a position dependent phase correction based on a precise mapping of the wall thickness.

\subsection{Expected signal} 

To analyse the data, we need to calculate the field $\varphi$ inside the cell for different helium pressure. 
The field satisfies the chameleon equation $\Delta \varphi = \mathcal{V}_{\rm eff}'(\varphi)$. 
Since no exact analytical solution exists for this problem, a numerical scheme was employed. 
To simplify the problem, we considered the chameleon equation in two dimensions $y,z$ transverse to the beam. 
Indeed, the longitudinal size of the cell ($l = 94$~mm) is longer than the transverse size ($a = 40$~mm), we can assume that the field is approximately uniform in the $x$ direction. 
We solve for the field in units of $M_\Lambda$, i.e. we introduce the dimensionless field $F(y,z) = \varphi(y,z) / M_\Lambda$. 

The numerical method adopted to calculate the solution is described in Appendix \ref{A1}. 
Figure \ref{bubble} shows selected results of such calculations.

\begin{figure}
\centering
\includegraphics[width=0.77\linewidth]{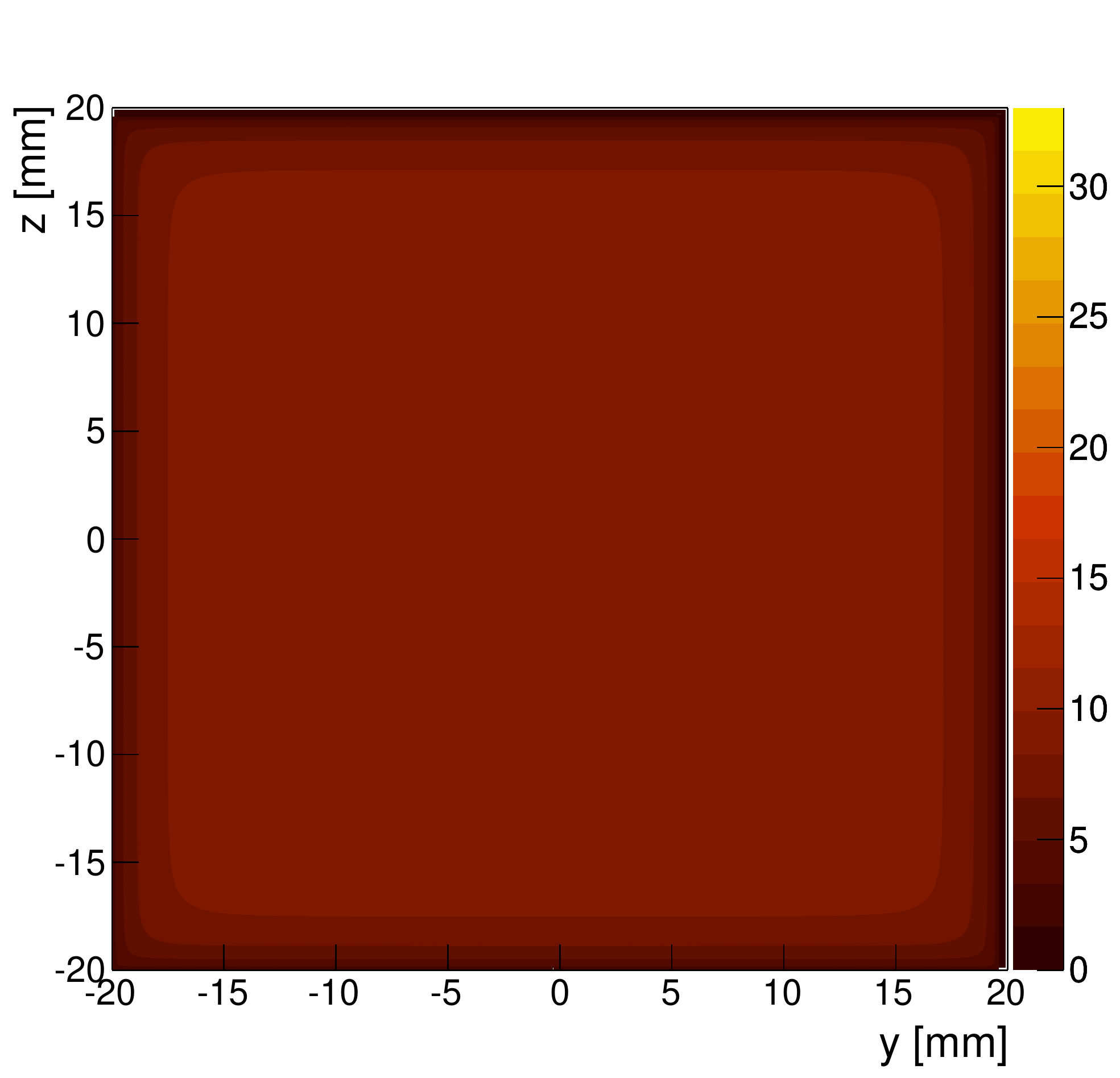}
\includegraphics[width=0.77\linewidth]{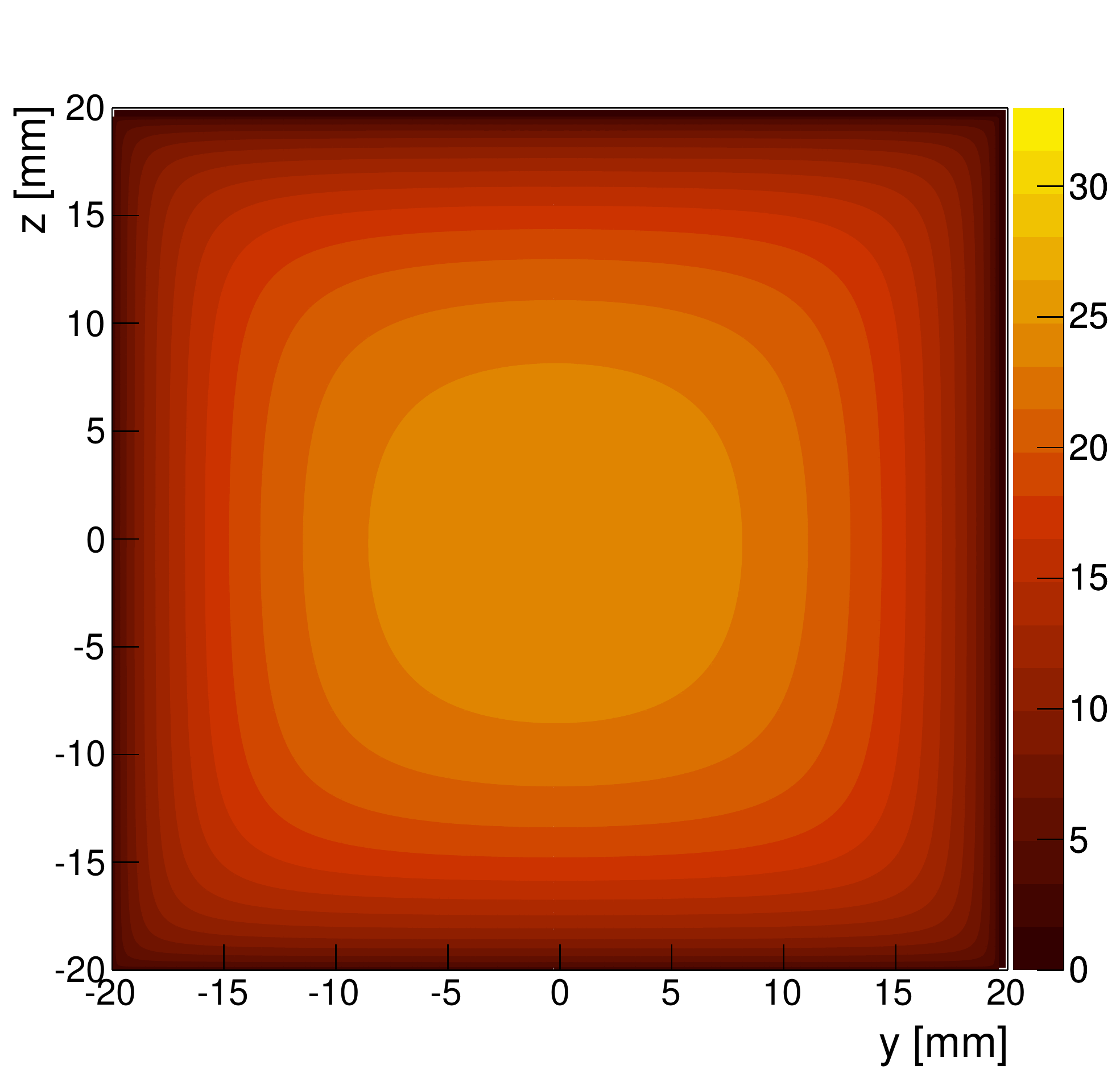}
\includegraphics[width=0.77\linewidth]{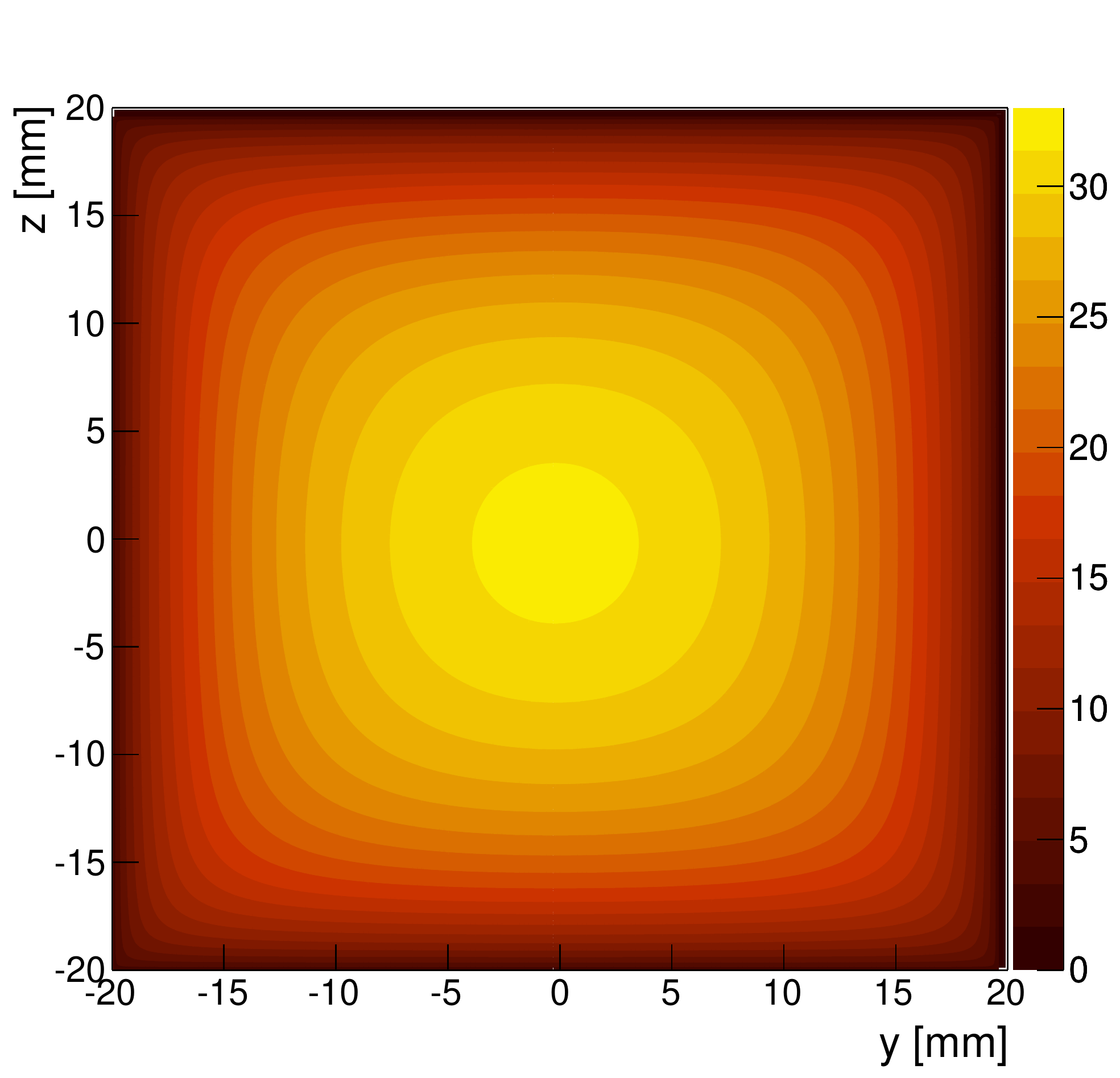}
\caption{
Calculations of the chameleon bubble profile transverse to the neutron beam for different helium pressure in the cell. 
We plot $F(y,z) = \varphi(y,z) / M_\Lambda$, for $n=1$ and $\beta = 5 \times 10^7$. 
Top: pressure $=10^{-2}$~mbar, middle: pressure $=10^{-3}$~mbar, bottom: in vacuum. 
 \label{bubble}
}
\end{figure}

\para{Expected chameleon phase-shift. }
From the calculations of the transverse profile $F$, we can extract the expected phase signal
\begin{eqnarray}
\xi & = & - \frac{m}{k \hbar^2}  \beta \frac{m}{m_{\rm pl}} \int M_\Lambda F \ dx \\
               & = & - \frac{m}{k \hbar^2}  \beta \frac{m}{m_{\rm pl}} M_\Lambda F \ l_{\rm eff}. 
\end{eqnarray}
We have introduced the effective length $l_{\rm eff}$ to take into account the edge effects in the direction longitudinal to the beam. 
To estimate the effective length of the bubble in the longitudinal direction, we performed a 3D lattice calculation of the bubble. 
The 3D calculation is much more demanding in terms of CPU time as compared to the 2D calculation, in particular it does not permit to use a fine grid neither to repeat the calculation for multiple change of parameters. 
Calculating with a grid size of $0.5$~mm the empty bubble for $n=1$ we found an effective length of $l_{\rm eff} = 84$~mm (compare to the physical length of $l = 94$~mm). 
At the end, we found the following numerical relation between the expected phase shift (in degree) of the bubble $\xi_{\rm exp}$ and the 2D profile $F$: 
\begin{equation}
\xi = - 4.7 \times 10^{-9} \ {\rm deg} \times \beta \times F. 
\end{equation}
The 2D profile $F$ is a function of the transverse position $y,z$ of the beam with respect to the cell, 
the helium pressure $p$ in the cell and the coupling $\beta$. 

\para{Nuclear phase-shift. }
There is in principle an expected phase shift proportional to the pressure due to the Fermi potential of the helium. 
Combining \eqref{FermiPot} and \eqref{phase} we find that this phase amounts to
\begin{equation}
\xi_{\rm nucl} = - \lambda \ l \, b \, n_{\rm He} = - 0.1 \ {\rm deg} \times \frac{P}{1 \ {\rm mbar}}, 
\end{equation}
where $\lambda = 0.27$~nm is the neutron wavelength, $l = 94$~mm is the longitudinal length of the chamber, $b = 3.26$~fm is the bound neutron scattering length of $^4$He and $n_{\rm He}$ is the number density of atoms. 
For pressures below $10^{-2}$~mbar, as in the actual experiment, this phase shift is too small to be detected. 
We will ignore this effect. 

\subsection{The results} 

\begin{figure}
\centering
\includegraphics[width=0.93\linewidth]{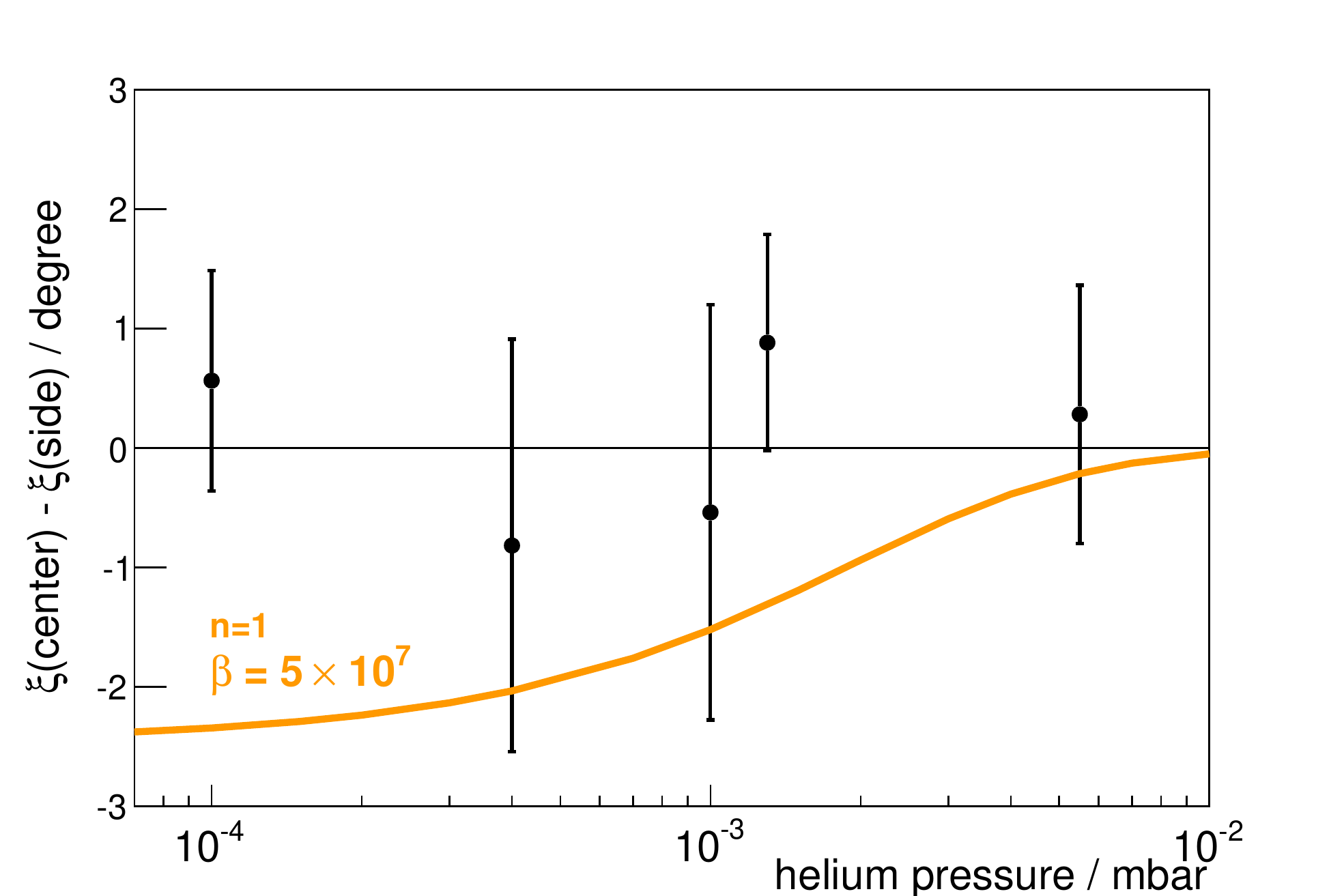}
\includegraphics[width=0.93\linewidth]{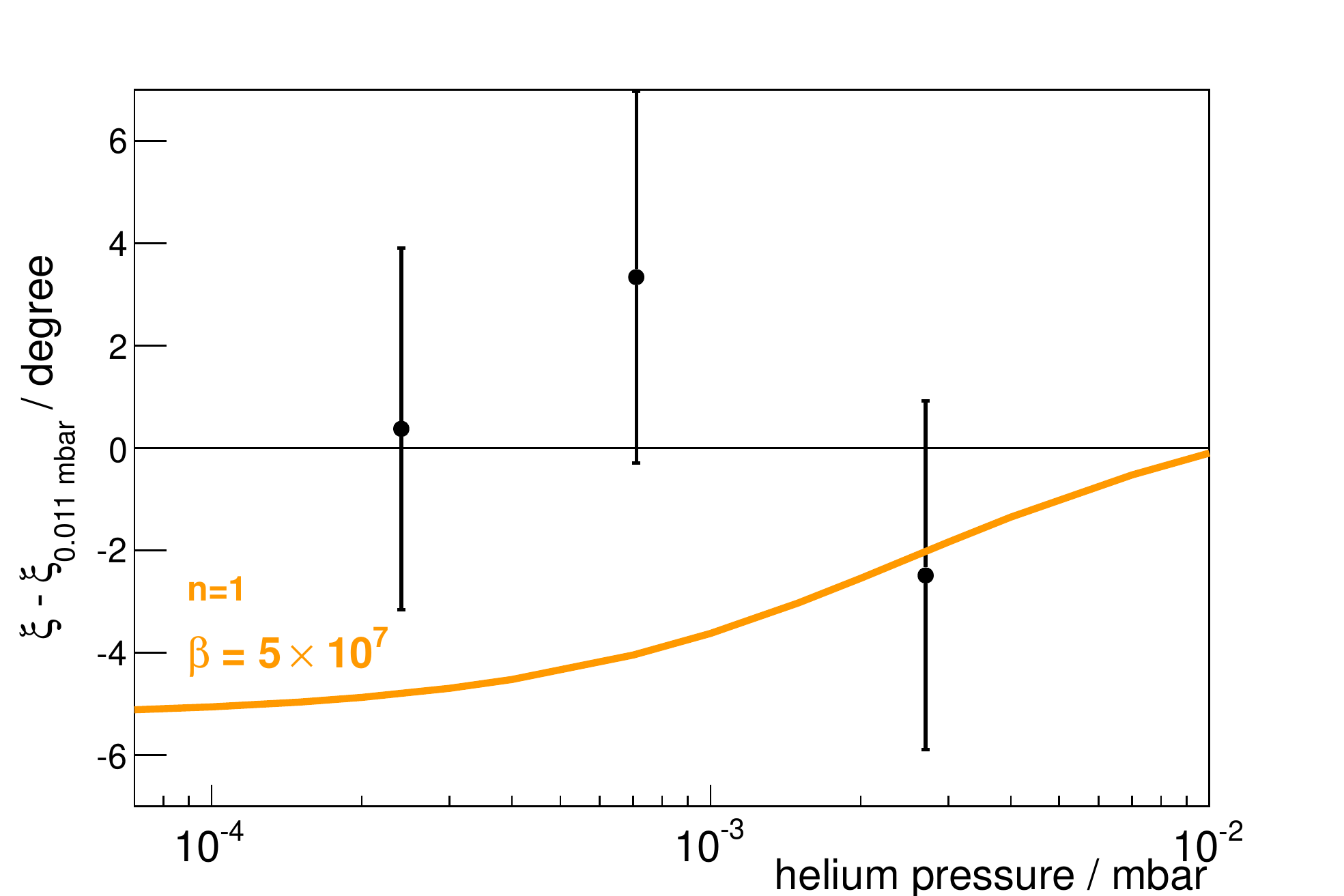}
\caption{
Results of the experiment at S18. 
Top: phase difference between center and side position in the cell as a function of the helium pressure (profile mode). 
Bottom: phase difference between low pressure and 0.011 mbar (pressure mode). 
In both cases the orange line represents the theoretical expectation for $n=1$ and $\beta = 5 \times 10^7$. 
 \label{resultsS18}
}
\end{figure}

The final results of the S18 experiment are shown in Fig. \ref{resultsS18}. 
In the profile mode, there is no significant phase difference between the center and side position of the beam, even when the cell is evacuated to $10^{-4}$ mbar of helium. 
The data does not show the bubble profile associated with a chameleon field. 
In the pressure mode, where the beam always probes the center of the cell, there is no significant phase difference when the helium pressure changes. 
The data is in agreement with the ``no chameleon'' hypothesis. 

\begin{figure}
\centering
\includegraphics[width=0.93\linewidth]{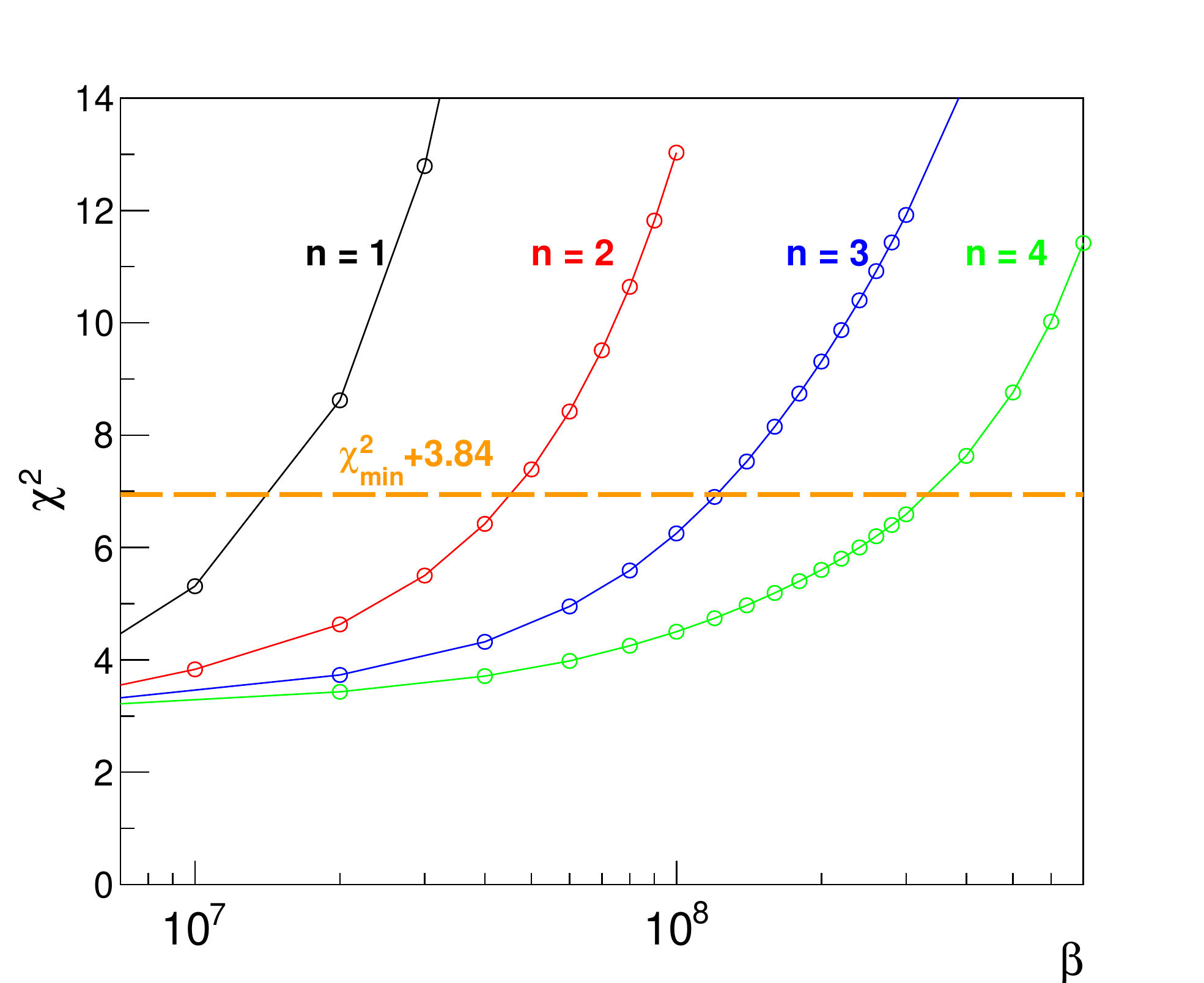}
\caption{
Calculated $\chi^2(\beta)$ (according to Eq. \eqref{chi2}) for $n=1,2,3,4$. 
The horizontal orange dashed line corresponds to the limit at a confidence level of 95 \%.
 \label{chi2S18}
}
\end{figure}

To derive a limit on the coupling $\beta$, we calculate the $\chi^2$: 
\begin{equation}
\label{chi2}
\chi^2(\beta) = \sum_i \frac{(\xi(\beta)_i - \zeta_i)^2}{\sigma_i^2}
\end{equation}
where the sum goes over all 8 data points $\zeta_i \pm \sigma_i$ shown in Fig. \ref{resultsS18} thus combining the profile and pressure modes. 
Figure \ref{chi2S18} shows $\chi^2(\beta)$ calculated for $n = 1, 2, 3, 4$. 
The minimum of $\chi^2(\beta)$ is always obtained for $\beta = 0$. 
To derive the exclusion limits (at a confidence level of $95$ \%) we use the standard criterion 
\begin{equation}
\chi^2(\beta_{\rm lim}) = \chi^2_{\rm min} + 1.96^2, 
\end{equation}
the limits are reported in Table \ref{limitsS18} and also in Fig. \ref{exclusion}. 
For $n=1$, the limit is already a factor of 30 better than the current limit obtained with Gravity Resonance Spectroscopy \cite{Jenke2014}. 

\separe

A new experiment with the interferometer is scheduled for summer 2015. 
By improving the contrast of the interferometer and increasing the statistics, we can gain at least a factor of 10 in sensitivity.

\begin{table}
\center
\caption{
Limits on the chameleon coupling $\beta$ (at 95 \% C.~L.) derived from the S18 experiment at ILL. 
}
\begin{tabular}{lll}
$n$  & ~ ~ & $\beta_{\rm lim}$ \\
\hline
1  & ~ ~  & $1.6 \times 10^7$ \\
2  & ~ ~  & $4.6 \times 10^7$ \\
3  & ~ ~  & $1.2 \times 10^8$ \\
4  & ~ ~  & $3.4 \times 10^8$ \\
\hline
\end{tabular}
\label{limitsS18}
\end{table}

\clearpage
\section{Bouncing neutrons with GRANIT}
\label{sectionGRANIT}

\subsection{Quantum states of bouncing neutrons}

If the kinetic energy of an incident neutron is less than the Fermi potential of a material wall (or floor, or ceiling) 
then the neutron will be reflected off the surface for any angle of incidence. 
This is the definition of an ultracold neutron. 
For example, glass has a Fermi potential of 90 neV. 
It reflects neutrons up to an energy of $E = 90$~neV, or a velocity of 4 m/s. 
The reflection is almost perfectly specular, i.e. mirror-like. 
The probability of non-specular (diffuse) reflection off a well-polished glass mirror can be less than $10^{-3}$ \cite{Nesvizhevsky2007}. 

The Fermi potential of glass corresponds to a height in the gravity field of $h = 88$~cm through the relation $E = m g h$, where $m$ is the neutron mass and $g = 9.806$~m/s$^2$ is the acceleration in the lab. 
It means that a neutron dropped from a height of less than a meter will bounce above a glass floor. 
The bouncing period $T = \sqrt{8 h/g}$ is of the order of a second, like any object bouncing with a height of about a meter. 

Now, if we consider a shorter bouncing height, or a higher bouncing frequency, the simple classical description of the bouncing ball might fail. 
We will see that when the bouncing frequency enters the audio range (above about 100 Hz), corresponding to a bouncing height of about $10 \ \micron$, a quantum description is needed. 

\para{The quantum bouncer. }
The vertical motion of a neutron bouncing above a mirror nicely realizes the academic problem of a particle confined in a potential well. 
The well consists in the gravitational potential $V(z) = m g z$ pulling the particle down and a perfect mirror $V(0) = + \infty$
\footnote{The mirror potential is the Fermi potential of glass ($\approx 100$~neV), it can be considered infinite because it is much larger than the energies of the quantum levels ($\approx 1$~peV).}
 pushing the particle up. 
According to the quantum description, the vertical motion has a discrete energy spectrum. 
The bound states energies $E_k$ and the corresponding wavefunctions $\psi_k(z)$ satisfy the stationary Schr\"odinger equation
\begin{equation}
\label{schrodinger}
- \frac{\hbar^2}{2m} \frac{d^2}{dz^2} \psi_k + m g z \psi_k = E_k \psi_k, \quad \psi_k(0) = 0.
\end{equation}
The problem of finding the bound states energies and wavefunctions is solvable and we are going to solve it. 
By dimensional analysis we define the characteristic height of the quantum bouncer: 
\begin{equation}
\label{z0}
z_0 = \left( \frac{\hbar^2}{2 m^2 g} \right)^{1/3} = 5.87 \ \micron. 
\end{equation}
Defining the dimensionless height $Z = z/z_0$, the stationary Schr\"odinger equation takes the form of the \emph{Airy equation}
\begin{equation}
\frac{d^2 \psi_k}{dZ^2} + (\epsilon_k - Z) \psi_k = 0, 
\end{equation}
where $\epsilon_k = E_k / mgz_0$. 
The general solution of this second order linear differential equation is a linear combination of the two Airy functions Ai and Bi: 
$\psi(Z) = C \ {\rm Ai}(Z-\epsilon) + D \ {\rm Bi}(Z - \epsilon)$. 
The function Bi diverges at $Z = + \infty$ therefore the physically acceptable solutions are those with $D=0$. 
Next, the condition $\psi_k(0) = 0$ at the boundary enforces the quantification of the energy levels ${\rm Ai}(-\epsilon_k) = 0$. 
Therefore the energy levels of the stationary states are given by 
\begin{equation}
E_k = mg z_0 \ \epsilon_k
\end{equation}
where 
\begin{equation}
\epsilon_k = \{ 2.338, 4.088, 5.521, 6.787, \cdots \}
\end{equation}
is the sequence of the negative zeros of the Airy function. 
Numerically the energy levels are in the range of $1 \ {\rm peV} = 10^{-12}$~eV ($E_1 = 1.41$~peV, $E_2 = 2.46$~peV $\cdots$), 
five orders of magnitude below the kinetic energy of ultracold neutrons. 

\begin{figure}
\centering
\includegraphics[width=0.93\linewidth]{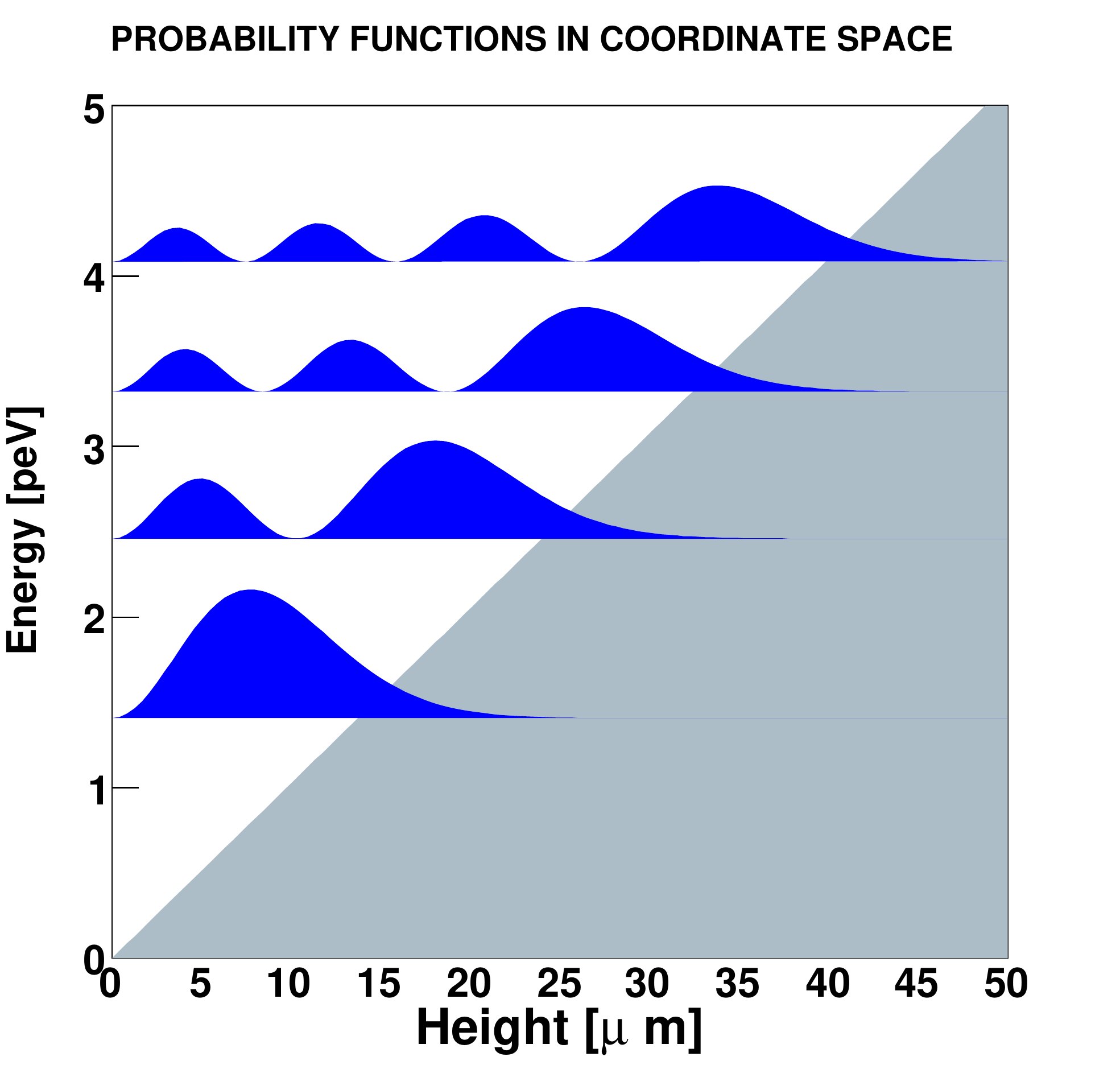}
\caption{
Representation of the probability density functions $|\psi_k(z)|^2, k = 1, 2, 3, 4$ for the first four energy states of the neutron bouncer. 
The four functions are translated in the vertical axis according to the energy of the state. 
The grey zone is classically forbidden. 
 \label{wavefunctions}
}
\end{figure}

To each quantum state $\ket{k}$ corresponds the wavefunction
\begin{equation}
\psi_k(z) = C_k \ {\rm Ai}( z/z_0 - \epsilon_k), 
\end{equation}
where $C_k$ is a normalization coefficient. 
The probability density functions $|\braket{z}{k}|^2 = |\psi_k(z)|^2$ associated with the wavefunctions are plotted in Fig. \ref{wavefunctions}. 
As in any 1D potential well, the number of nodes (the places where the wavefunction vanishes) of the $k^{\rm th}$ quantum state is $k$. 

Let us compare the neutron quantum bouncer to the hydrogen atom, a more familiar case of a bound system. 
The typical energy of the neutron bouncer is 13 orders of magnitude smaller than 13.6~eV, the binding energy of the electron in the ground state. 
The typical size of the wavefunction is 5 orders of magnitude larger than 0.05~nm, the size of the hydrogen atom. 
These very unusual scales arise due to the extraordinary weakness of gravity as compared to the Coulomb force. 
Another important difference concerns the stability of the excited levels. 
An excited state of the hydrogen atom decays by spontaneous emission of a photon with a time scale of the order of 1~ns. 
The neutron bouncer could also in principle decay by spontaneous emission of gravitons. 
However, \textcite{Pignol2007} estimated the timescale of this effect to $10^{77}$~s, i.e. 60 orders of magnitude longer than the age of the Universe. 
Again, this is due to the extreme weakness of gravity. 

\para{Observing the quantum states. }
It is usually believed that the weakness of gravity is not the experimentalist's friend. 
In the case of the neutron bouncer, however, it is a great asset. 
It makes the wavefunctions very large, almost visible to the naked eye. 
This property has been used in a series of experiments started in 1999 at the PF2 ultracold neutron source of the ILL reactor 
\cite{Nesvizhevsky2002,Nesvizhevsky2003,Nesvizhevsky2005}. 

\begin{figure}
\centering
\includegraphics[width=0.93\linewidth]{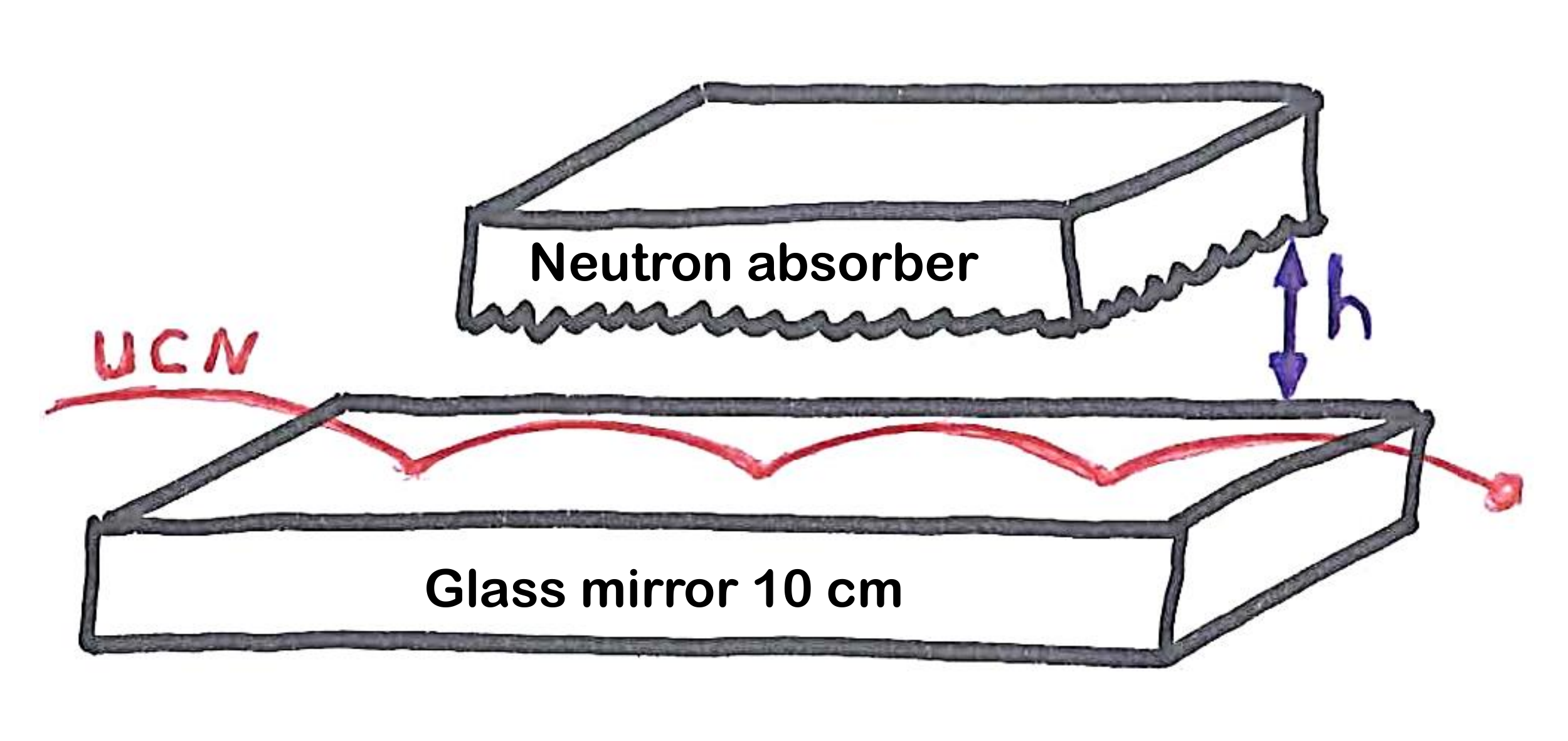}
\includegraphics[width=0.93\linewidth]{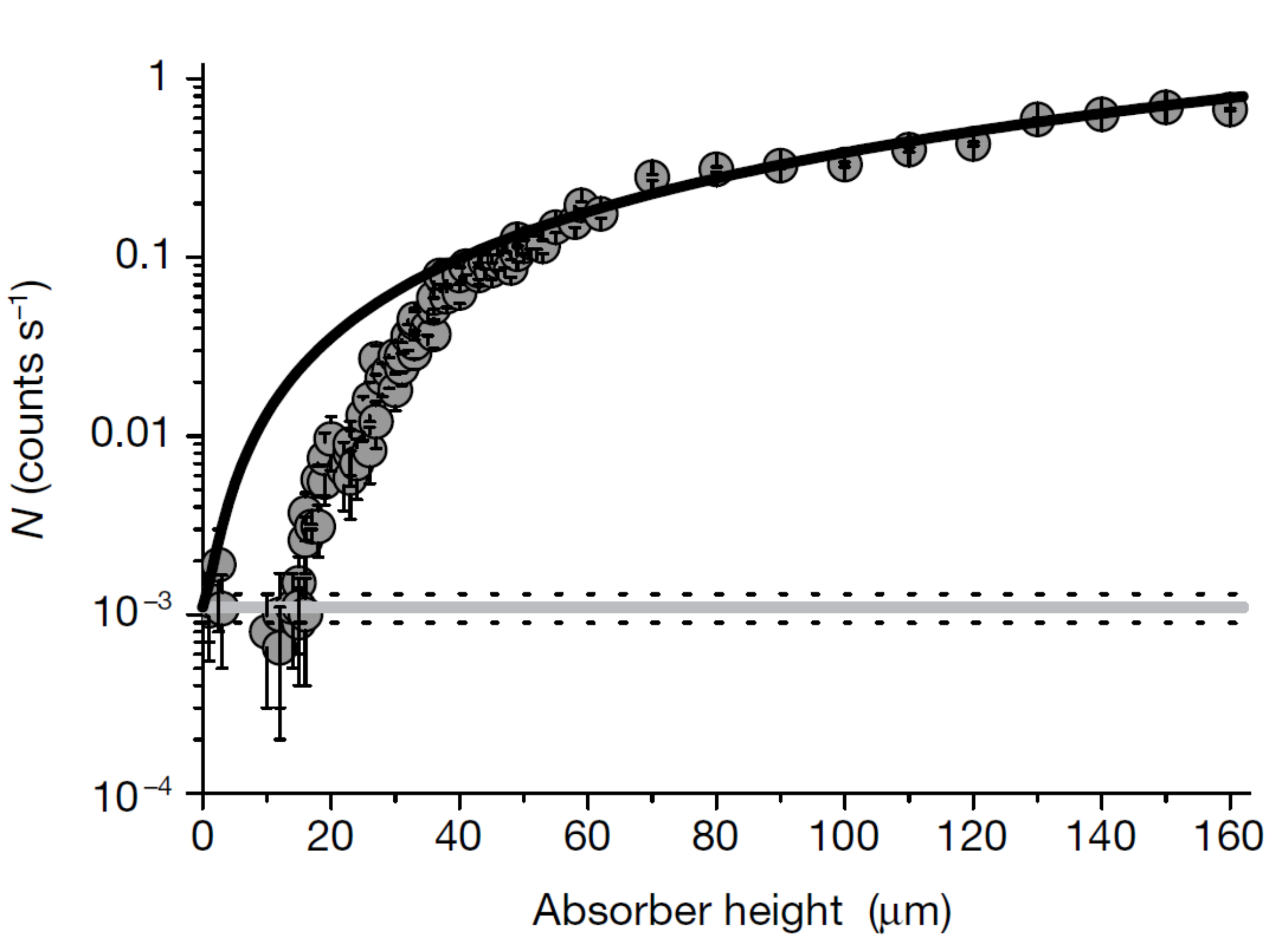}
\caption{
Top: scheme of the method used to discover the quantum states of the neutron bouncer. 
Bottom: Figure taken from \textcite{Nesvizhevsky2002}, reporting the flux in the detector as a function of the height of the slit. 
The solid curve is the classical fit to these data. 
 \label{DiscoveryILL}
}
\end{figure}

The setup which served to observe for the first time the quantum states of the neutron bouncer is sketched in Fig. \ref{DiscoveryILL}. 
It is a flow-through experiment: ultracold neutrons with horizontal velocity are coming from the left, they bounce on top of a horizontal, well-polished glass mirror. 
They are then counted in a gaseous detector filled with $^3$He. 
A flat neutron absorber is installed on top of the horizontal mirror, parallel to it. 
It forms a slit, with a height $h$ that can be precisely regulated with a micrometer accuracy. 
The experiment consists in recording the neutron flux passing through as a function of the height of the slit. 
The result obtained in this experiment is shown in Fig. \ref{DiscoveryILL}. 
When the slit height is $h < 15 \ \micron$, no neutron passes through the slit, contrary to the classical expectation. 
In the quantum picture, the neutron wavefunction is a linear combination of the stationary quantum states. 
A given state ``decays'' in the slit if its wavefunction overlaps significantly with the absorber. 
When the absorber is lower than $15 \ \micron$, every state, including the ground state $\ket{1}$, overlaps with the absorber and the slit becomes completely opaque to neutrons. 
This is precisely what was observed.

A quantitative analysis of the transmission curve allows to extract the spatial parameters of the quantum states. 
The difficulty of the analysis lies in the choice of a model for the absorber. 
\textcite{Nesvizhevsky2005} parametrized the transmission curve by introducing the critical height of the first quantum state $h_1 = \epsilon_1 z_0$. 
It corresponds to the turning point of a classical bouncer with vertical energy $E_1$. 
They report $h_1^{\rm exp} = (12.2 \pm 1.8_{\rm syst} \pm 0.7_{\rm stat}) \micron$, the systematic uncertainty takes into account the error in the slit size calibration and the finite accuracy of the absorber model. 
In terms of the parameter $z_0$, the extracted value $z_0^{\rm exp} = (5.2 \pm 0.8) \micron$ agrees with the theoretical value $z_0^{\rm th} = 5.87 \ \micron$. 

There is a more direct way to access the spatial structure of the quantum states, by taking a photograph. 
Instead of just counting the neutrons at the exit of the mirror, a position sensitive detector with a micrometric resolution could be used. 
As proposed by \textcite{Nesvizhevsky2000}, one should aim at observing the node of the second quantum state 
(at a height of $10 \ \micron$) which provides a clear signature, instead of looking at the ground state which does not display a specific pattern. 
Selecting the first few levels could be done with a slit: a horizontal mirror and an absorber on top. 
At the exit of the slit, we place a second horizontal mirror (no absorber on top). 
The second mirror is placed $13.5 \ \micron$ below the first mirror, in order to increase the vertical energy of the neutron bouncing on the second mirror. 
This way, the ground state population will be suppressed and the population of the second quantum state enhanced. 

\begin{figure}
\centering
\includegraphics[width=0.93\linewidth]{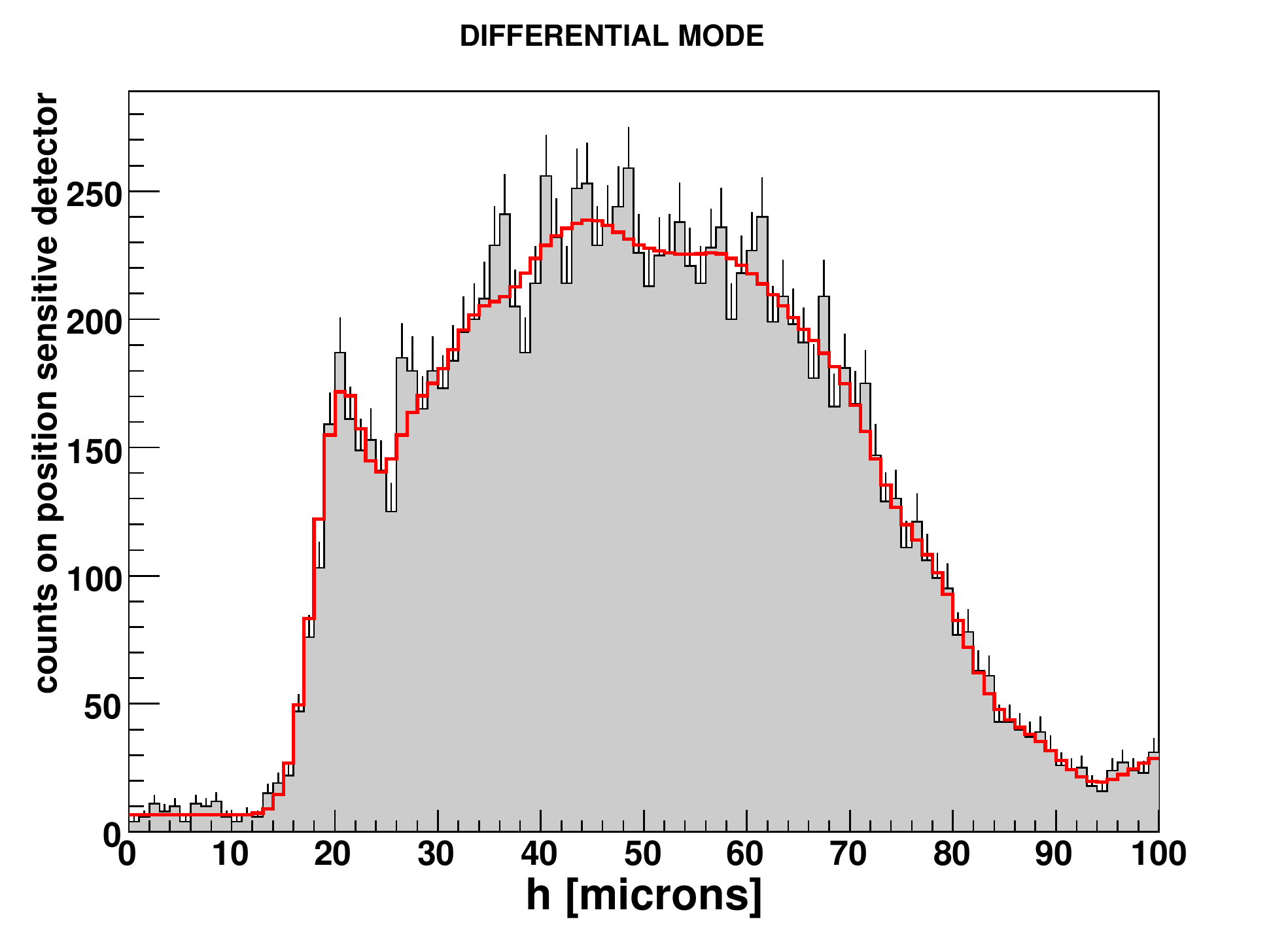}
\caption{
Histogram of the neutron vertical position measured with the nuclear track position sensitive detector. 
A dip corresponding to the first node of the excited wavefuctions is clearly visible. 
The figure is taken from \textcite{Pignol2009}.
 \label{photo}
}
\end{figure}

A first version of this experiment has been performed at the PF2 ultracold neutron source in 2005. 
A plastic nuclear track detector coated with a thin $^{235}$U conversion layer was used. 
Absorption of an ultracold neutron in the uranium induces a fission. 
One of the two fission products is emitted in the plastic and the ionization produces a latent track. 
After removing the conversion layer and etching the plastic with NaOH, the tracks become visible with an optical microscope. 
The result of this measurement is shown in Fig. \ref{photo}. 
A fit of the data allows to extract the populations of the quantum states, the spatial resolution of the detector, the background level, the height offset and the $z_0$ parameter \cite{Pignol2009}. 
The result is $z_0 = (6.0 \pm 0.2) \ \micron$, it is more accurate than the transmission-through-the-slit method. 
This first measurement was not ideal because the populations of quantum states up to $\ket{10}$ was quite high. 
Improvements are definitely possible, by refining the purity of the selection of the state $\ket{2}$. 
We aim at taking better pictures with the GRANIT instrument. 

\para{Gravity Resonance Spectroscopy. }
Imagine we prepare an initial quantum state $\ket{k}$. 
It is possible to induce a transition $k \rightarrow l$ to the state $\ket{l}$ 
with a periodic excitation of frequency $f_{kl} = (E_k - E_l)/ 2 \pi \hbar$. 
We define\footnote{Notice that a classical ball dropped from the height $z_0$ bounces with a frequency of $\pi \times f_0$. }
\begin{equation}
f_0 = \frac{m g z_0}{2 \pi \hbar} \approx 145 \ {\rm Hz}, 
\end{equation}
the transition frequencies write
\begin{equation}
f_{k l} = f_0 ( \epsilon_k - \epsilon_l ). 
\end{equation}
For the first few quantum states, the transition frequencies lie in the audio-range, 
for example we expect a resonance associated to the $1 \rightarrow 2$ transition at the frequency $f_{21} = 254$~Hz. 
Conveniently, the audio frequency range is easily accessible by electrical as well as mechanical oscillators. 
This allows direct experimental access to the energy spectrum of the neutron quantum bouncer, a technique called \emph{Gravity Resonance Spectroscopy} (GRS). 

Let us recall the quantum mechanics of a Rabi resonance, at the basis of the GRS. 
Assume we apply the external harmonic perturbation $V(t) = \hat{V} \cos(\omega t)$, 
with the excitation frequency $\omega/2 \pi$ close to a transition frequency $f_{k l}$. 
It means that the detuning $\delta \omega = \omega - 2 \pi f_{k l}$ should be small compared to the angular frequencies of the transitions in the spectrum. 
Assume also that the bouncer is prepared initially in the sate $\ket{\psi(0)} = \ket{k}$. 
In this case, the two level approximation holds and the system can be described by $\ket{\psi(t)} = a(t) \ket{k} + b(t) \ket{l}$. 
Solving the time-dependent Schr\"odinger equation
\begin{equation}
i \hbar \frac{d}{dt} \ket{\psi(t)} = \left( \frac{\hat{p}^2}{2m} + mg \hat{z} \right) \ket{\psi(t)} + \hat{V} e^{i \omega t} \ket{\psi(t)}
\end{equation}
we deduce the quantum amplitudes $a(t)$ and $b(t)$. 
Finally the transition probability $P_{k \rightarrow l}(t) = |b(t)|^2$ is given by the \emph{Rabi formula}: 
\begin{equation}
P_{k \rightarrow l}(t) = \frac{\sin^2 (\sqrt{\delta \omega ^2 + \Omega^2} \ t /2)}{1 + \delta \omega ^2 / \Omega^2}, 
\end{equation}
where
\begin{equation}
\Omega = \frac{1}{\hbar} \bra{l} \hat{V} \ket{k}
\end{equation}
is the Rabi angular frequency that characterizes the strength of the excitation. 
The duration $t$ available for the excitation is an important parameter. 
A longer duration is desirable because it corresponds to a narrower resonance line. 
In flow through mode, neutrons pass a transition region of typically 16~cm at a speed of 4~m/s, corresponding to a duration of 
$t_0 = 40$~ms (these parameters are those of the system built for the first stage of GRANIT). 
With the duration of the excitation fixed, one should set the excitation strength such that $\Omega t_0 = \pi$. 
Therefore, the transition probability is maximum at resonance: $P_{k \rightarrow l}(t_0) = 1$ when $\delta \omega = 0$. 
The measurement consists in recording the presence of the state $\ket{l}$ (or the disappearance of the state $\ket{k}$) for different values of the excitation frequency $\omega/2 \pi$. 
It forms a resonance curve with a peak at $\omega/2 \pi = f_{k l}$. 
The width (FWHM) of the peak is $\Delta f = \Omega / \pi = 1 / t_0$. 

There are two practical possibilities to do the excitation (i) using a vibrating mirror (ii) using a magnetic force. 
GRS with vibrating mirror has been realized at the PF2 UCN beamline by \textcite{Jenke2011,Jenke2014}. 
The technique with a magnetic excitation is in preparation with the GRANIT instrument. 
We will come back to this subject in detail later. 

\para{Testing the equivalence principle. }
We learn at school that it is impossible to tell the mass of a body by observing how it falls down. 
Newton's law of motion for a falling body is $m_i a = m_g g$, 
where $a$ is the acceleration of the body, 
$g$ is the gravity strength, 
$m_i$ is the inertial mass, $m_g$ is the gravitational mass. 
If we assume that both masses are equal then we find that the acceleration of all bodies is the same, equal to $g$. 
The hypothesis ($m_i = m_g$) is called the {\bf weak equivalence principle}, 
the observable consequence is called {\bf the universality of free fall}. 

Surprisingly, the weak equivalence principle does not imply the universality of free fall in the case of the quantum bouncer. 
This is because the mass in the Schr\"odinger equation \eqref{schrodinger} does not simplify, contrary to Newton's law of motion. 
By measuring the properties of the quantum bouncer we can tell the mass. 
Indeed, from Eq. \eqref{z0} we derive $m = \hbar/\sqrt{2 g z_0^3}$. 
Using the result of the photograph Fig. \ref{photo} we measure the neutron mass: 
$m c^2 = 910 \pm 50 \ {\rm MeV}$. 

\begin{figure}
\centering
\includegraphics[width=0.93\linewidth]{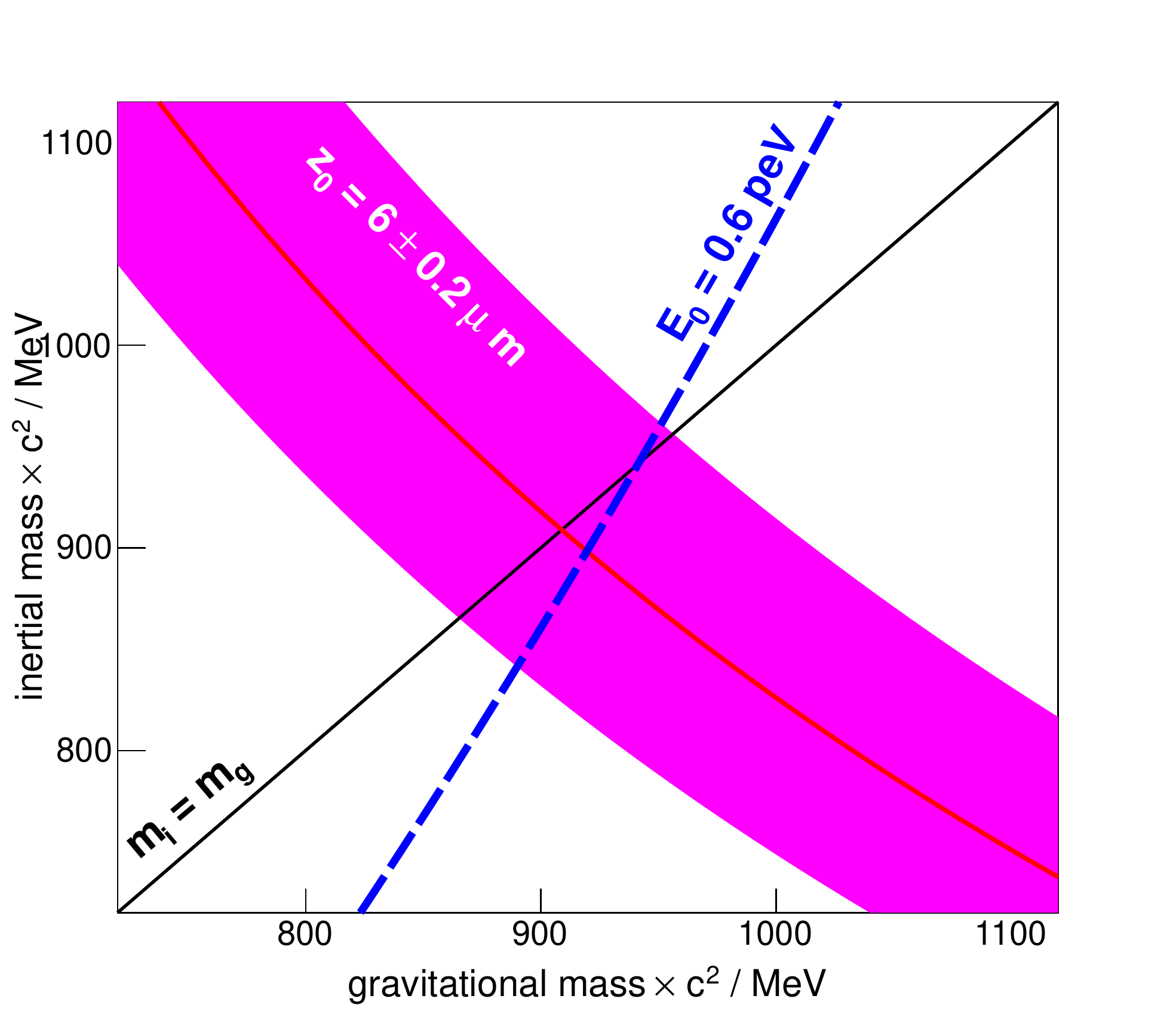}
\caption{
Inertial mass versus gravitational mass of the neutron. 
The diagonal line corresponds to the weak equivalence principle $m_i = m_g$. 
The pink band represents the $1 \sigma$ region deduced from the analysis of the photograph of the quantum states. 
The blue dashed line is an iso-$E_0$ line. 
 \label{EPtest}
}
\end{figure}

It is therefore of interest to investigate the question of the equivalence principle with the quantum bouncer in more details. 
Distinguishing the inertial and gravitational masses, the Schr\"odinger equation  becomes
\begin{equation}
\label{schrodinger2}
- \frac{\hbar^2}{2m_i} \frac{d^2}{dz^2} \psi_k + m_g g z \psi_k = E_k \psi_k, \quad \psi_k(0) = 0.
\end{equation}
In this framework, the size of the wavefunctions is governed by the parameter
\begin{equation}
z_0 = \left( \frac{\hbar^2}{2 m_i m_g g} \right)^{1/3}. 
\end{equation}
It means that the measurements of the spatial properties of the quantum states are sensitive to the product $m_i m_g$. 
Figure \ref{EPtest} interprets the result of the first photograph in the $m_i, m_g$ plane. 
Next, the energies of the quantum levels are $E_k = E_0 \epsilon_k$ with 
\begin{equation}
E_0 = m_g g z_0 = \left( \frac{m_g^2}{m_i} \frac{g^2 \hbar^2}{2} \right)^{1/3}. 
\end{equation}
It is remarkable that the measurement of the energy levels is sensitive to another combination of the masses, namely $m_g^2/m_i$. 

Then, not only can we tell the mass of the neutron by observing the quantum bouncer, 
but we can measure \emph{separately} the inertial mass $m_i$ and gravitational mass $m_g$. 
For this, we need to measure the spatial feature of the quantum states to extract $z_0$ on the one hand 
and to perform the spectroscopy of the quantum states with GRS to extract $E_0$ on the other hand.

\subsection{Effects of the chameleon field}

Since the energy difference between the first quantum states is as low as about $1$~peV, 
the neutron quantum bouncer is a sensitive system to search for hypothetical new forces acting on the neutron at the vicinity of the mirror. 
Such a new force could be induced by the presence of the chameleon field. 
Since this force is attractive, it would shrink the wavefunctions of the quantum states and enlarge the energy spectrum. 

We have argued in \textcite{Brax2011} that extremely large coupling $\beta > 10^{11}$ 
would lead to new bound states at a distance of less than $2 \ \micron$ 
which is ruled out by the first transmission-through-the-slit experiment. 
The limit $\beta < 10^{11}$ is practically independent of the Ratra-Peebles index $n$. 
Next, \textcite{Jenke2014} analyzed their experiment with the vibrating mirror to derive limits on the chameleon coupling. 
They quote the limit $\beta > 6 \times 10^8$, also practically independent of $n$. 

Let us now analyse quantitatively how the properties of the quantum bouncer are modified in the presence of the chameleon field. 
As we have seen in part \ref{Chameleon}, if the chameleon is strongly coupled with matter ($\beta \gg 1$), 
a scalar field $\varphi(z)$ will be generated above the mirror. 
We have seen that the field $\varphi(z)$ is saturated: it is independent of $\beta$ as long as $\beta \gg 1$, it is also independent of the mass density of the mirror. 
When no absorber is placed above the mirror the scalar field is given by Eq. \eqref{OnePlate}. 
Then, the neutron feels the potential
\begin{equation}
V(z) = m g z + \beta \frac{m}{m_{\rm Pl}} \varphi(z). 
\end{equation}
The chameleon term in the potential has the form
\begin{equation}
\delta V (z) = \beta \frac{m}{m_{\rm Pl}} \varphi(z) = \beta v_n ( z / \lambda )^{\alpha_n}
\label{deltaV}
\end{equation}
with
\begin{equation}
v_n = \frac{m}{m_{\rm Pl}} M_\Lambda \left( \frac{2+n}{\sqrt{2}} \right)^{\alpha_n} \approx 
\left( \frac{2+n}{\sqrt{2}} \right)^{\alpha_n} \times 10^{-21} \ {\rm eV}, 
\end{equation}
where $\lambda = \hbar c / M_\Lambda = 82 \ \micron$ and $\alpha_n = 2/(2+n)$. 

The task is now to calculate the shift of the energy states 
and the shrinking of the wavefunctions induced by the perturbation $\delta V$. 
We skip the details of the calculations (see appendix \ref{A2}) and we discuss the results. 

\begin{figure}
\centering
\includegraphics[width=0.93\linewidth]{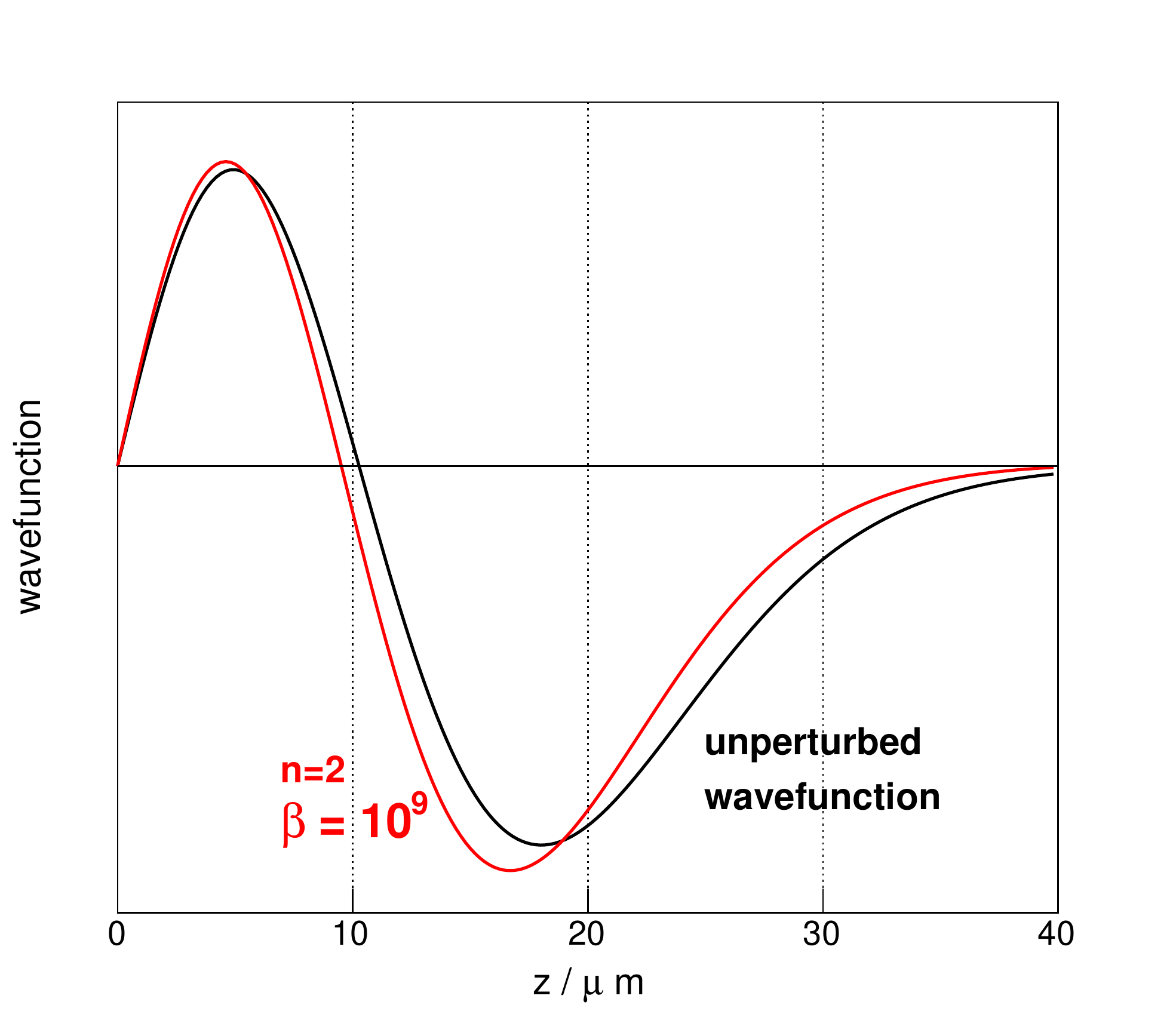}
\caption{
Wavefunction of the second quantum state $\psi_2(z)$. 
The black curve corresponds to the unperturbed case (no chameleon). 
The red curve is the wavefunction in the presence of the chameleon with $n=2$ and $\beta = 10^9$, calculated at the first order of perturbation theory. 
 \label{wavefunctionsChameleon}
}
\end{figure}

\para{Shrinking of the wavefunctions. }
At first order in perturbation theory, the correction of the wavefunctions is
\begin{equation}
\psi_k(z) = \psi_k^{(0)}(z) + \beta v_n (z_0/\lambda)^{\alpha_n} \sum_{l \neq k} \frac{ O_{kl}(\alpha_n) }{E_k - E_l} \psi_l^{(0)}(z). 
\end{equation}
In Fig. \ref{wavefunctionsChameleon}, we plot the modification of the second state for the coupling $\beta = 10^9$ 
using the results in the appendix \ref{A2} (the sum goes up to $l=5$). 
The effect of the chameleon is a shrinking of the wavefunction, as expected for an attractive force. 
The node of the wavefunction $h^0_2$ (which we will try to measure in GRANIT) is displaced to lower $z$. 

More quantitatively, a numerical study in the case $n=2$ shows that the displacement of the node is a linear function of $\beta$ for $\beta < 10^9$: 
\begin{equation}
h^0_2 = ( 10.3 - \beta \times 7.7 \times 10^{-10} ) \ \micron
\end{equation}
A precision of $0.1 \ \micron$, which is what we could reasonably get in forthcoming measurements with GRANIT, 
will correspond to $\beta \approx 10^8$. 
This method is less sensitive than the gravity resonance spectroscopy, as we will see. 

\para{Shifts of the transition frequencies. }
At first order in perturbation theory, the correction of the energy levels is 
\begin{equation}
\delta E_k = \beta v_n (z_0/\lambda)^{\alpha_n} O_{kk}(\alpha_n), 
\end{equation}
and the correction of the transition frequencies is
\begin{eqnarray}
\nonumber
\delta f_{kl} & = & \frac{\delta E_k - \delta E_l}{2 \pi \hbar} \\ 
              & = & \beta \frac{v_n}{2 \pi \hbar} (z_0/\lambda)^{\alpha_n} ( O_{kk}(\alpha_n) - O_{ll}(\alpha_n) ). 
\end{eqnarray}
For example, consider the shift of the resonance line $2 \rightarrow 1$ (expected at $f_{21} = 254$~Hz) due to the chameleon with the Ratra-Peebles index $n=2$. 
We find numerically
\begin{equation}
\delta f_{21} = \beta \times 3.7 \times 10^{-8} \ {\rm Hz}. 
\end{equation}
As we will see later, we aim at a precision better than $1$~Hz for the measurement of the $2 \rightarrow 1$ transition frequency 
with the flow-through resonance setup in GRANIT. 
This measurement will therefore be sensitive to a chameleon coupling of $\beta \approx 2 \times 10^{7}$. 

It is possible in principle to improve the precision on the transition frequency by increasing the interaction time $t$. 
We recall that the width of the resonance is given by $\Delta f = 1/t$, 
and the precision is approximately given by $\delta f = \Delta f / \sqrt{N}$, where $N$ is the number of counts in the resonance. 
Instead of a flow-through setup, one needs to trap the neutrons in quantum states. 
Ultimately, the precision is limited by the beta decay lifetime of the neutron $\tau_n \approx 880$~s. 
Therefore the ultimate precision is $\delta f \approx 1/(\tau_n \sqrt{N}) \approx 10^{-5}$~Hz (taking $N = 10000$). 
In conclusion, the measurement of the spectrum of the quantum states could potentially be sensitive to chameleon couplings down to $\beta = 10^2$ (a long way to go, though...).

\subsection{The GRANIT instrument}

\begin{figure}
\centering
\includegraphics[width=0.93\linewidth]{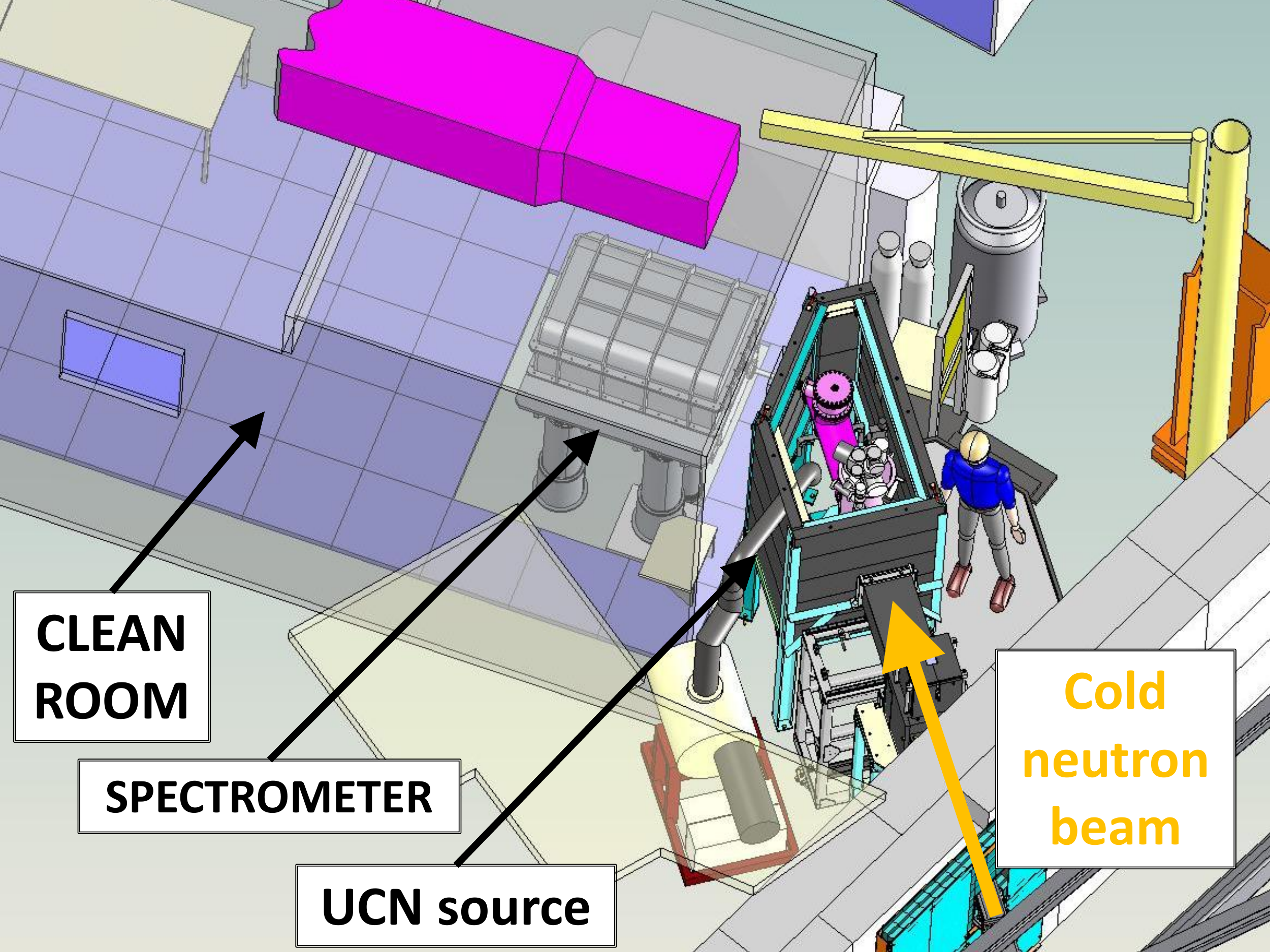}
\includegraphics[width=0.93\linewidth]{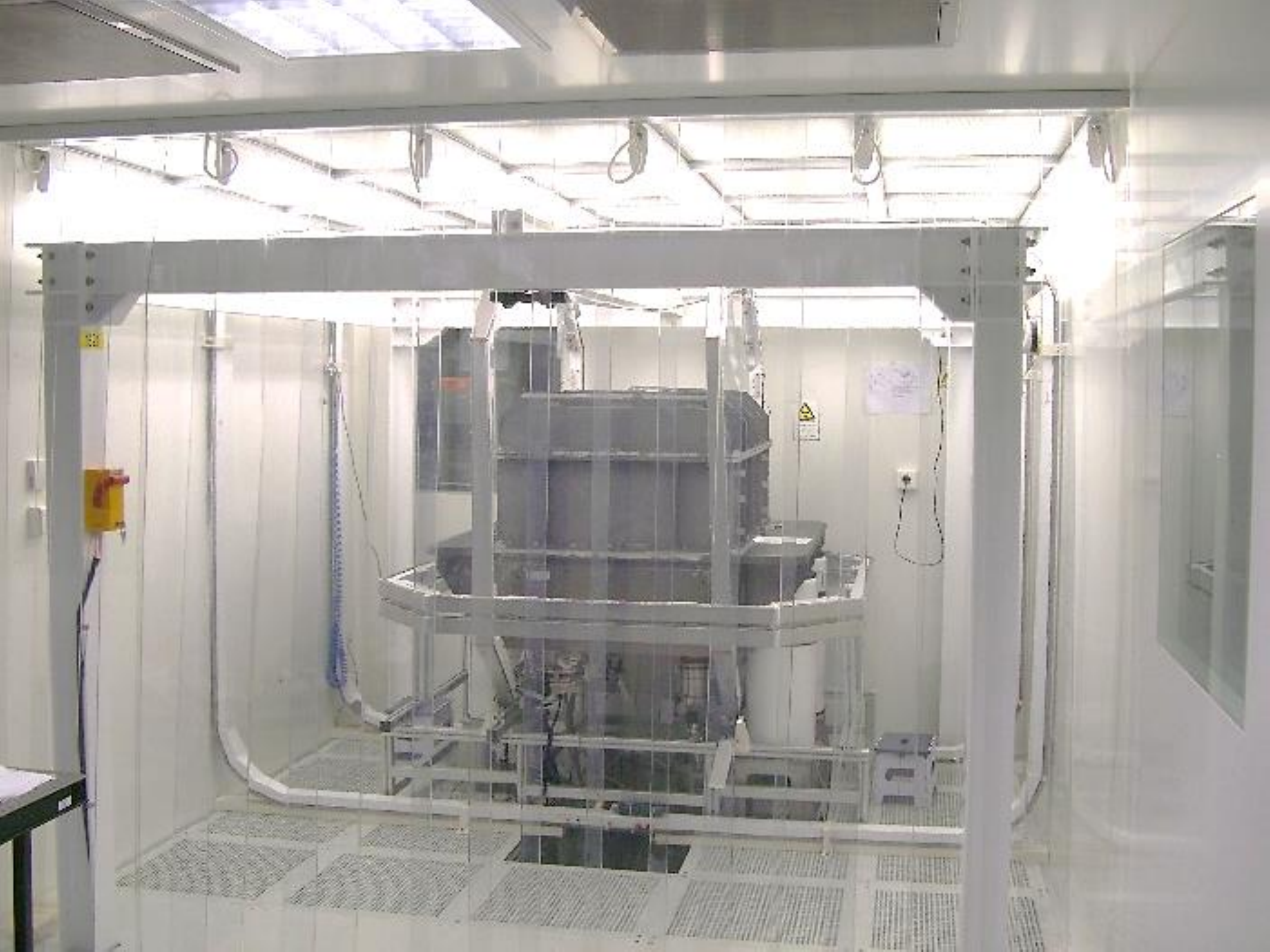}
\caption{
Top: schematics of the GRANIT instrument at the ILL H172A beam line. 
Bottom: picture of the spectrometer inside the clean room. 
 \label{GRANIT}
}
\end{figure}

GRANIT is an instrument dedicated to the study of the neutron quantum bouncer, now in the commissioning phase. 
For a description of the instrument see \textcite{SchmidtWellenburg2009,Baessler2011a,Baessler2011b,Roulier2015,Roulier2015b}. 
GRANIT is located at the level C of the ILL reactor in Grenoble. 
As shown in Fig. \ref{GRANIT}, the instrument comprises a superthermal $^4$He source 
(fed by the monochromatic cold neutron beam H172A), 
producing ultracold neutrons to be used in the spectrometer located in a ISO 5 class clean room. 

No attempt to describe the instrument in technical details will be made here, except for a few selected items. 
We will first describe the ultracold neutron source. 
Next, the setup for the magnetic excitation of resonant transitions will be explained. 

\para{The ultracold neutron source. }
Previous experiments observing the neutron quantum bouncer were all performed at the PF2 instrument, 
the ``first generation'' ultracold neutron source situated at the level D of the ILL reactor. 
``Next generation'' sources can potentially deliver many more ultracold neutrons, 
and GRANIT operates now a source based on the SUN1 prototype described in \textcite{Zimmer2011,Piegsa2014}. 

The concept of the source has been proposed by \textcite{Golub1977}: 
a volume filled with superfluid $^4$He is irradiated with a beam of cold neutrons. 
Neutrons can undergo inelastic scattering in the medium, losing energy while emitting a phonon. 
Given the dispersion relation of the phonons in superfluid helium, 
it turns out that a neutron having a wavelength of exactly $\lambda^* = 0.89$~nm 
transfers its entire energy to the phonon and comes at rest in the superfluid. 
Neutrons with a wavelength not too far from $\lambda^*$ come almost at rest. 
The process produces ultracold neutrons, with a volumic production rate (number of UCN produced per unit volume and unit time) of
\begin{equation}
P = \left. \frac{d \phi}{d \lambda} \right|_{\lambda^*} \times (3.0 \pm 0.2) \times 10^{-9} \ {\rm nm} \ {\rm cm}^{-1}, 
\end{equation}
where $d \phi / d \lambda$ is the differential flux in the cold neutron beam. 
This theoretical estimate is valid for neutrons created in the velocity range $0 < v < 5.6 \ {\rm m/s}$ \cite{Piegsa2014}, 
corresponding to the storage capability of a trap made with stainless steel walls
\footnote{
Our trap is made of BeO walls with a Fermi potential of $257$~neV, 
but a small portion is made of stainless steel with a Fermi potential of $184$~neV only. 
Since superfluid helium has a Fermi potential of 18~neV, the threshold energy for trapped neutrons is $E < 184 - 18$~neV, 
corresponding to a critical velocity of 5.6~m/s. 
}. 

When the reactor power is $58$~MW, H172A delivers $3 \times 10^9$ neutrons per second with a wavelength of 0.89~nm 
(the differential flux has a Gaussian shape with a RMS $\sigma = 0.021(1)$~nm), 
in a square section of $7 \times 7 \ {\rm cm}^2$ \cite{Roulier2015b}. 
In the continuity of the guide, the source itself consists in a tube with square section (inner dimensions $7\times7\times100 \ {\rm cm}^3$) with BeO walls filled with superfluid helium. 
The differential flux inside this production volume is reduced by absorption of cold neutrons in windows ($\times 0.9$) and by the divergence of the beam ($\times 0.8$), 
resulting in an effective flux of $d\phi/d\lambda |_{\rm eff} = 8 \times 10^8 \ {\rm cm}^{-2} {\rm s}^{-1} {\rm nm}^{-1}$. 
The UCN production inside our source is then expected to occur at a rate of $P \approx 2.3$~UCN/cm$^3$/s. 
In total the 4900~cm$^3$ should produce about $10,000$ ultracold neutrons per second. 

Those ultracold neutrons produced in the superfluid helium are stored by the BeO walls of the production volume. 
To extract the neutrons, the volume is connected to a chimney (a stainless steel tube) going up to a UCN valve. 
When the valve is closed UCNs accumulate in the volume. 
Above the valve there is a UCN guide going up, then a 90 degree bent, followed by an horizontal stainless steel guide connected to the spectrometer inside the clean room. 

\begin{figure}
\centering
\includegraphics[width=0.9\linewidth]{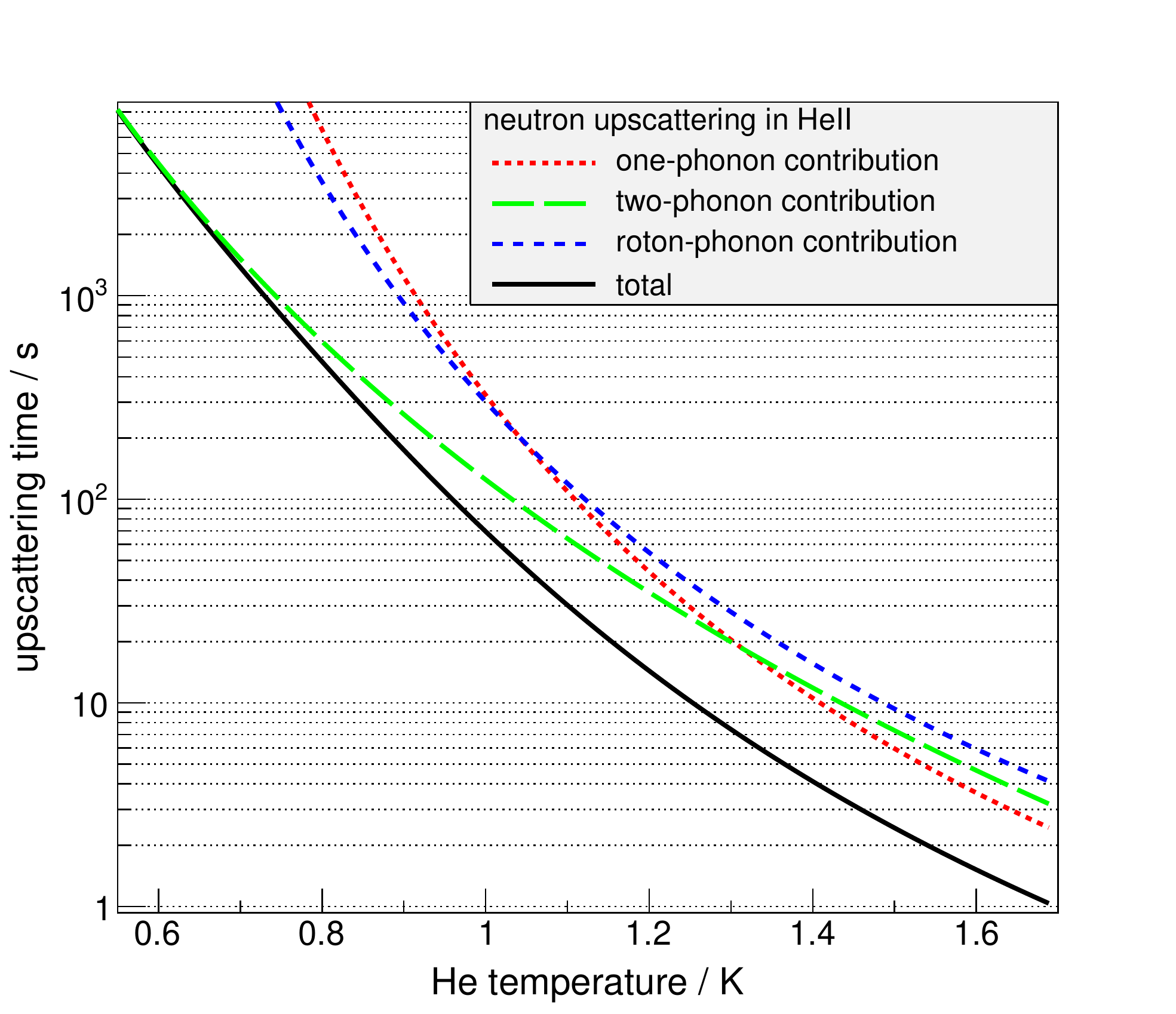}
\includegraphics[width=0.96\linewidth]{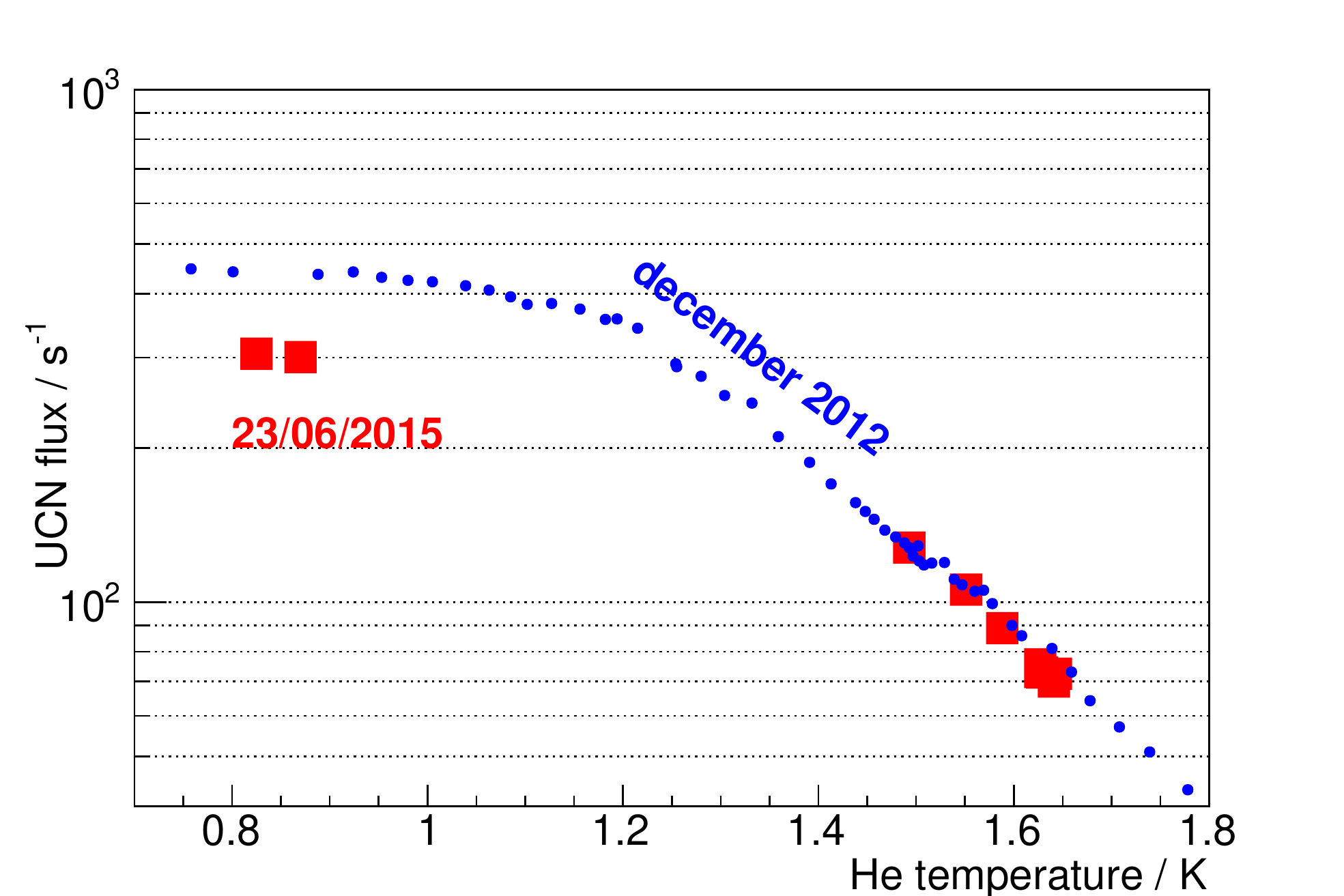}
\caption{
Top: Theoretical estimation of the lifetime due to upscattering $\tau_{\rm up}$ in superfluid helium, 
as a function of the temperature. 
Bottom: Flux of UCN at the measured at the exit of the horizontal extraction guide (a tube with inner diameter 3~cm) 
as a function of the temperature of the helium (with the UCN valve completely open). 
 \label{upscattering}
}
\end{figure}

For efficient accumulation of UCNs inside the source, the storage time $\tau$ should be as long as possible 
in order to maximize the UCN density $\rho = P \tau$ at saturation. 
The loss rate $1/\tau$ inside the source has several contributions: 
\begin{equation}
1/\tau = 1/\tau_n + 1/\tau_{\rm wall} + 1/\tau_{^3{\rm He}} + 1/\tau_{\rm up}
\end{equation}
where $\tau_n \approx 880$~s is the beta decay lifetime, 
$\tau_{\rm wall}$ corresponds to the losses at wall collision, 
$\tau_{^3{\rm He}}$ is due to the absorption by residual traces of $^3$He in the bath of normally pure $^4$He, 
and $\tau_{\rm up}$ corresponds to upscattering by thermal excitations in superfluid helium. 
Inconveniently, this last contribution forces us to operate the source at temperatures below $1$~K. 
According to the theoretical estimation by \textcite{Golub1979}, the temperature dependent upscattering rate is
\begin{equation}
1/\tau_{\rm up} = A e^{-E^*/T} + B \ T^7 + C T^{3/2} e^{- \Delta / T}
\end{equation}
where the first term corresponds to one-phonon absorption with $A = 500 \ {\rm s}^{-1}$, 
the second term corresponds to two-phonon process with $B = 0.008 \ {\rm s}^{-1}$, 
the last term corresponds to roton-phonon process with $C = 18 \ {\rm s}^{-1}$. 
Helium temperature $T$ is expressed in Kelvin, 
$E^* = 12$~K is the phonon energy corresponding to $\lambda^*$ and $\Delta = 8.6$~K. 
The time $\tau_{\rm up}$ is plotted as a function of the temperature in Fig. \ref{upscattering} (top). 
We see the dramatic effect of the temperature in Fig. \ref{upscattering} (bottom): 
we measured the flux of UCNs at the exit of the source as a function of the temperature. 
The flux is significantly reduced when the temperature is above 1~K. 
Even at low temperature we only count 400 UCNs per second. 
Some losses are due to absorption of neutrons in the vacuum separation window (a $15 \ \micron$ thick titanium foil situated in the horizontal guide) 
and the entrance window of the detector (another $15 \ \micron$ thick titanium foil). 
Also, the storage time at low temperature is only about $30$~s, which is not very long compared to the emptying time of $13$~s \cite{Roulier2015}. 
There are also certainly some more losses which are being investigated. 

\begin{figure}
\centering
\includegraphics[width=0.99\linewidth]{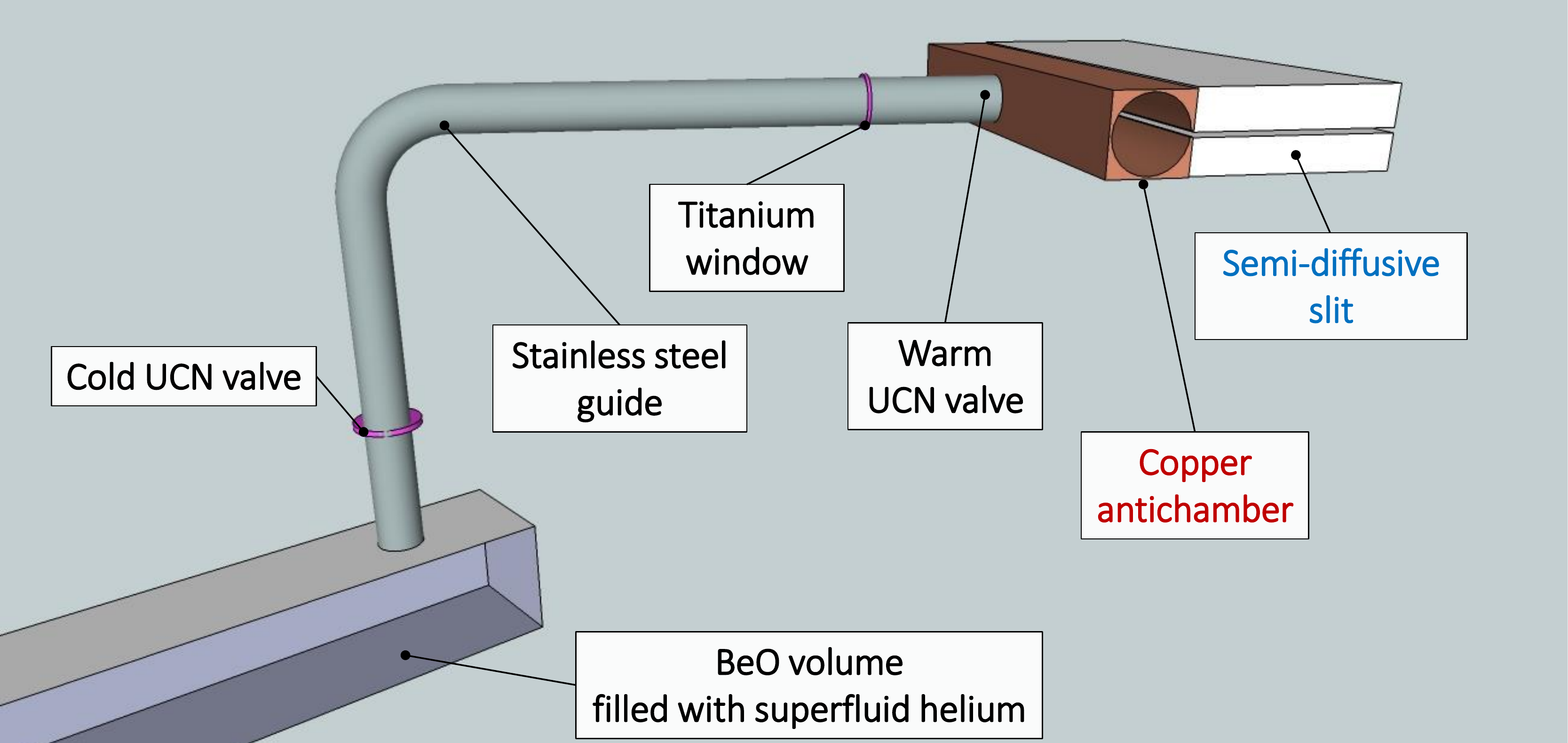}
\caption{
Scheme of the connection of the source to the spectrometer. 
Not shown: container of the superfluid, thermal screens, vacuum chamber of the source, 
lead biological shield, wall of the clean room, vacuum chamber of the spectrometer, support of the copper antichamber and semi-diffusive slit. 
 \label{GRANIT3}
}
\end{figure}

Let us describe a little bit what's inside the spectrometer. 
As shown in Fig. \ref{GRANIT3}, the horizontal guide is connected 
to a cylindrical storage volume labeled ``copper antichamber'' 
(in the test mentioned before, a detector was placed instead of the antichamber). 
The $30$~cm long cylinder has a diameter of $4$~cm, it is made of pure copper (Fermi potential $168$~neV). 
A small aperture in the antichamber leads to a semi-diffusive slit made of a bottom reflective mirror and a top diffusing mirror. 
Both mirrors are coated with diamond like carbon. 
The idea is that neutrons with too large vertical velocity in the slit are diffused back in the antichamber. 
On the contrary, neutrons with small vertical velocity, those bouncing on the bottom mirror without touching the roof, are transmitted. 
In a first test of this system conducted in December 2014, the opening of the semi-diffusive slit was set to $127 \ \micron$, 
preparing the first 20 quantum states. 
A $^3$He counter with an aluminium entrance window was placed immediately at the exit of the slit. 
Unfortunately due to a cryogenic failure it was only possible to get to $T = 1.3$~K for the temperature of the source. 
In this configuration, we counted $0.13$ neutrons per second out of the semi-diffusive slit. 
This is not yet as good as the experiments performed at PF2 (see Fig. \ref{DiscoveryILL}). 
By fixing the cryogeny and using a detector with improved efficiency for soft UCNs (i.e. with a titanium entrance window for example) things will get better. 
Progress is under way to deliver a decent flux of ultracold neutrons at the exit of the slit. 

\para{The magnetic excitation. }
Now I describe how we plan to measure the resonant transitions using the magnetic excitation as proposed in 
\textcite{Pignol2009,Kreuz2009,Pignol2014,Baessler2015}. 
An oscillating magnetic field gradient will generate the vertical force on the neutron to excite the system. 
Indeed, the motion of a neutron in a magnetic field ${\bf B}$ is governed by the potential $s \mu |{\bf B}|$
\footnote{This is valid when the variation of the magnetic field is slow compared to the Larmor frequency. 
In this case, called \emph{the adiabatic regime}, the spin follows the direction of the magnetic field and the neutron trajectory and the spin dynamics are decoupled.}, 
where $\mu = 60$~neV/T is the magnetic moment and $s=1$ for ``spin up'' neutrons and $s=-1$ for ``spin down'' neutrons. 
Classically, a vertical force is applied on the neutron by the field gradient $\partial_z |{\bf B}|$ 
(an homogeneous field does not exert any force, it acts only on the spin dynamics). 
Let us now assume a magnetic excitation of the form $|{\bf B}| = \beta z \cos \omega t$, 
the corresponding quantum mechanical potential reads
\begin{equation}
\hat{V}(t) = s \mu \beta \hat{z} \cos \omega t. 
\end{equation}
This is what we need to excite resonant transitions between the quantum states of the neutron bouncer. 

\begin{figure}
\centering
\includegraphics[width=0.93\linewidth]{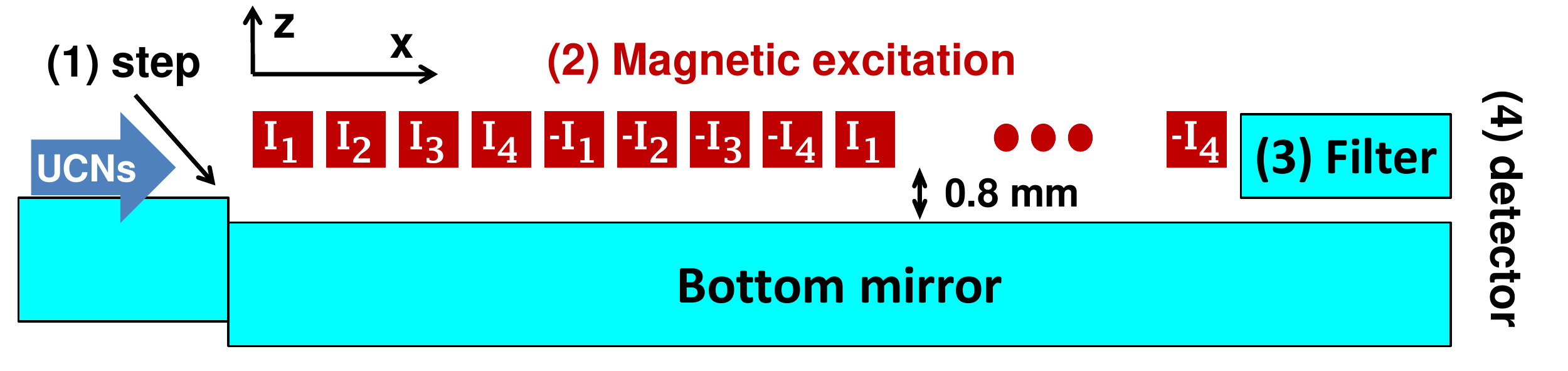}
\caption{
Sketch of the flow through setup. 
UCNs enter from the semi-diffusive slit on the left ; they fall down the step to depopulate the ground state (1), 
16~cm long transition region (2), 
9~cm long filter (3), and detector (4). 
 \label{sketchMagneticTransition}
}
\end{figure}

The sketch of the setup is shown in Fig. \ref{sketchMagneticTransition}. 
The spectroscopy method is conceptually the same as the famous 1939 Rabi resonance experiment. 
It is performed with four steps
(1) state preparation, 
(2) resonant transition, 
(3) state analysis, and
(4) detection of the transmitted flux. 

\begin{enumerate}
\item UCNs are first prepared in an excited state by going down 
a step (1) of height $15 \ \micron$. 
The population of the first state $\ket{1}$ is suppressed as compared to the populations of the excited states $\ket{2}, \ket{3}, \cdots$. 
\item Next, transitions between quantum states are excited with a periodic magnetic field gradient. 
The length of the transition region $L = 16$~cm corresponds to an average passage time $t_0 = 40$~ms
\footnote{We expect the $v_x$ velocity along the beam to be distributed with a mean value of 4~m/s and a standard deviation of 1.5~m/s.}. 
Two different schemes could be implemented in principle: the AC excitation ad the DC excitation. 
In the DC mode, the field gradient is static and spatially oscillating in the $x$ direction with a period of $d = 1$~cm. 
In this case, only neutrons with specific horizontal velocities meet the resonance condition. 
The deexcitation $2 \rightarrow 1$ is expected to occur at a frequency of $f_{21} = 254$~Hz corresponding to the resonant horizontal velocity of $v_{21} = d f_{21} = 2.54$~m/s (in the $3 \rightarrow 1$ case we expect $v_{31} = 4.62$~m/s). 
In the AC mode, the field gradient is spatially uniform and oscillating in time. 
One would then find the resonances by directly scanning the excitation frequency. 
\item A second horizontal mirror above the main mirror serves as a state filter. 
For a slit opening of about $25 \ \micron$, only neutrons in the ground state are accepted and higher quantum states are absorbed. 
\item Neutrons are finally detected at the exit of the filter. 
In the AC mode one should see a resonance in the transmitted flux as a function of the excitation frequency. 
In the DC mode, one needs to measure the horizontal velocity of transmitted neutrons. 
This is achieved by measuring the height of the neutrons after a 30~cm long free fall. 
\end{enumerate}

\begin{figure}
\centering
\includegraphics[width=0.93\linewidth]{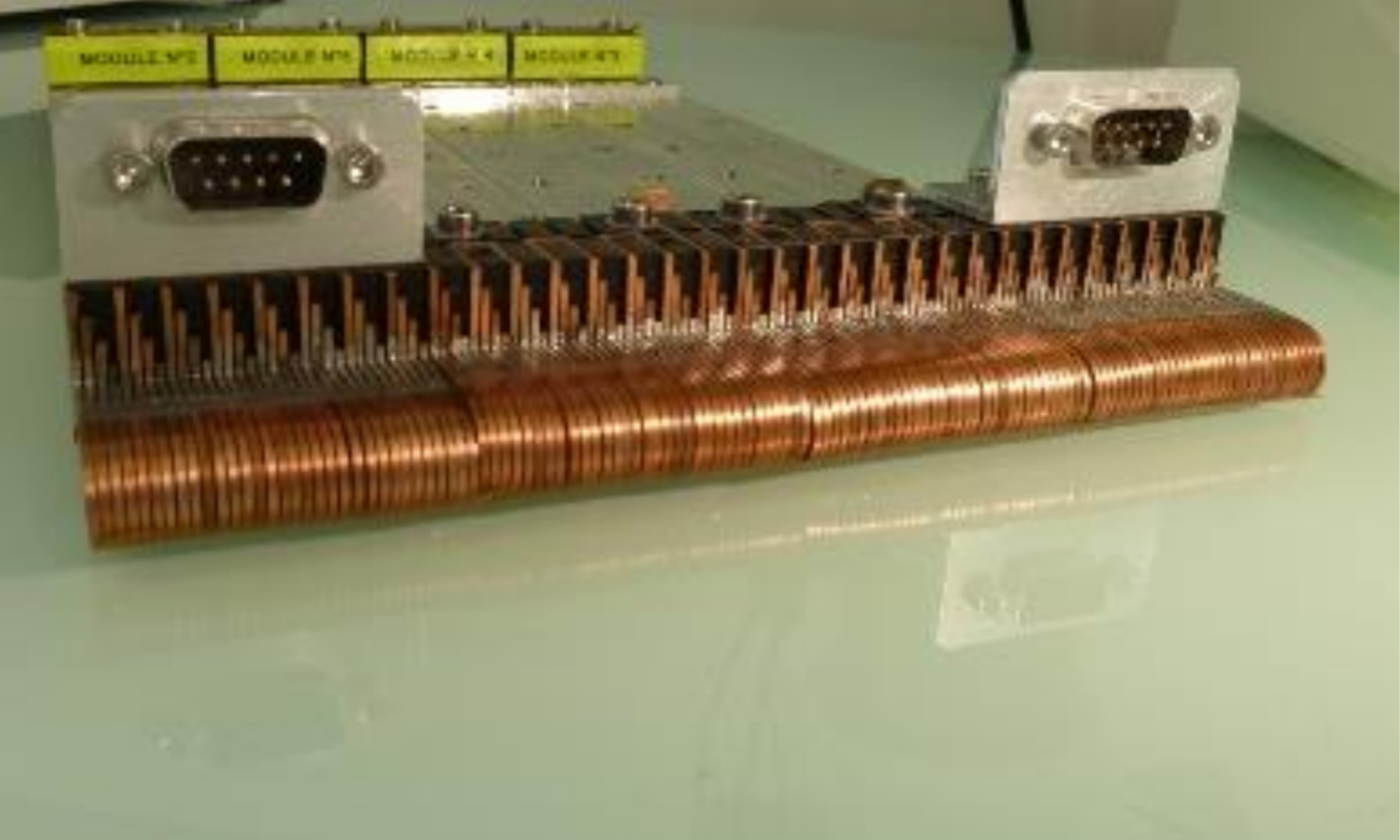}
\caption{
Picture of the wire array for the magnetic excitation. 
It is made of four modules, each one holding 32 adjacent wires with a section of 1~mm$^2$ each and a length of 30~cm in the $y$ direction. 
Adjacent wires are separated by a gap of 0.25~mm. 
 \label{wireArray}
}
\end{figure}

\begin{figure}
\centering
\includegraphics[angle=90,width=0.93\linewidth]{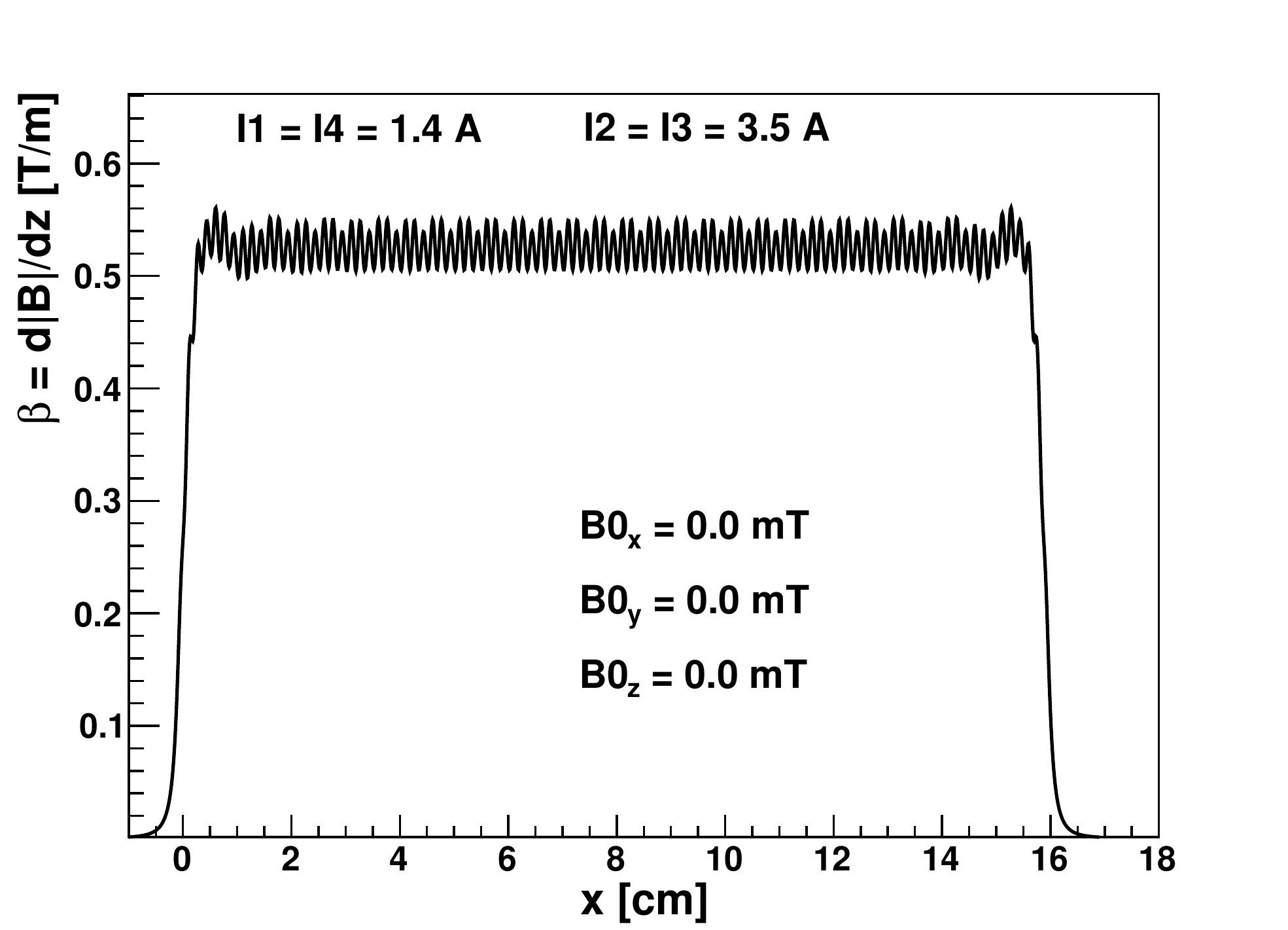}
\caption{
Calculation of the magnetic field gradient $\partial_z |{\bf B}|$ produced at the surface of the main mirror 
with currents set to $I_1 = I_4 = 1.4$~A, $I_2 = I_3 = 3.5$~A and without any external field applied. 
 \label{gradient}
}
\end{figure}

We need to apply an oscillating gradient with an amplitude of $\beta = 0.22$~T/m to induce the $2 \rightarrow 1$ transition and $\beta = 0.74$~T/m for $3 \rightarrow 1$ \cite{Pignol2014}. 
This is done with an array of 128 copper wires with square section arranged as shown in Fig. \ref{sketchMagneticTransition}. 
The array will be placed above the horizontal mirror in region (2) at a distance of 0.8~mm from the mirror. 
The real system shown in Fig. \ref{wireArray}. 
Electrical connectors are arranged so that the following 8-periodic pattern current could be applied: 
$I_1, I_2, I_3, I_4, -I_1, -I_2, -I_3, -I_4, I_1, \dots$ and thus the magnetic field produced by the array will be 1~cm periodic in the $x$ direction. 
It is possible to tune the currents in the 4 circuits $I_1, I_2, I_3, I_4$ to create a homogeneous gradient at the surface of the mirror, 
as shown in Fig. \ref{gradient}. 
The configuration of the gradient depends on the external magnetic field: 
by applying a strong external field (of the order of 2~mT) on top of the field created by the wire array, 
one generates an oscillating gradient in the $x$ direction. 
The wire array is therefore a versatile device to generate the field gradient that can be used for the AC as well as for the DC excitation modes. 

We will apply currents up to 5~A in the circuits, corresponding to Joule heating of up to 25~W 
(the resistance of each circuit is $0.3 \ \Omega$). 
Since the wire array will be operated in vacuum, an active cooling system is necessary to dissipate the heat. 
We built a heat exchanger with circulating cold gaseous nitrogen and tested it successfully in vacuum. 

\begin{figure}
\centering
\includegraphics[angle=90,width=0.93\linewidth]{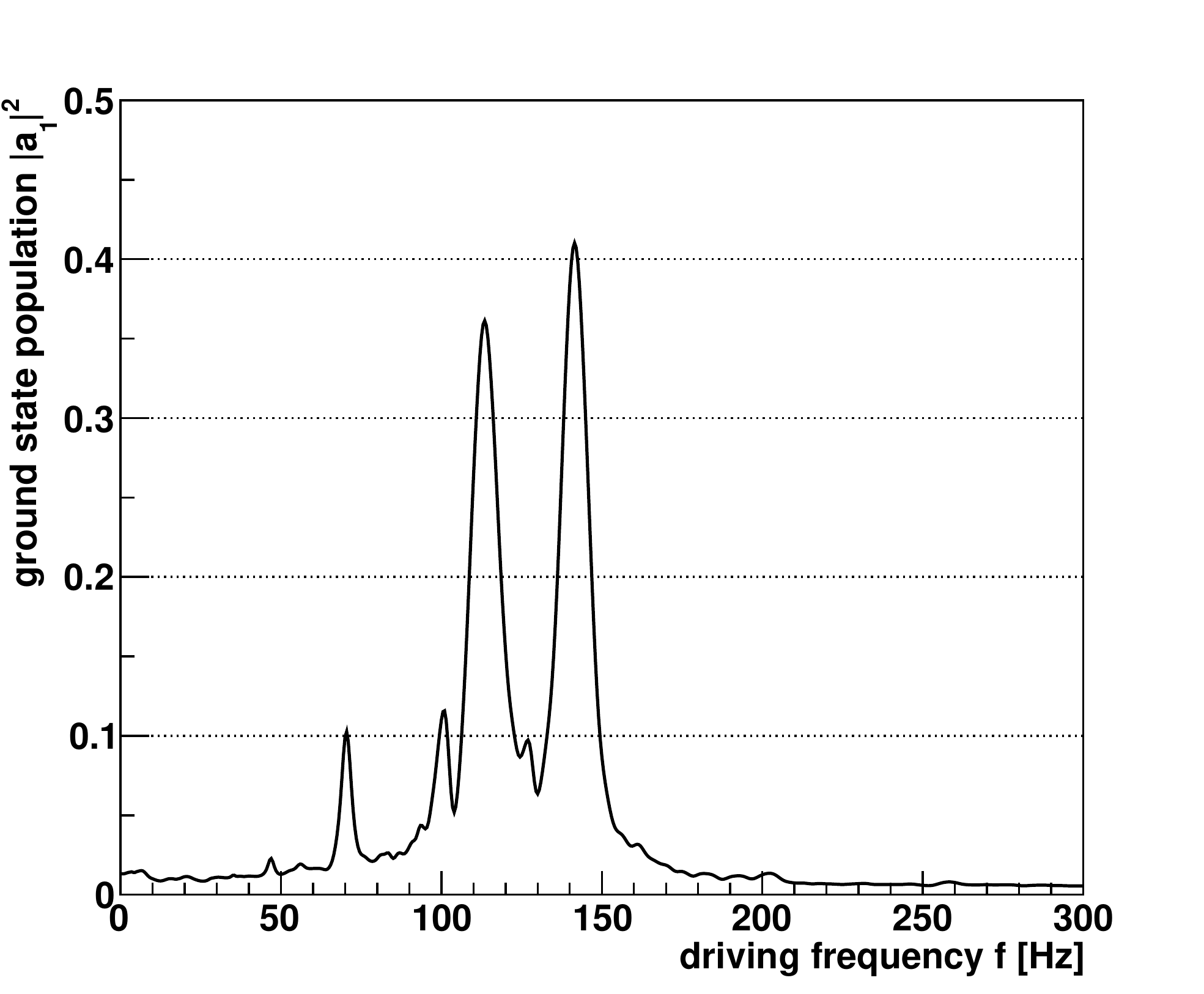}
\caption{
Calculation of the expected resonance curve in the AC mode of excitation. 
The probability of the $2 \rightarrow 1$ transition is plotted as a function of the driving frequency. 
 \label{resonance}
}
\end{figure}

In the AC mode, we apply AC currents in the wires. 
Note that the gradient $\beta$ is not a linear function of the currents. 
In particular, the gradient always satisfies $\beta > 0$, because the field $|{\bf B}|$ is always weaker away from the wire. 
As a result, the excitation frequency (i.e. the frequency of $\beta(t)$ seen by the neutrons) 
is twice the driving frequency (i.e. the frequency of the AC current in the wires). 
The complete system of excitation has been simulated in \textcite{Pignol2014} by solving the time-dependent Schr\"odinger equation 
with realistic function $\beta(t)$ taking into account the non-linearities. 
The expected resonance curve is shown in fig. \ref{resonance}. 
We see a double resonance, one corresponding to spin up neutrons, the other to spin down. 
This is a shift called \emph{Stern-Gerlach shift} \cite{Baessler2015} caused by the unavoidable constant (time-independent) component of $\beta(t)$. 
Fortunately this shift cancels when averaging the two resonant frequencies. 

With a statistics of 10,000 neutrons counted during a scan of the driving frequency between $50$~Hz and $200$~Hz, 
it will be possible to extract the resonance frequency $f_{21}$ with a precision better than $1$~Hz.

\separe

In the long term, increasing the precision of the transition frequencies can in principle be achieved by storing the neutrons in the quantum states 
for long times compared with the flow-through arrangement. 
One would use a trap consisting of an horizontal mirror and vertical walls, the excitation (vibration or magnetic) could be applied during a longer storage period. 
The control of the environmental noises such as vibrations is an issue because it could provoke unwanted transition between quantum states, 
but the lifetime of the quantum states in the trap could reach a few tens of seconds \cite{Pignol2009,Codau2012}. 
The filling and emptying of such a trap with limited losses of neutrons remains an open problem. 

Another interesting route for the future is the development of position sensitive detectors with micrometric resolution to ``see'' the quantum states. 
We have mentioned an attempt to take pictures of the quantum states using a uranium-coated plastic nuclear track detector. 
It achieves a good resolution of $2 \ \micron$, but it lacks real-time readout. 
More recently, pixel detectors based on CCD (charge coupling device) coated with $^6$Li or $^{10}$B have shown to offer both a good spatial resolution 
and real-time readout capability \cite{Kawasaki2010}. 
Pursuing in this way will certainly lead to clean measurements of the wavefunctions. 

In conclusion, progress are foreseen both in the spectroscopy of the quantum bouncer 
and in the measurement of the spatial features of the quantum states. 
These will permit to probe new forces acting on the neutron and
to test the equivalence principle in a quantum context.

\newpage
\section{Conclusion}

It is fascinating that experiments at low energy with neutrons can address three big questions about the Universe: 
\begin{enumerate}
\item Does the standard theory of primordial nucleosynthesis predict the correct amount of helium and deuterium? 
\item Was the asymmetry between matter and antimatter generated during the electroweak phase transition? 
\item Is Dark Energy a dynamical field interacting with matter? 
\end{enumerate}

The measurement of the neutron lifetime helps to answer question 1. 
Experiments will continue until the inconsistencies of the present data are resolved. 

The answer to the second question is either yes or no. 
By improving the measurement of the neutron electric dipole moment we could be able to reach a definitive answer in the next decade. 

\begin{figure}
\centering
\includegraphics[width=1.\linewidth]{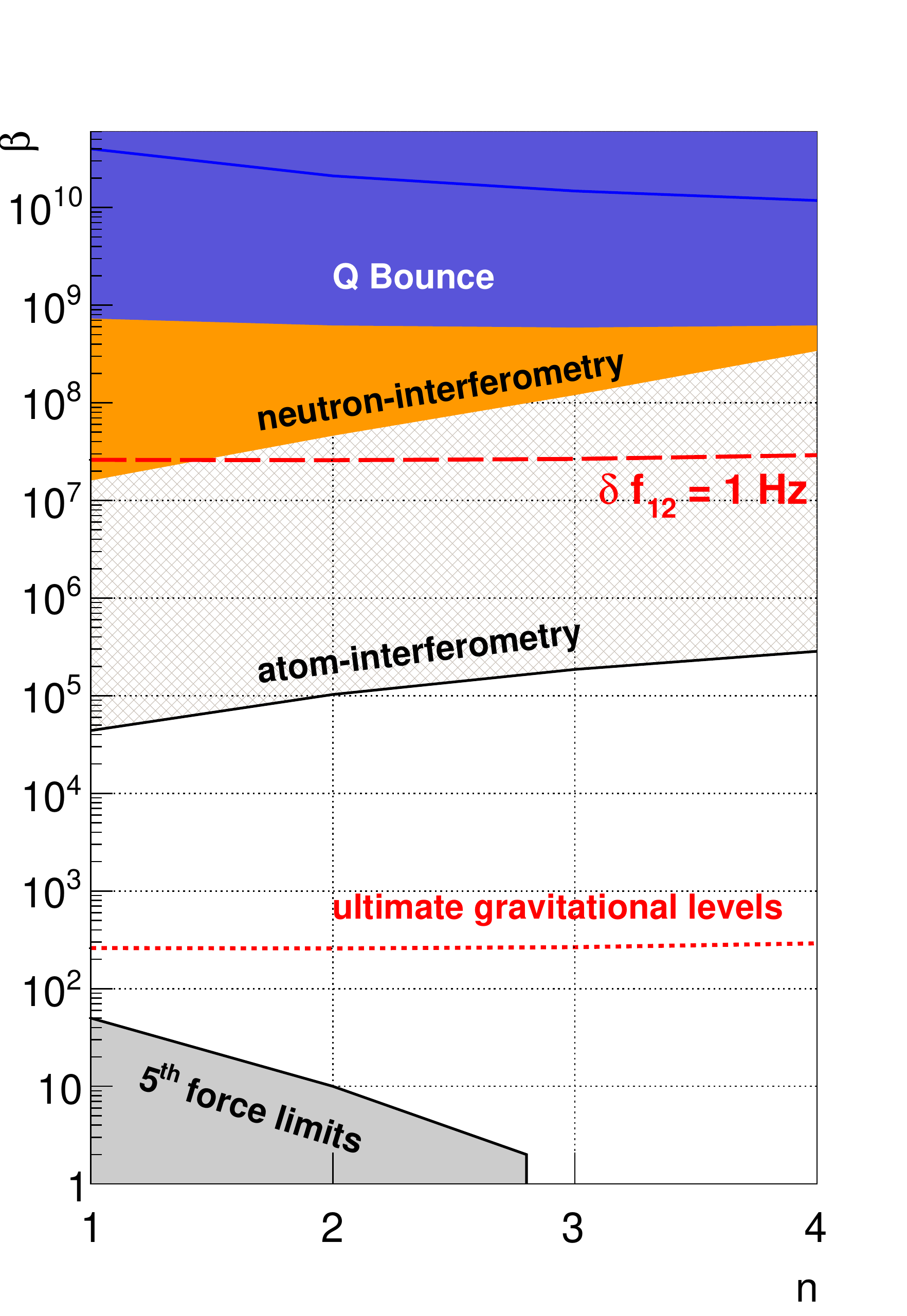}
\caption{
Exclusion regions (95 \% C.L.) in the chameleon parameter space (Ratra-Peebles index $n$ and matter coupling $\beta$). 
The blue zone is excluded by gravitational resonance spectroscopy of the quantum bouncer \cite{Jenke2014}. 
The red dashed line corresponds to the sensitivity of the flow-though setup in GRANIT, 
the red dotted line corresponds to the ultimate sensitivity. 
The orange zone is excluded by the neutron-interferometry experiment \cite{Lemmel2015}. 
The grey zone is excluded by the Eot-Wash experiment to search for a short range force \cite{Upadhye2012}. 
The hatched region is excluded by the atom-interferometry experiment \cite{Hamilton2015}. 
 \label{exclusion}
}
\end{figure}

The third question is the main focus of this work. 
We have looked in details at the strongly coupled Khoury-Weltman chameleon. 
The present constraints are compiled in Fig. \ref{exclusion}. 
We have identified two methods to probe the chameleon with neutrons. 
The first is neutron interferometry. A pilot experiment performed in 2013 was described in this work. 
The second is using the quantum states of bouncing neutrons. 
Ultimately, it has the potential to explore a significant part of the parameter space. 

Searching for Dark Energy in the lab is a new and rapidly developing field of research. 
Less than ten years after the chameleon was proposed, 
a major part of the parameter space is already excluded by experiments searching for deviations of the inverse square law of gravity at short distances, 
neutron experiments, and very recently atom-interferometry. 
Besides the chameleon, there are other theoretical ideas related to Dark Energy to be explored in the lab. 
For sure there are other neutron experiments to be invented. 

\clearpage

\appendix
\section{Lattice calculation of the chameleon field}
\label{A1}
We describe here the calculation of the chameleon field in the rectangular cell used in the neutron-interferometry experiment. 
The dimensionless field $F = \varphi / M_\Lambda$ satisfies the equation
\begin{equation}
\frac{\partial^2 F}{dy^2} + \frac{\partial^2 F}{dz^2} = K^2 \left( - \frac{n}{F^{n+1}} + \frac{\beta \rho \hbar^3 c^5}{M_\Lambda^3 m_{\rm Pl}} \right)
\end{equation}
with $K = \frac{M_\Lambda}{\hbar c} = (0.082 \ {\rm mm})^{-1}$. 
The second term on the right hand side of the equation is proportional to the helium pressure ($\rho$ is the mass density of helium in the cell). 
In addition, we assume that the field is saturated in the walls of the cell, therefore we impose $F = 0$ at the boundary of the cell. 

Let us now describe the numerical method adopted to calculate the solutions. 
We used a finite difference on a 2D grid of step $h$. 
The Laplace operator on the grid is taken to be
\begin{equation}
\left( \Delta F \right)_{i,j} = \frac{1}{h^2} \left( F_{i+1,j} + F_{i,j+1} + F_{i-1,j} + F_{i,j-1} - 4 F_{i,j} \right), 
\end{equation}
then the discrete chameleon equation becomes
\begin{equation}
4 F_{i,j} = N_{i,j} - h^2 S_{i,j}
\end{equation}
where 
\begin{eqnarray*}
N_{i,j} & = & F_{i+1,j} + F_{i,j+1} + F_{i-1,j} + F_{i,j-1} \\
S_{i,j} & = & K^2 \left( - \frac{n}{F_{i,j}^{n+1}} + \frac{\beta \rho (\hbar c)^3}{M_\Lambda m_{\rm Pl}} \right). 
\end{eqnarray*}
We used the iterative Jacobi relaxation method with SOR (Successive Over-Relaxation). 
The algorithm starts from an initial field $F^{(0)}$ (in our case the initial field is constructed from the 1D profile \eqref{TwoPlates}) and iterates according to
\begin{equation}
F_{i,j}^{(k+1)} = (1- \omega) F_{i,j}^{(k)} + \frac{\omega}{4} \left( N_{i,j}^{(k)} - h^2 S_{i,j}^{(k)} \right). 
\end{equation}
The boundary condition ($F_{i,j}=0$ for $i,j$ at the boundary) is enforced at each iteration. 
If the sequence converges, the fixed point necessarily satisfies the discrete chameleon equation. 
The SOR parameter $\omega$ usually serves to accelerate the convergence (in this case one sets $\omega > 1$). 
In our case, we observed numerical instabilities spoiling the convergence when calculating the bubble for high helium pressure. 
The numerical instabilities were fixed by setting the SOR parameter to $\omega = 0.8$, at the price of a slower convergence. 
Finally we used a step of $h = 0.2$~mm and iterate the algorithm 30000 times. 
A calculation takes about 10 min in a normal computer. 
Figure \ref{bubble} shows selected results of such calculations. 

\section{Perturbation of the quantum states}
\label{A2}
We use the Rayleigh-Sch\"odinger perturbation theory to calculate the effect of the chameleon field on the quantum states of the bouncing neutron. 
The perturbation $\delta V$ is given by Eq. \eqref{deltaV}. 
The shift $\delta E_k$ of the energy level $E_k$ is given at first order by
\begin{equation}
\delta E_k = \bra{k} \delta V(z) \ket{k}
\end{equation}
where $\ket{k}$ are the unperturbed stationary states. 
Likewise, the modification of the wavefunction at first order of the perturbation theory is
\begin{equation}
\psi_k(z) = \psi_k^{(0)}(z) + \sum_{l \neq k} \frac{\bra{l} \delta V(z) \ket{k} }{E_k - E_l} \psi_l^{(0)}(z). 
\end{equation}
where $\psi_k^{(0)}(z) = \braket{z}{k}$ are the unperturbed wavefunctions. 

It is useful to define the matrix elements of the power law operator as: 
\begin{equation}
O_{kl} (\alpha) = \bra{l} (z/z_0)^\alpha \ket{k}
 = \int_0^\infty \psi_l(z) \psi_k(z) (z/z_0)^\alpha dz. 
\end{equation}
These matrix elements are calculated numerically for the first five quantum states and for the Ratra-Peebles indices $n=1,2,3,4$. 
We provide the result in the form of four symmetric $5 \times 5$ matrices: 
\begin{eqnarray}
O(\alpha_1 = 2/3) & = &
\left( \begin{array}{ccccc}
1.31 & 0.36 & 0.13 & 0.08 & 0.05 \\
     & 1.89 & 0.47 & 0.17 & 0.09 \\
     &      & 2.31 & 0.56 & 0.20 \\
     &      &      & 2.65 & 0.63 \\
     &      &      &      & 2.94
\end{array} \right) \\
O(\alpha_2 = 1/2) & = &
\left( \begin{array}{ccccc}
1.22 & 0.25 & 0.10 & 0.06 & 0.04 \\
     & 1.59 & 0.31 & 0.12 & 0.07 \\
     &      & 1.85 & 0.35 & 0.14 \\
     &      &      & 2.05 & 0.38 \\
     &      &      &      & 2.22
\end{array} \right) \\
O(\alpha_3 = 2/5) & = &
\left( \begin{array}{ccccc}
1.16 & 0.19 & 0.08 & 0.05 & 0.03 \\
     & 1.44 & 0.23 & 0.09 & 0.06 \\
     &      & 1.62 & 0.25 & 0.10 \\
     &      &      & 1.76 & 0.27 \\
     &      &      &      & 1.88
\end{array} \right) \\
O(\alpha_4 = 1/3) & = &
\left( \begin{array}{ccccc}
1.13 & 0.15 & 0.07 & 0.04 & 0.03 \\
     & 1.35 & 0.18 & 0.08 & 0.05 \\
     &      & 1.49 & 0.20 & 0.08 \\
     &      &      & 1.60 & 0.21 \\
     &      &      &      & 1.68
\end{array} \right). 
\end{eqnarray}

\clearpage
\bibliographystyle{lemieuxdumonde}
\bibliography{biblioHDR}

\end{document}